\documentclass[acmlarge, nonacm]{acmart}

\usepackage{balance}
\usepackage{placeins} 
\usepackage{color, colortbl, xcolor}
\usepackage{url}
\usepackage{subcaption}
\usepackage{textcomp}
\usepackage{soul}
\usepackage{multirow}
\usepackage{enumitem}
\usepackage{mathtools}
\usepackage{siunitx}
\usepackage{tabularx}

\usepackage{booktabs} 
\usepackage{longtable,booktabs}

\usepackage{multicol}
\usepackage{natbib}  

\usepackage{arydshln}
\usepackage[breakable]{tcolorbox} 

\definecolor{linkColor}{RGB}{6,125,233}
\definecolor{green}{rgb}{0.0, 0.65, 0.31}
\definecolor{bleudefrance}{rgb}{0.19, 0.55, 0.91}
\definecolor{ceruleanblue}{rgb}{0.16, 0.32, 0.75}
\definecolor{grey}{HTML}{969696}
\definecolor{violet}{HTML}{756bb1}
\definecolor{dgrey}{HTML}{01665e}
\definecolor{lgrey}{HTML}{5ab4ac}
\definecolor{dgreen}{HTML}{005a32}
\definecolor{purple}{HTML}{ae017e}
\definecolor{soothinggreen}{HTML}{C1FFC1}

\definecolor{editCol}{HTML}{0000FF}
\definecolor{maskCol}{HTML}{c51b7d}
\definecolor{lrColor}{HTML}{8856a7}
\definecolor{trColor}{HTML}{d01c8b}
\definecolor{ctColor}{HTML}{4dac26}
\definecolor{brickred}{HTML}{f03b20}

\definecolor{improveCol}{HTML}{4dac26}
\definecolor{worsenCol}{HTML}{d01c8b}

\definecolor{DarkBlue}{HTML}{00008B}
\definecolor{mscolor}{HTML}{01665e}
\definecolor{nmscolor}{HTML}{bf812d}
\definecolor{lgreen}{HTML}{ccece6}
\definecolor{dolive}{HTML}{308014}

\colorlet{tablerowcolor4}{gray!50} 

\newcommand*{\textlabel}[2]{%
  \edef\@currentlabel{#1}
  \phantomsection
  #1\label{#2}
}

\colorlet{tableheadcolor}{gray!25} 
\colorlet{tablerowcolor}{gray!10} 
\colorlet{tablerowcolor2}{gray!45} 
\colorlet{tablerowcolor3}{gray!12} 
\colorlet{tablerowgreen}{soothinggreen!75} 

\newcolumntype{a}{>{\columncolor{tablerowcolor}}r}
\definecolor{aicolor}{HTML}{018571}
\definecolor{occolor}{HTML}{ff7799}

\definecolor{aicolor}{HTML}{fc8d62}
\definecolor{occolor}{HTML}{253494}

\newif{\ifhidecomments}
\hidecommentsfalse 
\ifhidecomments
    \newcommand{\soorya}[1]{}
    \newcommand{\pragya}[1]{}
    \newcommand{\shivam}[1]{}
    \newcommand{\chandrayee}[1]{}
    \newcommand{\talayeh}[1]{}
    \newcommand{\koustuv}[1]{}
\else
    \newcommand{\soorya}[1]{\textbf{\small\sffamily{\textcolor{DarkBlue}{[#1 -- Soorya]}}}}
    \newcommand{\pragya}[1]{\textbf{\small\sffamily{\textcolor{orange}{[#1 -- Pragya]}}}}
    \newcommand{\shivam}[1]{\textbf{\small\sffamily{\textcolor{violet}{[#1 -- Shivam]}}}}
    \newcommand{\chandrayee}[1]{\textbf{\small\sffamily{\textcolor{dgreen}{[#1 -- Chandrayee]}}}}
    \newcommand{\talayeh}[1]{\textbf{\small\sffamily{\textcolor{blue}{[#1 -- Talayeh]}}}}
    \newcommand{\koustuv}[1]{\textbf{\small\sffamily{\textcolor{brickred}{[#1 -- Koustuv]}}}}
  \fi

\renewcommand{\textrightarrow}{$\rightarrow$}

\colorlet{tableheadcolor}{gray!25} 

\definecolor{neutralCol}{HTML}{dd1c77}
\definecolor{neutralGreen}{HTML}{31a354}
\definecolor{NewBlue}{HTML}{1879ba}
\definecolor{bleudefrance}{rgb}{0.19, 0.55, 0.91}  
\definecolor{AfTrColor}{HTML}{0868ac}  
\definecolor{BfTrColor}{HTML}{a8ddb5}  

\definecolor{AfCtColor}{HTML}{b10026}  
\definecolor{BfCtColor}{HTML}{fd8d3c}

\graphicspath{ {figures/} }

\newcommand{\para}[1]{\vspace{0.2em}\noindent\textbf{\textit{#1}~}}

\usepackage{booktabs}
\usepackage{tabularx}
\usepackage{lipsum} 
\usepackage{multirow}

\AtBeginDocument{%
  \providecommand\BibTeX{{%
    \normalfont B\kern-0.5em{\scshape i\kern-0.25em b}\kern-0.8em\TeX}}}

\copyrightyear{2026}
\acmYear{2026}
\setcopyright{acmlicensed}
\acmJournal{IMWUT}

\begin{document}

\title{MEM-STRESS: Longitudinal Field Study of a Wearable-Informed AI Chatbot for Stress Management}

\author{Soorya Ram Shimgekar}
\orcid{0000-0002-1110-9699}
\affiliation{%
 \institution{University of Illinois Urbana-Champaign}
 \city{Urbana}
 \state{IL}
 \country{USA}}
 \email{sooryas2@illinois.edu} 

\author{Pragya Singh}
\orcid{0000-0003-3933-2224}
\affiliation{%
 \institution{Aalto University}
 \city{Espoo}
 \country{Finland}}
 \email{pragya.singh@aalto.fi} 

\author{Shivam Sharma}
\orcid{0009-0006-0732-5870}
\affiliation{%
 \institution{Aalto University}
 \city{Espoo}
 \country{Finland}}
 \email{shivam.sharma@aalto.fi} 

\author{Chandrayee Roy}
\orcid{0000-0002-5517-1056}
\affiliation{%
 \institution{Aalto University}
 \city{Espoo}
 \country{Finland}}
 \email{chandreyee.roy@aalto.fi}

\author{Koustuv Saha}
\orcid{0000-0002-8872-2934}
\affiliation{%
 \institution{University of Illinois Urbana-Champaign}
 \city{Urbana}
 \state{IL}
 \country{USA}}
 \email{ksaha2@illinois.edu}

 \author{Antti Oulasvirta}
\orcid{0000-0002-2498-7837}
\affiliation{%
 \institution{Aalto University}
 \city{Espoo}
 \country{Finland}}
 \email{antti.oulasvirta@aalto.fi}

\author{Talayeh Aledavood}
\orcid{0000-0002-0110-5694}
\affiliation{%
 \institution{Aalto University}
 \city{Espoo}
 \country{Finland}}
 \email{talayeh.aledavood@aalto.fi}

\renewcommand{\shortauthors}{Shimgekar et al.} 



\begin{abstract}

Stress is an increasingly prevalent concern, creating a need for accessible and personalized support. 
Stress-management-chatbots primarily rely on what users report, which may not capture their physiological changes. 
Wearables capture these changes, but chatbots rarely use them as memory across conversations. 
We therefore present MEM-STRESS, a stress-support chatbot that turns wearable ring's physiological-data into a layered memory of daily records, baseline deviations, LLM-generated summaries, and uses it to drive multi-session conversations grounded in Dialectical Behavior Therapy(DBT). In a four-week, non-randomized study with 60 adults, we compared MEM-STRESS(Treatment) with a chatbot remembering only past conversations(Control). 

\para{The memory shaped conversations:} The Treatment chatbot referenced participants' physiological-features (e.g., activity in 71.8\% of sessions) and suggested more body-related DBT-skills than Control (14.5\% vs. 5.7\%), with Treatment participants also using slightly more body-related words (1.76\% vs. 1.45\%). 
Treatment participants praised personalization three-times as often (44.4\% vs. 14.3\%), more frequently reported reflecting differently on stress (69.0\% vs. 46.4\%, not significant), and described gaining insight into their stress signals. Their perceived stress fell 14.2\% from baseline versus 3.6\% for Control, though the group difference was not significant. 
Grounding stress-support chatbots in layered physiological memory is thus feasible and may improve perceived personalization, motivating larger randomized studies.

\end{abstract}



\maketitle


\section{Introduction}

Stress is among the most common everyday health concerns~\cite{apa2023stress}: academic, financial, and social pressures accumulate over weeks, and persistent stress erodes sleep, cardiovascular and immune function, and performance~\cite{mcewen1998protective, cohen2007psychological, pascoe2020impact}. Yet most stressed people never receive support: professional care is limited by cost, stigma, and provider shortages~\cite{kazdin2011rebooting, clement2015stigma}, and many prefer to cope alone~\cite{ebert2019barriers, eisenberg2007helpseeking}. The UbiComp (Ubiquitous Computing) and HCI (Human-Computer Interaction) communities have therefore designed low-barrier, self-guided everyday tools~\cite{wang2014studentlife, nahumshani2018justintime}.

Chatbots are a promising form of such support: from early cognitive-behavioral agents~\cite{fitzpatrick2017woebot, inkster2018wysa, fulmer2018tess} to empathic ones built on LLMs (Large Language Models)~\cite{stade2024large, sharma2023human}, AI (Artificial Intelligence) chatbots reduce distress~\cite{li2023systematic, heinz2025therabot}. Yet users trust them only with some topics~\cite{purohit2026conditional}, report generic or inconsistent answers and forgetfulness~\cite{ma2023understanding, siddals2024happened, song2025typing}, and LLMs risk misinterpreting and mirroring users~\cite{obradovich2024opportunities, sharma2024towards}. At root, such chatbots \emph{know only what the user types, and only while the conversation lasts}. Memory-augmented agents retain information across sessions~\cite{packer2023memgpt, zhong2024memorybank, chhikara2025mem0, rasmussen2025zep}, and long-term memory increases health-related self-disclosure~\cite{jo2024understanding}, but they recall what users \emph{said}, not how their bodies \emph{have been responding}: a chatbot cannot know its user slept only four hours last night unless the user notices and mentions it.

Consumer wearables such as smart rings continuously record this missing context: sleep, heart rate, HRV (Heart Rate Variability), physical activity, and recovery. These signals are tied to stress: stress and sleep influence each other~\cite{akerstedt2006psychosocial, kalmbach2018impact}, heart rate rises under acute stress~\cite{campbell2012acute}, and reduced HRV marks stress and diminished emotion regulation~\cite{kim2018stress, thayer2012metaanalysis}. UbiComp research has mostly used such signals to \emph{detect} stress~\cite{hovsepian2015cstress, sano2018identifying, martinez2022alignment, 10.1145/3699755, smets2018largescale, schmidt2018wesad}. Detection alone, however, does not tell a person what to do next~\cite{mishra2020reproducibility}.
LLMs have begun using wearable data for health and well-being: they answer personal health questions~\cite{kim2024healthllm, merrill2026phia}, coach sleep and fitness~\cite{khasentino2025phllm, jorke2025gptcoach, wang2025exploring}, support reflection on fitness-tracker data~\cite{stromel2024narrating}, and turn multimodal sensing into narratives about how a person is doing~\cite{xu2026lens}. These systems, however, treat wearable data as one-off input and were studied in benchmarks, expert evaluations, or small deployments, not stress-support conversations.

These limitations highlight key gaps in both theory and practice. Theoretically, stress is both perceived and embodied~\cite{cohen1983global, mcewen1998protective}. However, conversational memory models capture only verbal disclosures, while passive-sensing research treats physiological signals as isolated inputs or detection targets rather than accumulating context. Practically, even memory-augmented chatbots remain blind to gradual physiological shifts. Beyond preliminary explorations~\cite{neupane2025wearable, dongre2026exploring}, no stress-support agent has maintained a persistent \emph{memory} of wearable physiological features across conversations, nor have such systems been evaluated in longitudinal field deployments to determine how this memory functions and what value it adds. Addressing these gaps is critical because everyday stress accumulates gradually and often unnoticed; support reliant solely on self-reports risks overlooking manifest bodily strain, such as poor sleep. Just as contextualizing text with passive physical sensing enhances person-centered psychological predictions~\cite{saha2021person}, pairing user accounts with physiological signals enables chatbots to deliver richer, person-centered support.


To address these theoretical and practical gaps, we introduce \textit{MEM-STRESS}, a conversational agent designed for multi-session stress support through a persistent \emph{memory of physiological features}. Using wearable sensor data (here, an Oura Ring), the chatbot tracks physiological metrics against individual baselines and maintains a three-layered memory structure---\emph{episodic}, \emph{pattern}, and \emph{semantic} (Section~\ref{sec:memory})---alongside past conversation summaries. Grounded in this contextual memory and the user's self-report, MEM-STRESS guides brief reflective dialogues and recommends actionable Dialectical Behavior Therapy (DBT) skills~\cite{linehan2015dbt}. We selected DBT because its brief, structured exercises are well-suited for self-guided practice outside clinical settings and have demonstrated efficacy in university populations and digital tools~\cite{rizvi2014pilot, schroeder2018pocketskills}.

Integrating physiological memory offers three distinct advantages. First, it can surface subtle, unmentioned physiological shifts, such as accumulating sleep loss. Second, it provides an objective longitudinal record independent of participant recall, potentially mitigating self-report recency and salience biases~\cite{shiffman2008ecological} as well as LLMs' tendency to mirror users~\cite{sharma2024towards}. Third, it allows the agent to tailor skill recommendations to current physiological states---for instance, suggesting body-calming techniques over problem-solving when recovery is compromised, addressing a personalization gap noted in existing tools~\cite{borghouts2021barriers, kocaballi2019personalization}. However, whether these physiological cues effectively transfer to and enhance conversational dialogue remains an open question.

To evaluate this approach, we conducted a four-week field study with 60 adults experiencing everyday stress within a university community (Section~\ref{sec:recruitment}). Participants were assigned to two non-randomized conditions: a \textit{Treatment} group interacting with MEM-STRESS and a \textit{Control} group using a chatbot that retained only conversation history. Every two days, participants discussed a recent stressor, rated their momentary affect on the Valence-Arousal-Dominance (VAD) scale~\cite{mehrabian1996pad} before and after chatting, and completed the 10-item Perceived Stress Scale (PSS-10)~\cite{cohen1983global} at screening and after every fourth session.

We address three RQs (Research Questions):

\begin{itemize}
\item \textbf{RQ1 (Stress and Emotional Well-being):} Does a chatbot with a memory of physiological features help people manage their stress more than a chatbot that remembers only conversations?

\item \textbf{RQ2 (Conversation and Intervention):} How does the memory of physiological features impact the stress-support conversations?

\item \textbf{RQ3 (User Experience):} How do people experience a conversational chatbot that remembers their physiological features?
\end{itemize}

Our contributions are:

\begin{enumerate}
\item \textbf{A memory of wearable physiological features for a stress-support chatbot.} MEM-STRESS turns continuous wearable physiological features into a hierarchical memory that, with summaries of prior conversations, grounds DBT-informed stress support across sessions.

\item \textbf{Longitudinal field study on physiological memory in stress support.} A 4-week study with 60 participants (30/group; 772 sessions) collected 1,538 pre/post affect ratings and >1,000 Treatment ring-data days, tracking memory accumulation. To our knowledge, this is the first longitudinal field study of a physiology-grounded stress-support chatbot.

\item \textbf{Design knowledge for chatbots grounded in physiological features.} We derive design, ethical, and methodological implications: how to present physiological features, connect them to users' own accounts, and combine physiological and conversational memory.
\end{enumerate}
\section{Related Work}\label{section:rw}

\subsection{Passive Sensing for Stress}\label{sec:rw-sensing}

Retrospective self-report such as the PSS-10~\cite{cohen1983global} misses within- and across-day fluctuations; EMA (Ecological Momentary Assessment) and experience sampling reduce recall bias with momentary ratings~\cite{shiffman2008ecological, csikszentmihalyi2014validity} but still depend on what people notice and report. Momentary affect is often rated on the VAD (Valence, Arousal, Dominance) dimensions~\cite{russell1980circumplex, mehrabian1996pad}, for example with the pictorial SAM (Self-Assessment Manikin), which is quick enough to repeat before and after a brief intervention~\cite{bradley1994measuring}. Passive sensing captures stress without asking: smartphone sensing tracks students' stress, mental health, resilience, and academic performance~\cite{wang2014studentlife, nepal2024capturing}, and wearable physiological features track day-to-day self-reported stress~\cite{smets2018largescale}, classify high versus low stress better than phone features~\cite{sano2013stress, sano2018identifying}, and, from a consumer ring, relate to perceived stress~\cite{bloomfield2024predicting}. Yet in-the-wild detection remains difficult~\cite{can2019stress, mishra2020reproducibility}, largely owing to individual differences~\cite{xu2022globem, adler2022generalization}. Physiological and emotional stress responses also correspond only weakly~\cite{campbell2012acute}, and people have limited insight into their bodily signals~\cite{garfinkel2015knowing}; physiological features may thus add what self-report misses but should be read against each person's baseline and alongside their own account.

Most such work \emph{detects} stress rather than supporting \emph{reflection}; in the stage-based model of personal informatics, data collection alone does not produce reflection or action~\cite{li2010stage, choe2014understanding}. Reflection tools include wearable-triggered stressor logging~\cite{neupane2024momentary}, Fitbit-grounded reflective dialogues that changed physical activity (Reflection Companion)~\cite{kocielnik2018reflection}, and LLM-generated narratives of fitness-tracker data (Narrating Fitness)~\cite{stromel2024narrating}. Closest to ours, MindScape paired smartphone-sensed behavior with LLM-generated journaling prompts, improving college students' self-reported affect and self-reflection over eight weeks~\cite{nepal2024mindscape}, but uses behavior rather than physiological features such as HRV or recovery, and one-directional prompts rather than multi-turn conversation.

\subsection{Interventions for Stress Management}

We build on DBT, which teaches concrete skills in four modules: mindfulness, distress tolerance, emotion regulation, and interpersonal effectiveness~\cite{linehan2015dbt}. Skills use mediates DBT outcomes~\cite{neacsiu2010dbt}, brief skills formats show promise with university students~\cite{rizvi2014pilot, uliaszek2016group}, and skills have been delivered via smartphone coaching~\cite{rizvi2011dbtcoach} and a conversational app whose users reported less depression and anxiety over four weeks~\cite{schroeder2018pocketskills}. We ask what changes when a conversational DBT coach draws on physiological features.

Digital stress interventions yield small-to-moderate benefits~\cite{heber2017benefit} that shrink considerably after correcting for small-study bias~\cite{linardon2024stressapps}, with low real-world retention~\cite{baumel2019objective}; lack of personalization often hinders engagement~\cite{borghouts2021barriers}, yet personalization is rarely tested rigorously~\cite{hornstein2023personalization, kocaballi2019personalization}, and context-insensitive support can mismatch changing needs~\cite{bhattacharjee2023context}. Ecological momentary interventions tailor support to momentary states~\cite{heron2010ecological, myingermeys2016ecological}, the JITAI (Just-in-Time Adaptive Intervention) framework formalizes \emph{when}, \emph{what}, and \emph{how much} to provide~\cite{nahumshani2018justintime}, and systems deliver stress support conversationally~\cite{mauriello2021suite} or trigger it from sensed stress~\cite{sarker2017markers, paredes2014poptherapy}; in a four-week workplace study, sensor-triggered prompts yielded more completed interventions than scheduled ones, without significantly greater stress reduction~\cite{howe2022design}. These efforts largely decide \emph{when} to intervene; MEM-STRESS instead uses physiological features to shape \emph{what} support says.

\subsection{Conversational Agents for Mental Health and Memory-Augmented LLMs}\label{sec:rw-agents}

Conversational agents offer scalable mental health support: early reviews found preliminary but favorable evidence~\cite{vaidyam2019chatbots, abdalrazaq2020effectiveness}, and meta-analyses of randomized trials report modest-to-moderate efficacy, larger for generative agents~\cite{he2023conversational, li2023systematic}. LLM systems now support journaling~\cite{kim2024mindfuldiary}, cognitive restructuring~\cite{sharma2024facilitating}, behavior change~\cite{bhattacharjee2023procrastination}, conversational self-report~\cite{wei2024leveraging}, and public-health check-ins~\cite{jo2023understanding}, but users report harmful or inconsistent responses, over-dependence, and forgetting~\cite{ma2023understanding, siddals2024happened, song2025typing}, and reviews warn of inappropriate responses, stigma, and sycophantic mirroring~\cite{stade2024large, moore2025expressing, sharma2024towards}. These risks motivate us to ground responses in measured data and to examine where this helped or misfired.

Memory architectures that summarize, retrieve, and update past conversations~\cite{packer2023memgpt, park2023generative, zhong2024memorybank, chhikara2025mem0, rasmussen2025zep, wu2026balms} improve multi-session dialogue~\cite{xu2022beyond} and increase self-disclosure to public-health chatbots~\cite{jo2024understanding}, although long-term memory remains challenging~\cite{maharana2024evaluating}. Yet in these systems, memory consists only of conversation.

A few systems give agents sensed context. EMMA used smartphone-sensed mood to suggest emotionally appropriate micro-activities~\cite{ghandeharioun2019emma}, and LLM-based systems use sensed data to deliver psychotherapeutic interventions~\cite{nie2026caiti}, curb problematic smartphone use~\cite{wu2024mindshift}, gather older adults' health information~\cite{yang2024talk2care}, answer personal health questions from wearable data~\cite{kim2024healthllm, merrill2026phia}, and coach sleep, fitness, and physical activity~\cite{khasentino2025phllm, fang2024physiollm, jorke2025gptcoach}, though their accuracy in interpreting passive-sensing data for mental health is modest~\cite{englhardt2024classification, xu2024mentalllm}. HealthGuru, an LLM sleep chatbot using smart-ring data, was compared with a baseline chatbot over eight weeks with 16 participants~\cite{wang2025exploring}, and a duoethnography pairing wearable-triggered stress events with LLM chatbots found only a fraction of flagged events warranted intervention~\cite{neupane2025wearable}. Most such systems are prototypes, benchmarks, or small deployments, and those addressing stress are preliminary; to our knowledge, none has tested whether a memory of physiological features changes stress-support conversations in a multi-week field deployment against a chatbot with conversational memory. MEM-STRESS fills this gap by combining a hierarchical memory of physiological features with a summary of prior conversations.
\section{MEM-STRESS Design}\label{sec:design}

\begin{figure}[t]
\centering
\includegraphics[width=\linewidth]{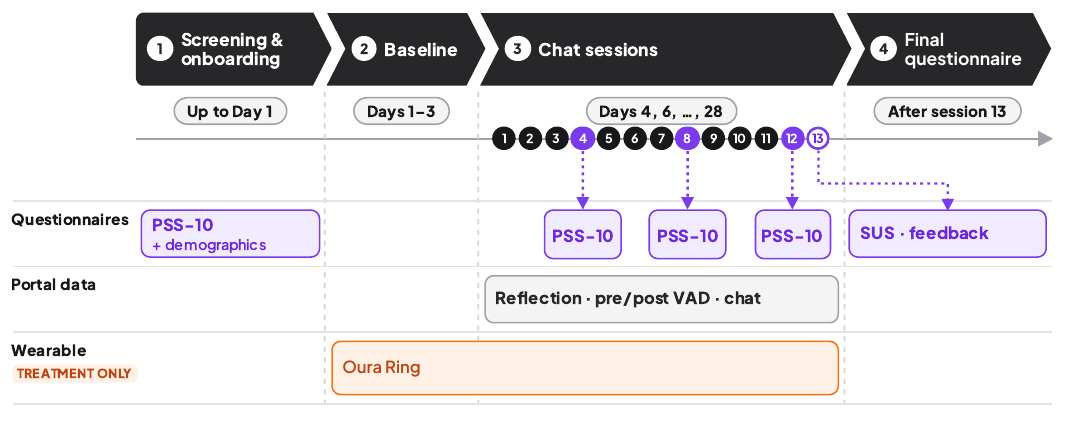}
\caption{Study design: (1) screening and onboarding (online PSS-10, demographics; ring setup in Treatment), (2) three-day baseline, (3) 13 alternate-day chat sessions (days 4--28), (4) final questionnaire after session 13.}
\label{fig:study_design}
\end{figure}

\begin{figure}[t]
\centering
\includegraphics[width=\linewidth]{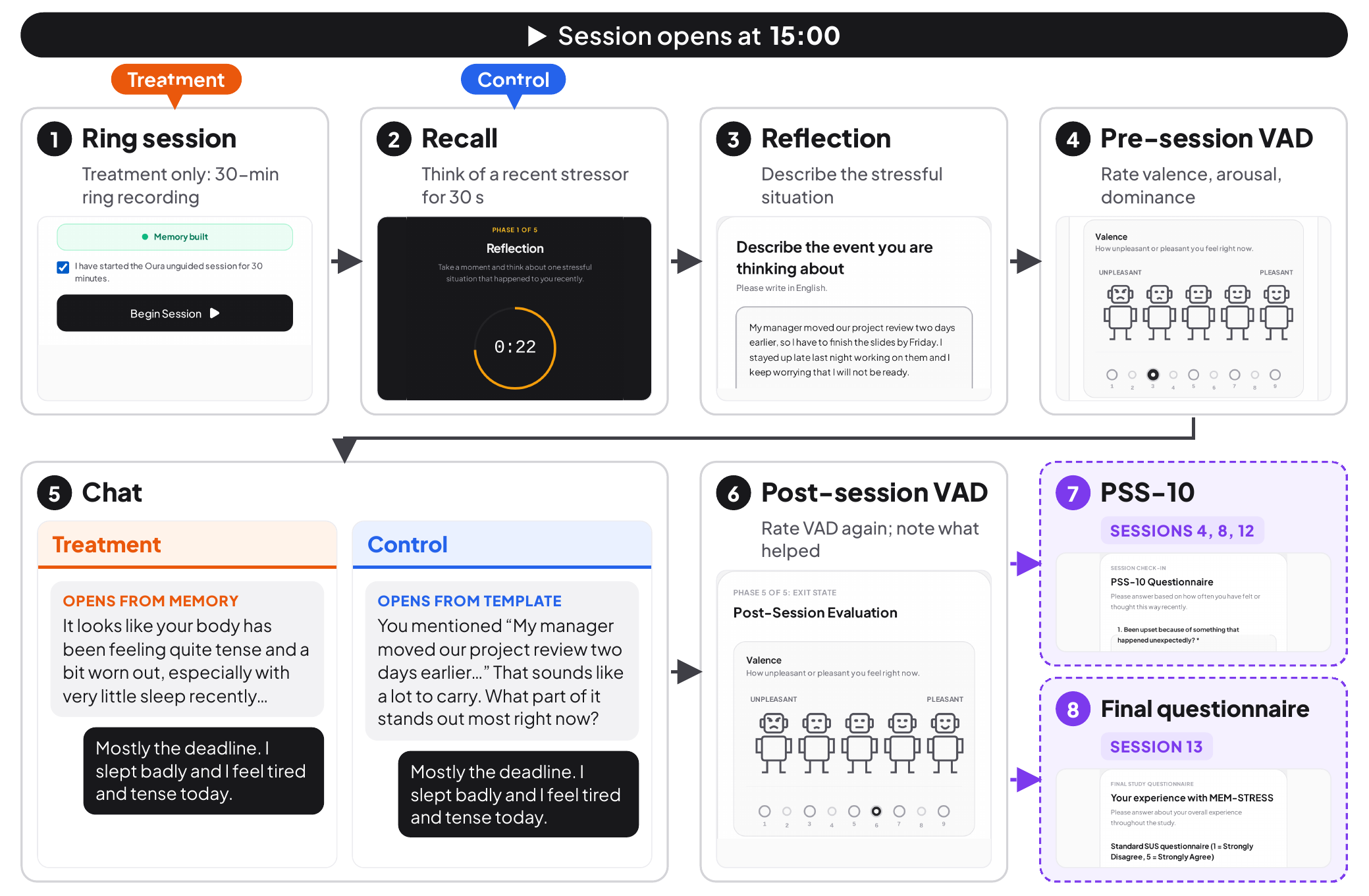}
\caption{Session flow (opening at 15:00): (1) Treatment only: confirming an unguided session (a self-started recording in the ring's app); (2) recalling a recent stressful situation; (3) written reflection; (4) pre-session VAD ratings; (5) chat, with example openings of the Treatment chatbot (from its memory of physiological features) and Control chatbot (fixed template); (6) post-session VAD ratings and a note on the chat's most helpful part; (7) PSS-10 after sessions 4, 8, and 12; (8) final questionnaire after session 13.}
\label{fig:session_design}
\end{figure}

MEM-STRESS is a web-based chatbot designed to provide multi-session stress management by integrating wearable sensor data into conversational memory. We evaluated the system in a four-week, between-subjects field deployment comparing two chatbot configurations. Both versions shared the same user interface, language model, session structure, and Dialectical Behavior Therapy (DBT) intervention framework. The \textit{Treatment} chatbot (referred to as MEM-STRESS) combined a layered memory of Oura Ring physiological features with past conversation summaries, whereas the \textit{Control} chatbot received conversation summaries alone. We hypothesized that supplementing conversation history with physiological data would result in greater stress reduction.
Across 13 alternate-day web sessions (\autoref{fig:study_design},~\autoref{fig:session_design}), participants wrote a brief stress reflection, interacted with the chatbot, rated their affect (Valence, Arousal, Dominance) using the Self-Assessment Manikin (SAM)~\cite{bradley1994measuring} before and after chatting, and periodically completed the PSS-10 questionnaire~\cite{cohen1983global}.

\subsection{Conversational Agent and DBT Intervention}\label{sec:chatflow}

Both chatbots use GPT-4.1-mini (2025-04-14) through our institution's Azure OpenAI gateway. Each reply draws on the system prompt, a hidden \emph{context message} (the reflection, pre-session VAD ratings, and group-specific memory), and the 12 most recent messages (prompts in Appendix~\ref{sec:app-prompts}, settings in~\autoref{tab:app-prompt-settings}). The shared system prompt frames the chatbot as an empathetic AI companion, not a therapist, giving brief, warm replies in everyday language; it never diagnoses or mentions study groups and switches to crisis guidance if a participant raises a safety concern such as self-harm (Section~\ref{sec:ethics}).
The conversation loosely follows Gibbs' Reflective Cycle~\cite{gibbs1988learning} without naming or forcing its stages, and participants can ask for advice at any time. When advising, the chatbot recommends exactly one DBT skill~\cite{linehan2015dbt} from a closed list of 48 skills spanning the four DBT modules (\autoref{tab:dbt_skills}): it identifies the participant's main difficulty, then the fitting module, then the skill, explaining why it fits and how to do it.

\subsection{Hierarchical Memory}\label{sec:memory}

A server retrieved each Treatment participant's ring data via the manufacturer's API (Application Programming Interface) nightly and just before each session. We call each measured quantity a \emph{physiological feature}; features span six domains (sleep, cardiac activity, readiness and recovery, activity, respiration, and self-entered activity tags) and include the ring's day-level \emph{readiness} (how recovered the body is) and daily \emph{stress} (minutes classified as high stress or high recovery; full list in~\autoref{tab:app-features}). As the ring computes most summaries per day, these data describe day-level change rather than moment-to-moment state.

MEM-STRESS organizes these features and the conversation history in three layers, each condensing the one beneath, so the model can describe them in everyday language rather than recite numbers:

\begin{itemize}
\item \textbf{Episodic memory: what was recorded.} A complete, time-stamped archive of the participant's physiological features, roughly 160 per day (storage in Appendix~\ref{sec:app-design}); the chatbot never reads it directly.

\item \textbf{Pattern memory: how the participant differs from usual.} For each feature, a rule-based procedure compares the day before the session with the participant's personal baseline (the mean of all earlier days from study day~1, not the screening PSS-10 baseline) and writes one plain-language statement of the change's direction and size (e.g., ``increased by X (+Y\%)''); changes below 5\% are labeled stable, and the participant's own tags are added where available.

\item \textbf{Semantic memory: what it may mean.} Before each session, GPT-4.1-mini generates two texts (prompts in Appendix~\ref{sec:app-prompt-memory}): a one-paragraph \emph{chat summary} of all previous conversations (recurring stressors, emotional trajectory, coping actions, and helpful or unhelpful responses) and a \emph{semantic description}: a cautious, non-diagnostic interpretation of the participant's recent state, drawn from the pattern memory and chat summary, linking changes across domains without claiming causality. The Control chatbot's only memory is a shorter (three-to-six-sentence) chat summary.

\end{itemize}

\para{Why layers.} Human memory separates specific experiences (episodic memory) from general knowledge about oneself (semantic memory)~\cite{tulving1972episodic}, and autobiographical memory is organized as a hierarchy, from specific events up to general patterns and life themes~\cite{conway2000construction}. LLM agents adopt the same split, storing raw experience and building higher-level summaries on top of it~\cite{sumers2024cognitive, park2023generative}. MEM-STRESS follows this structure: the episodic layer holds the raw ring records, the pattern layer captures how today differs from the person's usual, and the semantic layer describes what those changes may mean. This structure suits our problem for three reasons. First, roughly 160 features a day cannot go into the prompt, and LLMs handle raw numbers poorly; condensing each layer from the one below keeps the context small, much as recursive summarization keeps long-term dialogue consistent~\cite{wang2025recursively}. Second, group-level norms often fail to describe individuals~\cite{fisher2018lack}, and stress detection does not generalize across people~\cite{mishra2020reproducibility}, so the pattern layer compares each feature with the person's own running baseline rather than population thresholds. Third, the semantic layer adds a cautious interpretation that the chatbot can put into everyday language, while the intact episodic archive keeps every statement traceable to the measured data, the same idea of storing observations and synthesizing reflections from them that underlies generative agents~\cite{park2023generative}.


The system also ranks pattern statements by keyword relevance to the reflection and VAD ratings and adds the top eight as a \emph{personalization priority} list, each with a hint on a fitting kind of coping step (e.g., ``Offer a body-calming step because cardiovascular or breathing patterns may reflect activation''; Appendix~\ref{sec:app-prompt-treatment}). Pattern memory, semantic memory, and this list form the Treatment chatbot's group-specific memory in the context message, from which it also generates its opening message; the Control chatbot instead opens with a fixed template restating the participant's reflection (Appendix~\ref{sec:app-prompt-control}). The Treatment prompt makes pattern memory the primary driver of personalization and DBT skill choice but prioritizes the participant's own account when it differs from the data.

\subsection{Participants and Procedure}\label{sec:recruitment}

We recruited adults from a university community via mailing lists, word of mouth, and two group-specific campus flyers advertising the same four-week study with EUR~60 compensation (the Treatment version inviting people to ``experience a smart wearable ring'' while chatting with an AI chatbot) and linking to an online screening questionnaire with the PSS-10. Eligible people were 18 or older, experienced everyday stress, communicated in English, and reported no clinical mental health diagnosis or substantial psychological distress; we applied no PSS-10 cutoff. Assignment was non-randomized, following the flyer participants responded to, ring availability, and quotas. After online consent and registration, Treatment participants without a ring received one and connected their ring account to the study.

Registration marked study day~1. The first three days formed a baseline period in which Treatment participants already wore the ring, giving the memory earlier days to compare against at the first session. Chat sessions then ran from day~4 to day~28 (\autoref{fig:study_design}), each following~\autoref{fig:session_design} (flyer, onboarding, scheduling, and session details in Appendix~\ref{sec:app-design}). The 30-minute unguided session prompts the Oura Ring to log granular sensor data every second, replacing its standard 5-minute sampling rate.

The analysis cohort comprises 60 participants, 30 per group (\autoref{tab:participant_demographics}), most aged 25--34 (40 of 60) and holding a master's (39) or doctoral (12) degree. Screening PSS-10 scores (range 11--31; mean 23.1, SD (Standard Deviation) 4.7; 3 low, 43 moderate, and 14 high stress) did not differ significantly between groups (Treatment 23.9 vs.\ Control 22.4; Welch's $t = 1.22$, Cohen's $d = 0.31$). The study had 80\% power only for between-group effects of about $d = 0.74$ (48\% for $d = 0.5$), so its results are initial evidence.


\subsection{Measures}\label{sec:measures}

The SAM~\cite{bradley1994measuring}, a non-verbal, pictorial instrument, measured momentary affect on the three dimensions of Mehrabian's VAD model~\cite{mehrabian1996pad}: \emph{valence} (how pleasant a feeling is), \emph{arousal} (physical and mental activation), and \emph{dominance} (feeling in control of, rather than controlled by, one's situation). Participants rated each from 1 to 9 (slider starting at 5), anchored from unpleasant to pleasant, calm to tense (not the standard calm to excited, so arousal ratings partly reflect tension), and low to high control. Following recall of and reflection on a recent stressor (\autoref{fig:session_design}), the pre-session rating reflects participants' state while it was salient.

The PSS-10~\cite{cohen1983global, cohen1988perceived}, our measure of change across the study, instead captures stress appraisal over weeks: how unpredictable, uncontrollable, and overloaded people have found their lives in the past month. Its ten items sum to a 0--40 total (scoring in Appendix~\ref{sec:app-design}), which by convention we treat as low (0--13), moderate (14--26), or high (27--40) stress. Participants completed it online at screening (the \emph{baseline}) and on the study website after their 4th, 8th, and 12th sessions. The in-session instructions asked how participants had felt recently rather than in the last month, and the form followed a conversation, which may have lowered follow-up scores in both groups.

After their last session, participants were asked to complete the final questionnaire (items verbatim in Appendix~\ref{sec:app-rq3}):
\begin{itemize}
\item \textit{Usability.} The ten-item SUS (System Usability Scale)~\cite{brooke1996sus}, scored 0--100, higher being better (scoring in Appendix~\ref{sec:app-design}).
\item \textit{Agreement.} Four custom items (not from a validated scale; 1 = Strongly Disagree to 5 = Strongly Agree) on whether the chatbot (Q1) understood the participant's situation accurately, (Q2) provided suggestions relevant to their daily life, (Q3) gave explanations for its suggestions that made sense, and (Q4) gave good recommendations to manage stress.
\item \textit{Behavior change.} Two items on whether the chatbot had changed how participants \emph{reflected} on their stress or emotions and their \emph{daily routine} or behavior (``No change'', ``A little'', ``Yes, somewhat'', ``Yes, a lot'', or ``Not sure'').
\item \textit{Open-ended questions.} Eight free-text questions (e.g., the chatbot's biggest advantage and disadvantage, suggested improvements), plus two Treatment-only questions on when the ring did and did not accurately reflect how participants felt or behaved.
\end{itemize}

\subsection{Collected Longitudinal Data}\label{sec:data}

The 60 participants completed 772 sessions (386 per group, averaging 12.9 of 13 scheduled;~\autoref{tab:engagement}), and all 30 Treatment participants connected the ring. One further Treatment session had a chat but was not completed, so RQ2's conversation analyses cover 387 Treatment sessions; 769 sessions had both pre- and post-session ratings (1,538 ratings). Participants also submitted 178 in-session PSS-10 questionnaires (at least one each); one lacked a session number and is excluded from RQ1's mixed-effects model (237 readings including screening). Control participants wrote significantly more per session (more turns and words, and longer reflections), whereas Treatment participants reported significantly higher pre-session valence, which we consider when interpreting affect outcomes (\autoref{tab:engagement}; tests in Appendix~\ref{sec:app-design}); stress types in the reflections did not differ significantly between groups.

\FloatBarrier 
\section{Results}\label{sec:results}

We report results by research question (RQ1--RQ3). Given this initial field study's modest sample, several effects were not statistically significant; we still report their direction and size, because consistent patterns across measures are informative and motivate larger confirmatory studies.
\subsection{Changes in Stress, Affect, and Physiological Features (RQ1)}\label{sec:rq1}
We ask whether perceived stress (PSS-10) over four weeks and within-session affect changed differently when the chatbot remembered participants' physiological features, and complement these self-reports with Treatment participants' heart rate around sessions and across days.

\subsubsection{Perceived Stress (PSS-10) over Four Weeks}\label{sec:rq1-method-pss}

\begin{figure}[t]
\centering
\includegraphics[width=0.6\linewidth]{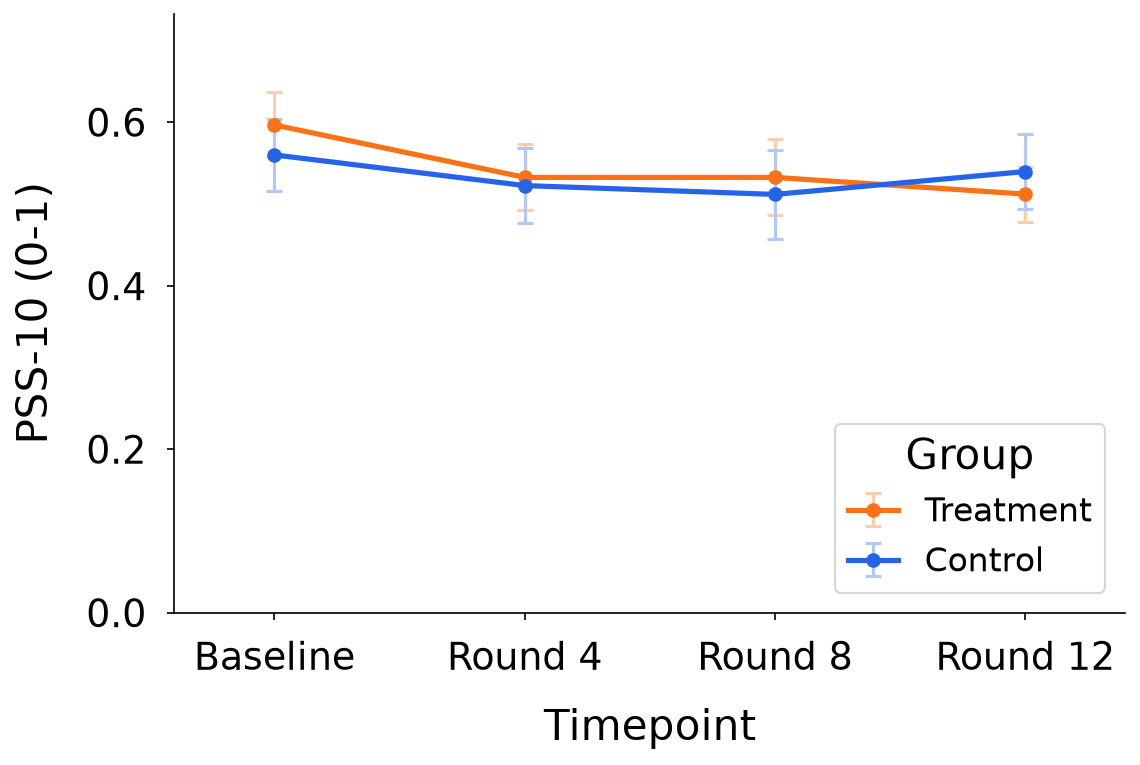}
\caption{Mean perceived stress (PSS-10, raw/40) for Treatment (orange) and Control (blue) with 95\% CIs (tests in~\autoref{tab:rq1-pss}).}
\label{fig:rq1-pss}
\end{figure}

All 60 participants had a screening PSS-10 (\emph{baseline}) and in-session PSS-10s at \emph{rounds} 4, 8, and 12. We normalized totals to 0--1 (raw/40) and plotted each group's mean with 95\% CIs (Confidence Intervals;~\autoref{fig:rq1-pss}). We compared groups at each timepoint and, within each group, each round with baseline and all four timepoints jointly (Friedman test;~\autoref{tab:rq1-pss}), and repeated the between-group tests within gender, age, and education subgroups (Appendix~\ref{sec:app-rq1}).

Because these tests treat each timepoint separately and the Friedman test uses only complete cases, we also compared trajectories with a linear mixed-effects model on all readings,
\begin{equation}\label{eq:rq1-mixed-interaction-pss}
\text{PSS-10} \sim \text{Group} \times \text{Session} + (1 \mid \text{Participant}),
\end{equation}
where Session counts completed sessions (0 at baseline), the random intercept absorbs stable between-participant differences, and Group$\times$Session, our most direct test, compares the groups' rates of stress change (\autoref{tab:rq1-mixed}).

\para{Stress fell below baseline in Treatment, but not significantly faster than in Control.} Over 12 sessions, stress declined about 2.3 raw PSS-10 points more in Treatment than in Control (95\% CI $-4.8$ to 0.1), a non-significant difference (Group$\times$Session $-0.0048$ per session, $z=-1.84$;~\autoref{tab:rq1-mixed}). Treatment's implied slope ($-0.0065$ per session) equals about 3.1 raw points over 12 sessions, versus 0.8 in Control ($-0.0017$, nearly flat). Both groups started at moderate levels (\autoref{fig:rq1-pss}). Treatment's mean fell from 23.9 to 20.5 by round~12, significantly below baseline at every round with medium effects ($t=-3.42^{**}$, $-2.92^{**}$, and $-3.61^{**}$; $d_z=-0.62$, $-0.53$, and $-0.67$; Friedman $\chi^2(3)=15.36^{**}$;~\autoref{tab:rq1-pss}). Control's mean dipped at rounds~4 and~8 but returned near baseline by round~12 (22.4 to 21.6), with no round differing significantly from baseline ($d_z$ from $-0.30$ to $-0.14$; Friedman $\chi^2(3)=1.10$). The lines crossed between rounds~8 and~12: no timepoint differed significantly between groups, but the small screening difference ($d=0.31$) reversed by round~12 ($d=-0.25$). Because Treatment started higher, part of its decline may reflect regression to the mean: it was most visible in the two subgroups where Treatment started significantly higher (women and master's-degree holders), which ended near Control; other subgroups showed similar trends (\autoref{fig:app-rq1-pss-demographics}). Our hypothesis that Treatment would reduce stress more than Control was thus not confirmed.

\begin{table}[t]
\centering
\small
\caption{Perceived stress (PSS-10, raw/40) by timepoint ($n=30$ per group). Between groups: Welch's $t$ with Cohen's $d$; within groups: paired $t$ vs.\ baseline with $d_z$; $^{a}$Mann--Whitney $U$ or Wilcoxon $W$ for non-normal data; Friedman tests across timepoints ($^{*}p<0.05$, $^{**}p<0.01$, $^{***}p<0.001$).}
\label{tab:rq1-pss}
\setlength{\tabcolsep}{4pt}
\begin{tabular}{@{}lcccccccc@{}}
\toprule
 & \multicolumn{2}{c}{\textbf{Mean}} & \multicolumn{2}{c}{\textbf{Between}} & \multicolumn{2}{c}{\textbf{Within Treatment}} & \multicolumn{2}{c}{\textbf{Within Control}} \\
\cmidrule(lr){2-3}\cmidrule(lr){4-5}\cmidrule(lr){6-7}\cmidrule(l){8-9}
\textbf{Timepoint} & \textbf{Treatment} & \textbf{Control} & \textbf{stat.} & $d$ & \textbf{stat.} & $d_z$ & \textbf{stat.} & $d_z$ \\
\midrule
Baseline & 0.597 & 0.560 & $1.22$ & 0.31 & -- & -- & -- & -- \\
Round 4 & 0.533 & 0.523 & $0.32$ & 0.08 & $-3.42^{**}$ & $-0.62$ & $-1.63$ & $-0.30$ \\
Round 8 & 0.533 & 0.512 & $487.0^{a}$ & 0.15 & $-2.92^{**}$ & $-0.53$ & $177.0^{a}$ & $-0.27$ \\
Round 12 & 0.512 & 0.540 & $-0.96$ & $-0.25$ & $-3.61^{**}$ & $-0.67$ & $-0.75$ & $-0.14$ \\
\midrule
Friedman & & & & & \multicolumn{2}{c}{$\chi^2(3)=15.36^{**}$} & \multicolumn{2}{c}{$\chi^2(3)=1.10$} \\
\bottomrule
\end{tabular}
\end{table}


\begin{table}[t]
\centering
\small
\caption{Mixed-effects models of perceived stress (\autoref{eq:rq1-mixed-interaction-pss}) and session affect (\autoref{eq:rq1-mixed-interaction-vad}), outcomes normalized to 0--1: coefficients (standard errors), with Control and pre-session as references; Wald $z$ tests ($^{*}p<0.05$, $^{**}p<0.01$, $^{***}p<0.001$).}
\label{tab:rq1-mixed}
\setlength{\tabcolsep}{4pt}
\begin{tabular}{@{}lcccc@{}}
\toprule
\textbf{Term} & \textbf{PSS-10} & \textbf{Valence} & \textbf{Arousal} & \textbf{Dominance} \\
\midrule
Intercept & $0.5439^{***}$ & $0.4860^{***}$ & $0.5554^{***}$ & $0.5190^{***}$ \\
 & (0.0207) & (0.0278) & (0.0274) & (0.0275) \\
Group: Treatment & $0.0373$ & $0.0888^{*}$ & $-0.0243$ & $-0.0225$ \\
 & (0.0292) & (0.0393) & (0.0388) & (0.0390) \\
Session & $-0.0017$ & $-0.0070^{**}$ & $-0.0045$ & $-0.0088^{**}$ \\
 & (0.0018) & (0.0027) & (0.0029) & (0.0027) \\
Group$\times$Session & $-0.0048$ & $-0.0019$ & $0.0046$ & $0.0061$ \\
 & (0.0026) & (0.0038) & (0.0041) & (0.0039) \\
\addlinespace
Phase: post & & $0.1014^{***}$ & $-0.0794^{*}$ & $0.0503$ \\
 & & (0.0296) & (0.0323) & (0.0302) \\
Phase$\times$Group & & $-0.0367$ & $0.0212$ & $0.0148$ \\
 & & (0.0419) & (0.0457) & (0.0428) \\
Phase$\times$Session & & $0.0074^{*}$ & $-0.0035$ & $0.0065$ \\
 & & (0.0038) & (0.0041) & (0.0038) \\
Phase$\times$Group$\times$Session & & $-0.0014$ & $-0.0002$ & $-0.0039$ \\
 & & (0.0053) & (0.0058) & (0.0054) \\
\midrule
Observations & 237 & 1,538 & 1,538 & 1,538 \\
Participants & 60 & 60 & 60 & 60 \\
\bottomrule
\end{tabular}
\end{table}

\subsubsection{Within-Session Affect}\label{sec:rq1-method-vad}

\begin{figure}[t]
\centering
\begin{subfigure}[t]{0.32\linewidth}
\centering
\includegraphics[width=\linewidth]{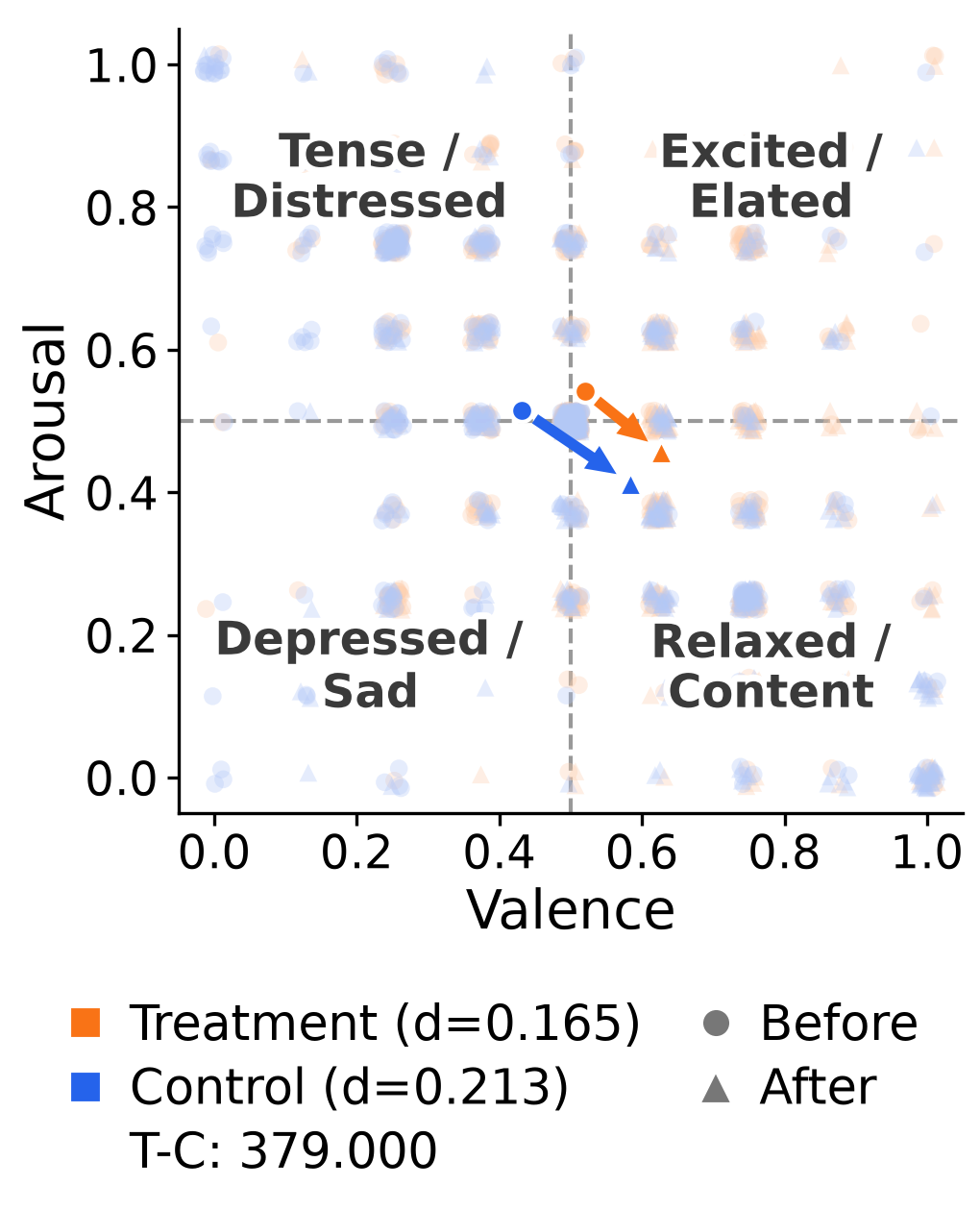}
\caption{Valence--arousal plane.}
\label{fig:rq1-ousigram-va}
\end{subfigure}\hfill
\begin{subfigure}[t]{0.32\linewidth}
\centering
\includegraphics[width=\linewidth]{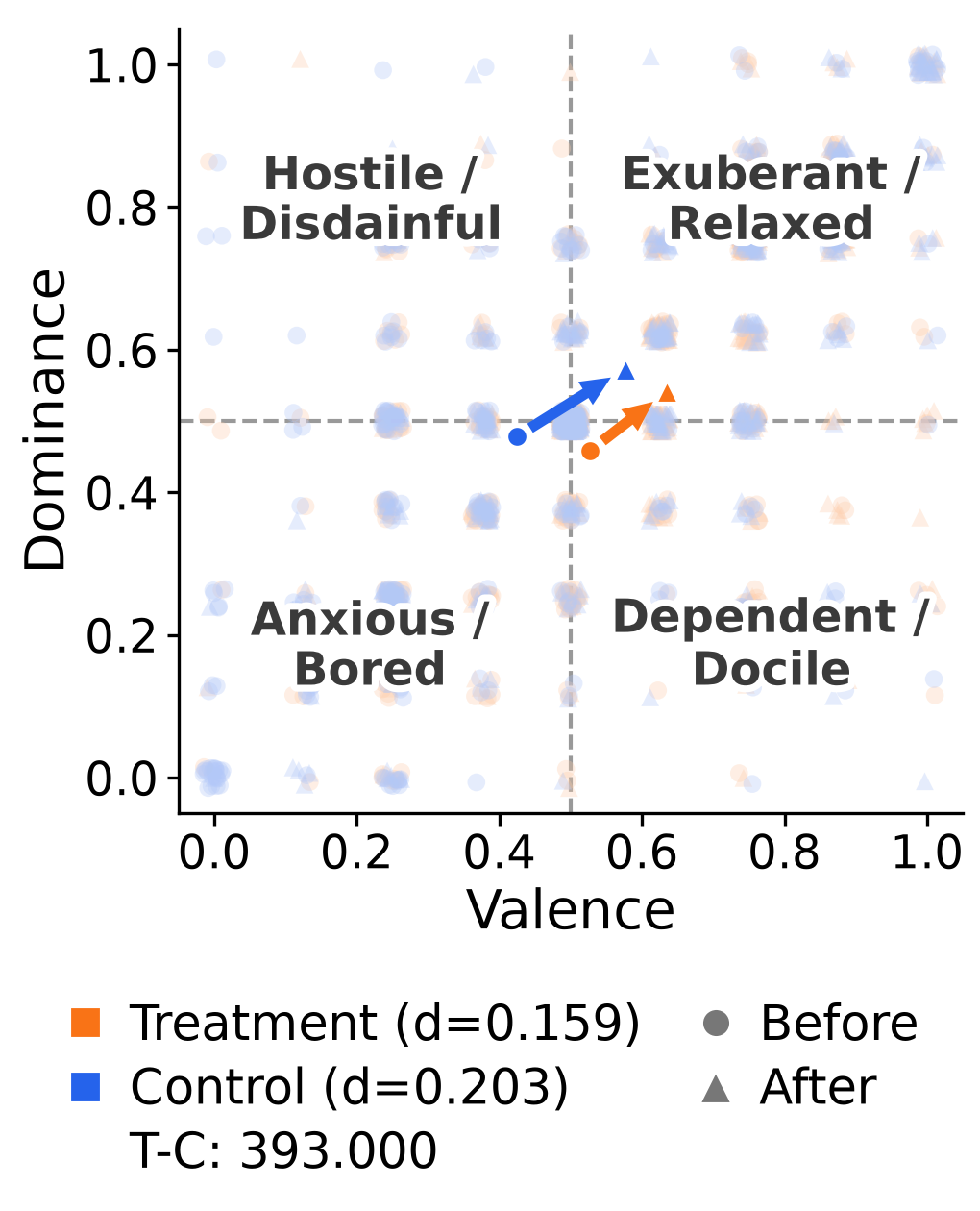}
\caption{Valence--dominance plane.}
\label{fig:rq1-ousigram-vd}
\end{subfigure}\hfill
\begin{subfigure}[t]{0.32\linewidth}
\centering
\includegraphics[width=\linewidth]{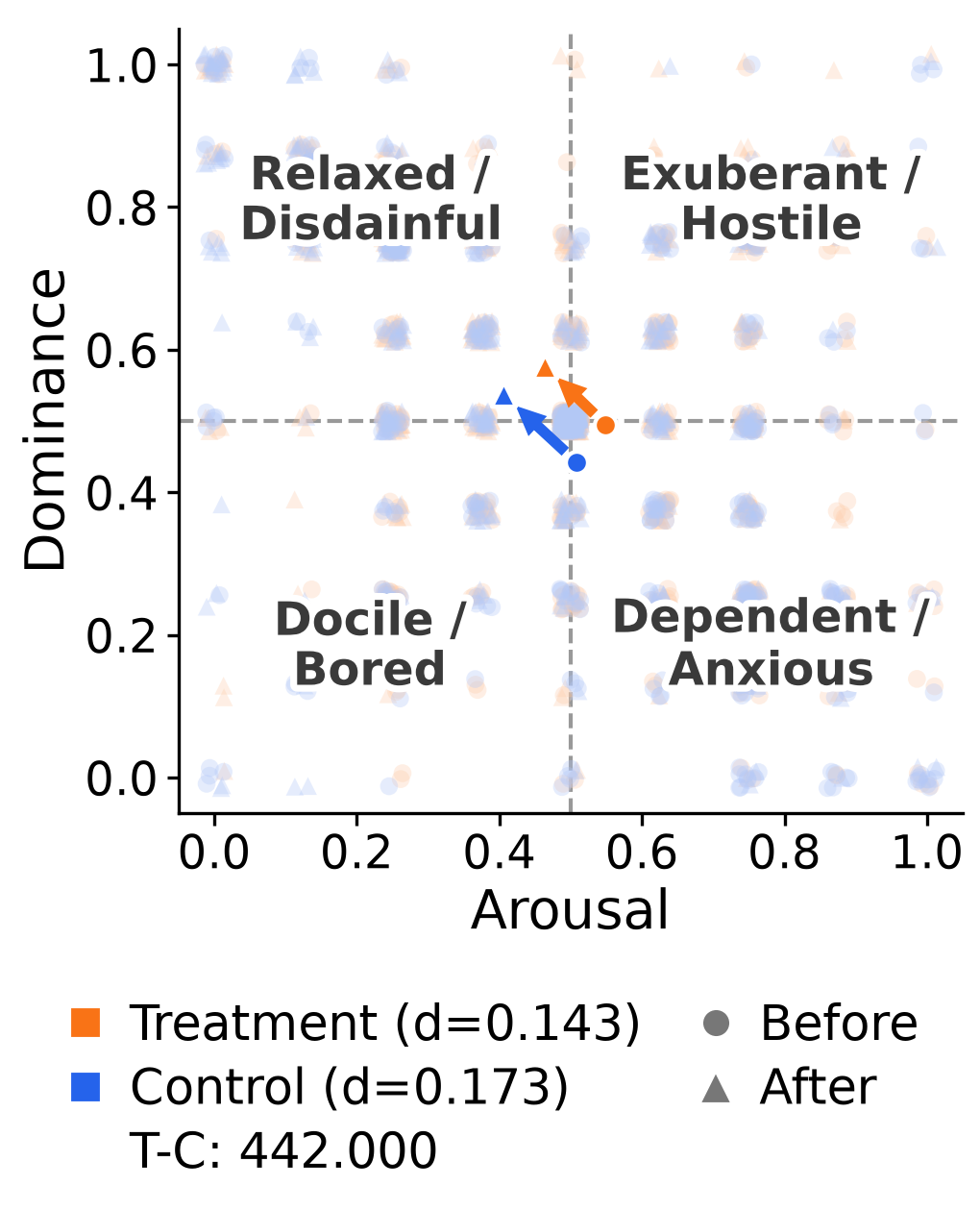}
\caption{Arousal--dominance plane.}
\label{fig:rq1-ousigram-ad}
\end{subfigure}
\caption{Ousigrams of within-session affect (SAM, 0--1) for Treatment (orange) and Control (blue): ratings before (circles) and after (triangles) the conversation, with arrows joining each group's mean positions. Legends give the mean shift distance (``d'', not Cohen's $d$) and a Mann--Whitney $U$ test between groups (``T-C'') ($^{*}p<0.05$, $^{**}p<0.01$, $^{***}p<0.001$).}
\label{fig:rq1-ousigram}
\end{figure}

To see each conversation's effect, we normalized the 1,538 SAM ratings given right before and after each one to 0--1 and plotted them as \emph{ousigrams}, our term for affect-plane plots with an arrow from each group's mean pre- to post-session position (\autoref{fig:rq1-ousigram}): one plane per pair of VAD dimensions, VA (Valence--Arousal), VD (Valence--Dominance), and AD (Arousal--Dominance), with quadrants labeled after Russell's~\cite{russell1980circumplex} and Mehrabian's~\cite{mehrabian1996pad} models (construction in Appendix~\ref{sec:app-rq1}). To quantify movement, we compared participants' \emph{shift distance} (Euclidean distance between mean pre- and post-session positions) between groups, overall and within demographic subgroups.

Because the ousigrams average over the whole study and the shift distance discards direction, we modeled the within-session change and its evolution as the chatbots' memories accumulated, separately for valence, arousal, and dominance,
\begin{equation}\label{eq:rq1-mixed-interaction-vad}
\text{VAD} \sim \text{Phase} \times \text{Group} \times \text{Session} + (1 \mid \text{Participant}),
\end{equation}
where Phase (pre- or post-session) captures the within-session change and Phase$\times$\allowbreak Session its evolution; Phase$\times$\allowbreak Group and Phase$\times$\allowbreak Group$\times$\allowbreak Session test whether these differ by group (\autoref{tab:rq1-mixed}).

\para{With both chatbots, affect shifted toward more pleasant and calmer states.} Individual ratings spanned all four quadrants, but in every plane both groups moved on average from near the center toward pleasant, calm, in-control states (\autoref{fig:rq1-ousigram}; per-plane quadrants in Appendix~\ref{sec:app-rq1}). The models agreed (\autoref{tab:rq1-mixed}): in Control (the reference), valence rose within sessions ($0.101^{***}$, about 0.8 SAM points at study start, as Session is uncentered), arousal fell ($-0.079^{*}$), and dominance rose non-significantly ($0.050$); the valence gain grew by $0.0074^{*}$ per session. Treatment's implied changes pointed the same way (valence $0.065$, about 0.5 SAM points; arousal $-0.058$; dominance $0.065$; valence gain $0.0060$ per session), and no Phase$\times$Group or Phase$\times$Group$\times$Session term was significant in any dimension.

\para{Shift size did not differ significantly between the chatbots.} Mean shift distances were numerically larger in Control (Treatment vs.\ Control: VA 0.165 vs.\ 0.213, $U=379.0$; VD 0.159 vs.\ 0.203, $U=393.0$; AD 0.143 vs.\ 0.173, $U=442.0$;~\autoref{fig:rq1-ousigram}), possibly because Control started at lower valence, leaving more room to improve, but did not differ significantly in any plane or in any of 24 gender, age, or education subgroup tests (\autoref{fig:app-rq1-ousigram-gender} and~\autoref{tab:app-rq1-vad-subgroups}).


\subsubsection{Complementary Analysis: Heart Rate Around and Across Sessions}\label{sec:rq1-method-hr}

Raw heart rate from the Oura Ring’s API offers a coarse check of whether participants’ bodies calmed down. For the 30 Treatment participants, we averaged heart rate in the hour before, during, and the hour after each conversation, and daily heart rate by nearest session (1,014 participant-days;~\autoref{fig:rq1-hr}; details in Appendix~\ref{sec:app-rq1}).

\begin{figure}[t]
\centering
\begin{subfigure}[t]{0.42\linewidth}
\centering
\includegraphics[width=\linewidth]{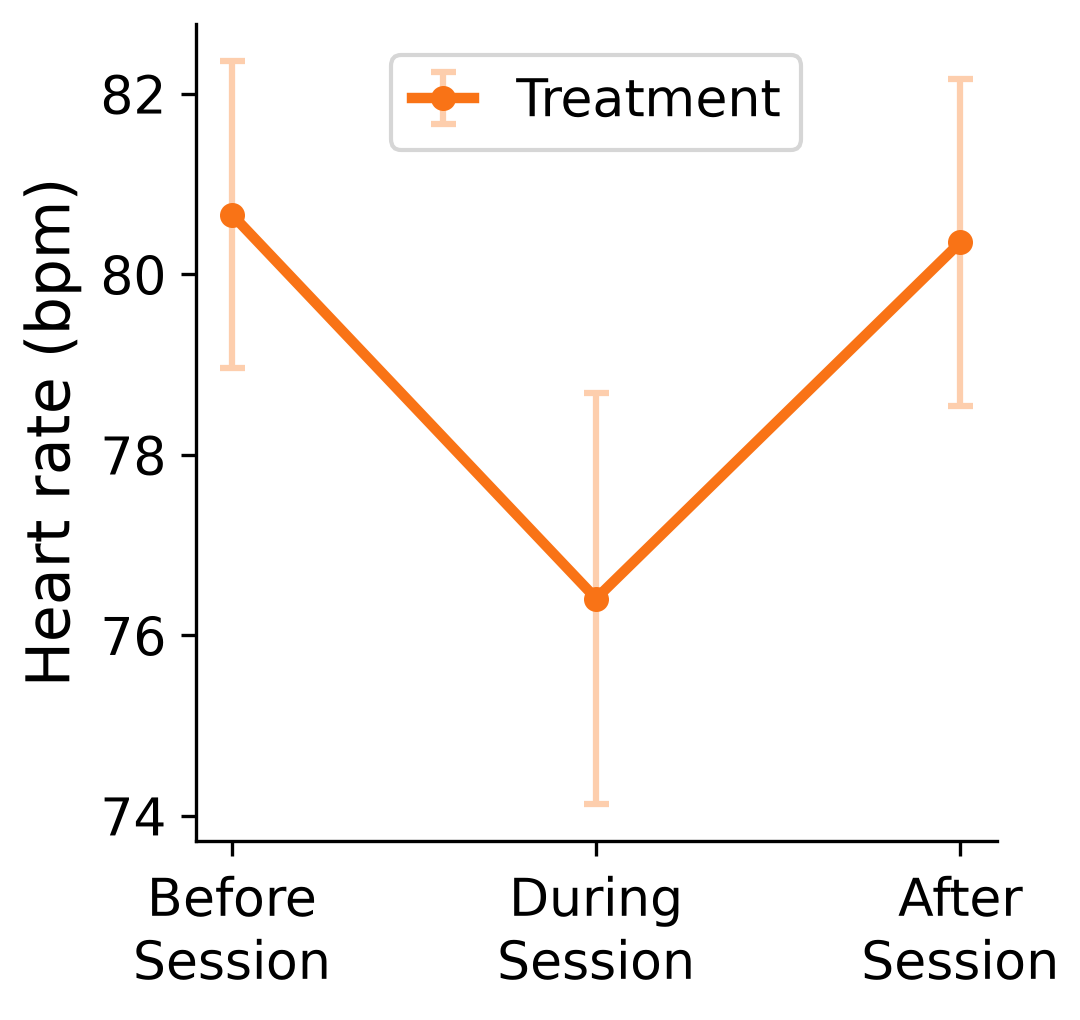}
\caption{Heart rate before, during, and after sessions.}
\label{fig:rq1-hr-window}
\end{subfigure}\hspace{2em}
\begin{subfigure}[t]{0.42\linewidth}
\centering
\includegraphics[width=\linewidth]{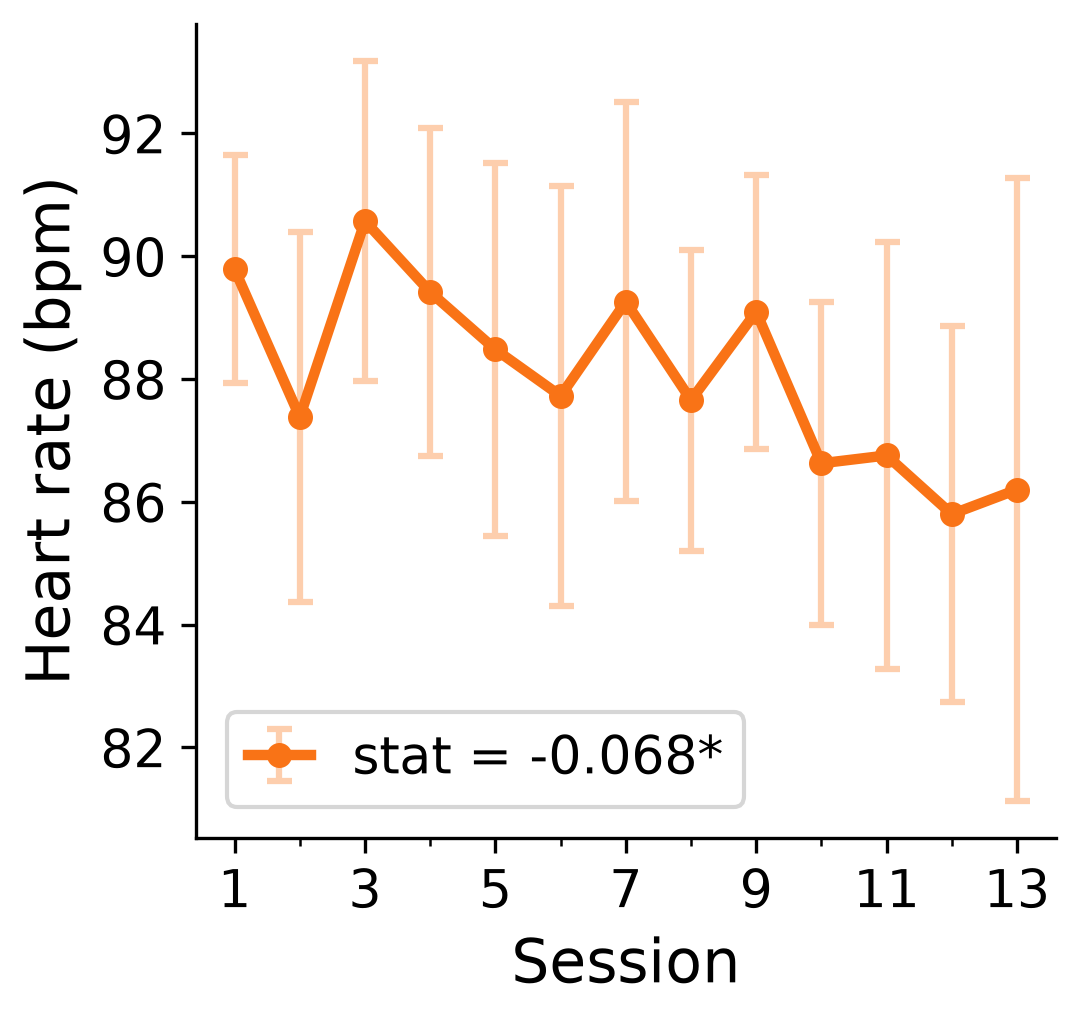}
\caption{Daily average heart rate across the study.}
\label{fig:rq1-hr-daily}
\end{subfigure}
\caption{Heart rate in Treatment (mean $\pm$ 95\% CI): (a) around conversations; (b) daily, by nearest session (y-axis truncated), with Spearman's $\rho$ (``stat'') in the legend ($^{*}p<0.05$, $^{**}p<0.01$, $^{***}p<0.001$).}
\label{fig:rq1-hr}
\end{figure}

\para{Heart rate was lower during conversations and drifted slightly downward.} Heart rate averaged $76.40\pm2.28$\,bpm (beats per minute) during conversations (107 sessions), 4.3 and 3.9\,bpm below the hour before ($80.66\pm1.70$\,bpm) and after ($80.35\pm1.81$\,bpm; 290 sessions each;~\autoref{fig:rq1-hr}, panel~\subref{fig:rq1-hr-window}).
Daily heart rate declined slightly from registration to the last session ($\rho=-0.068^{*}$, 959 days with heart rate; panel~\subref{fig:rq1-hr-daily}), with bin means of 87.4--90.6\,bpm through session~9 and 85.8--86.8\,bpm in sessions~10 to~13. High-stress and high-recovery minutes rose after the first sessions (\autoref{fig:app-rq1-oura-daily}), possibly partly because the ring calibrates personal baselines over the first weeks of wear. Because the windows were unpaired and not adjusted for activity, the trends lack a Control comparison, and the Spearman tests treat participant-days as independent, we do not attribute these patterns to the chatbot.

\para{Takeaway.} Perceived stress fell significantly below baseline within Treatment and declined more than in Control, though not significantly. Affect improved similarly in both chatbots' sessions, perhaps through their shared guided reflection and brief DBT skills, and heart rate was lower during conversations and drifted slightly downward, neither attributable to the chatbot. RQ2 and RQ3, which examine how the memory of physiological features changed the conversations and how participants experienced it, may help explain these results.

\FloatBarrier 

\subsection{Memory Use, Language, and DBT Skills in Conversations (RQ2)}\label{sec:rq2}
If the memory works, physiological features should reach the chatbot's messages, shape its language and DBT skill choices, and leave a trace in participants' writing.

\subsubsection{Memory Use in Conversations}\label{sec:rq2-method-memory}\label{sec:rq2-memory}\label{sec:rq2-examples}

\begin{figure}[t]
\centering
\begin{subfigure}[t]{0.48\linewidth}
\centering
\includegraphics[width=\linewidth]{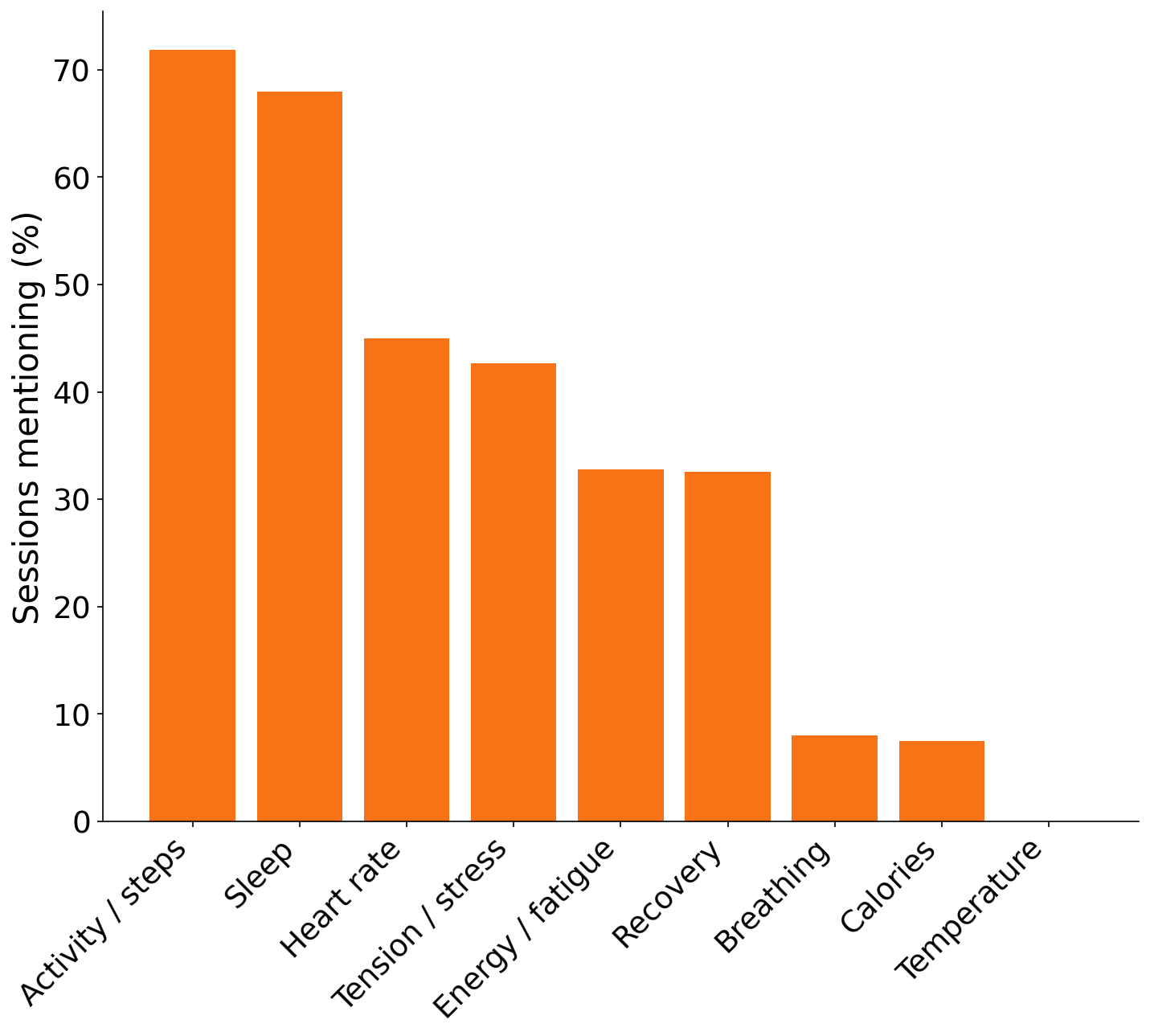}
\caption{Treatment: physiological-feature components ($n=387$ sessions).}
\label{fig:rq2-memory-use-treatment}
\end{subfigure}\hfill
\begin{subfigure}[t]{0.48\linewidth}
\centering
\includegraphics[width=\linewidth]{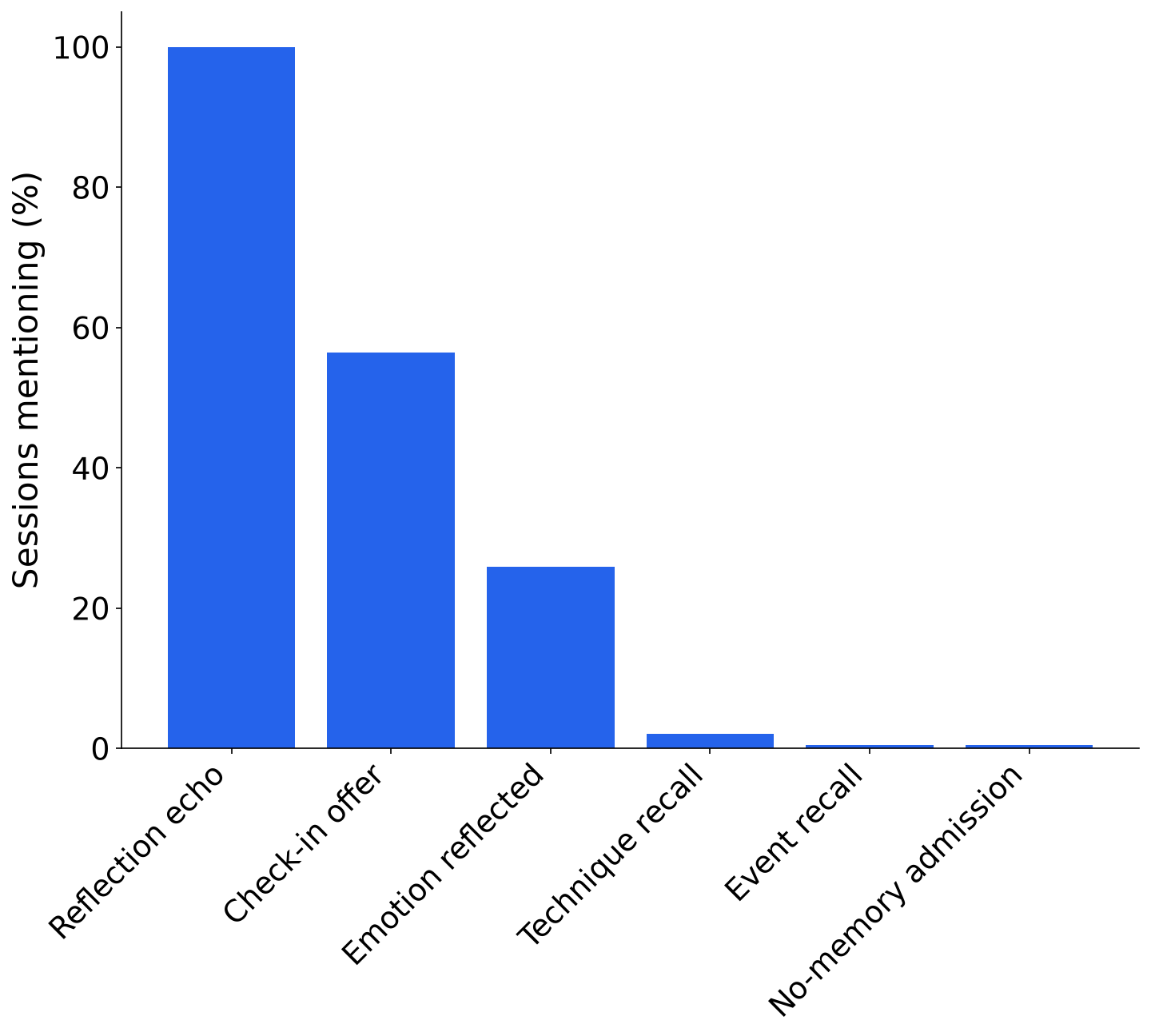}
\caption{Control: conversational components ($n=386$ sessions).}
\label{fig:rq2-memory-use-control}
\end{subfigure}
\caption{Percentage of sessions whose chatbot messages drew on each memory component (30 participants per group); the panels use different components and are not directly comparable.}
\label{fig:rq2-memory-use}
\end{figure}

To see what each chatbot drew on from memory, we joined each session's chatbot messages, including the opener, into one document (387 Treatment and 386 Control sessions); an LLM labeled the memory components each drew on, identically in both groups, and co-authors confirmed the labels (\autoref{fig:rq2-memory-use}). Treatment components were nine domains of physiological features, among them energy / fatigue and tension / stress, the chatbot's everyday terms for readiness and daily-stress features; Control components were six uses of the reflection, VAD ratings, and chat summary.

Because counts cannot show how a conversation unfolds, two co-authors (one per group) also fully read one random session from each of 15 randomly selected participants per group, commented on every participant and chatbot turn, and compared the groups from their combined notes. Excerpts are labeled T-P$n$ (Treatment) and C-P$n$ (Control) by sampled participant; further examples are in Appendix~\ref{sec:app-rq2-close}.

The memory of physiological features reached every Treatment participant's conversations, partly by design (the opening prompt asks for a physiological summary; Appendix~\ref{sec:app-prompt-treatment}). Activity (71.8\% of sessions) and sleep (68.0\%) were referenced most often (\autoref{fig:rq2-memory-use}, panel~\subref{fig:rq2-memory-use-treatment}), then heart rate (45.0\%), tension or stress (42.6\%), energy or fatigue (32.8\%), and recovery (32.6\%), each reaching at least 29 participants; breathing (8.0\%) and calories (7.5\%) were rare, and temperature was never referenced. The Control chatbot mostly mirrored participants' own words (panel~\subref{fig:rq2-memory-use-control}): its templated opener echoed the reflection in every session and it offered further check-ins in 56.5\%, but recalled an earlier technique in only 2.1\% and an earlier event in 0.5\%.

In the close reading, the memory made Treatment sessions more personal from the start: they opened from the participant's own physiological features, tied to their stressor (e.g., T-P4's difficulties with their manager), unlike the Control template. The memory also gave the Treatment chatbot a factual basis: the 15 sampled conversations contained no physiological statements unsupported by the data, although participants reported misreadings (RQ3). The Control chatbot, holding only the reflection and a short chat summary, could only imply continuity and recalled earlier sessions vaguely, as when C-P4 asked what they had discussed about their father.

Both chatbots handled the basics of a supportive conversation alike: they named the participant's stressor, tailored the skill to the moment at their best, and adapted when a suggestion was declined (the Control chatbot mostly); both sometimes offered several skills in one session or replied generically to ambiguous issues. Each had its own weaknesses: the Control chatbot reused a near-identical grounding script across participants, and several Treatment sessions repeated formulaic framing around physiological features or cited them despite little reported stress.

The Treatment chatbot used the memory in two ways. First, it linked participants' accounts to their physiological features to choose a fitting step: T-P15, anxious about the future, was offered STOP because their body showed signs of activation, and afterwards wrote, ``I felt my heartbeat slow down''. Second, it explained participants' feelings physiologically: having noted T-P3's higher activity but lower recovery and HRV, it replied to their self-blame for a few unproductive days, ``With your body showing signs of stress and less recovery, it makes sense that your mindset feels more fragile right now.'' T-P3 explained that the energy ``was spent on travel and tourism with my partner rather than work'', matching the measured rise in activity.

\subsubsection{Language of Participants and Chatbots}\label{sec:rq2-method-language}

To see whether the memory changed the conversations' language, we split all chat turns (not reflections) into four corpora by speaker and group and found each corpus's most distinctive words with SAGE (Sparse Additive Generative Model)~\cite{eisenstein2011sparse}, which scores each word by its deviation ($\eta$) from a background pooled over all corpora (each chatbot's top 20 words in~\autoref{tab:rq2-sage}).

To test whether the memory also changed how participants expressed stress, we used LIWC (Linguistic Inquiry and Word Count)~\cite{pennebaker2015liwc,tausczik2010psychological}, a standard, transparent dictionary of psychological word categories (LIWC2015-style); shifts in these categories during writing about stressful experiences predict later health improvements~\cite{pennebaker1997writing}. We scored each participant's reflections and chat turns, pooled over all sessions, as the percentage of words per category, and compared groups (not change over time) on 17 stress-relevant categories (\autoref{tab:rq2-liwc17}) and nine composites for broader constructs such as distress or coping (\autoref{tab:rq2-liwc-groups}; definitions in~\autoref{tab:app-rq2-liwc-composites}).

\begin{table}[t]
\centering
\small
\caption{Psycholinguistic composites in participants' reflections and chat turns (mean percent of words $\pm$ 95\% CI), with Cohen's $d$ and Welch's $t$ or $^{a}$Mann--Whitney $U$ tests ($^{*}p<0.05$, $^{**}p<0.01$, $^{***}p<0.001$).}
\label{tab:rq2-liwc-groups}
\setlength{\tabcolsep}{4pt}
\begin{tabular}{@{}lcccc@{}}
\toprule
\textbf{Composite} & \textbf{Treatment} & \textbf{Control} & \textbf{stat.} & $d$ \\
\midrule
Distress & 3.39 $\pm$ 0.16 & 3.39 $\pm$ 0.16 & $0.00$ & 0.00 \\
Emotional & 3.72 $\pm$ 0.25 & 3.67 $\pm$ 0.22 & $0.27$ & 0.07 \\
Coping & 4.38 $\pm$ 0.18 & 4.51 $\pm$ 0.14 & $-1.08$ & $-0.28$ \\
Cognitive reappraisal and reflection & 4.33 $\pm$ 0.30 & 4.69 $\pm$ 0.26 & $-1.78$ & $-0.46$ \\
Agency and perceived control & 1.87 $\pm$ 0.08 & 1.89 $\pm$ 0.10 & $-0.34$ & $-0.09$ \\
Social support and connectedness & 2.59 $\pm$ 0.21 & 2.57 $\pm$ 0.16 & $0.14$ & 0.04 \\
Temporal orientation and rumination & 7.53 $\pm$ 0.27 & 7.33 $\pm$ 0.33 & $509.0^{a}$ & 0.24 \\
Somatic and health-related distress & 1.76 $\pm$ 0.23 & 1.45 $\pm$ 0.17 & $2.18^{*}$ & 0.56 \\
Engagement and linguistic style & 6.48 $\pm$ 0.28 & 6.88 $\pm$ 0.31 & $-1.90$ & $-0.49$ \\
\bottomrule
\end{tabular}
\end{table}

\begin{table}[t]
\centering
\small
\caption{Stress-relevant LIWC categories in participants' reflections and chat turns (mean percent of words), with Welch's $t$ or $^{a}$Mann--Whitney $U$ tests ($^{*}p<0.05$, $^{**}p<0.01$, $^{***}p<0.001$).}
\label{tab:rq2-liwc17}
\begin{tabular}{@{}lccc@{}}
\toprule
\textbf{Category} & \textbf{Treatment} & \textbf{Control} & \textbf{stat.} \\
\midrule
Personal pronouns & 12.20 & 12.39 & $-0.34$ \\
\emph{I} & 10.64 & 10.61 & $0.05$ \\
\emph{We} & 0.53 & 0.39 & $469.0^{a}$ \\
Affect & 10.96 & 10.94 & $432.0^{a}$ \\
Positive emotion & 8.06 & 8.10 & $-0.08$ \\
Negative emotion & 2.77 & 2.70 & $0.26$ \\
Affiliation & 1.57 & 1.33 & $1.21$ \\
Social & 4.35 & 4.61 & $-0.54$ \\
Friend & 0.31 & 0.25 & $517.5^{a}$ \\
Negation & 2.49 & 2.86 & $384.0^{a}$ \\
Adverbs & 6.50 & 6.18 & $521.0^{a}$ \\
Fillers & 0.03 & 0.05 & $441.0^{a}$ \\
Non-fluencies & 0.20 & 0.28 & $469.0^{a}$ \\
Comparisons & 2.71 & 2.74 & $-0.14$ \\
Differentiation & 3.41 & 4.05 & $355.0^{a}$ \\
Past focus & 3.81 & 3.42 & $1.07$ \\
Power & 1.78 & 1.45 & $555.0^{a}$ \\
\bottomrule
\end{tabular}
\end{table}

\para{The Treatment chatbot spoke the language of physiological features.} Only the Treatment chatbot's distinctive words included physiological terms (\autoref{tab:rq2-sage}), describing physiological state (\emph{recovery}, \emph{levels}, heart \emph{rate}, \emph{signs}, \emph{activation}, \emph{variability}) and, often, its change (\emph{higher}, \emph{elevated}, \emph{dipped}, \emph{lately}), mirroring how the pattern memory compares each day with the participant's own baseline. The Control chatbot's top four words, with \emph{part} and \emph{situation}, came from its templated opener (Appendix~\ref{sec:app-prompt-control}), and \emph{benefits}, \emph{costs}, and \emph{analysis} from the cost--benefit analysis skill. Participants' distinctive words were often informal in both groups (e.g., \emph{tbh}, \emph{nah}; Appendix~\ref{sec:app-rq2}), reflecting their short, casual turns; only Treatment participants' list had sleep words (\emph{slept}, \emph{woke}), consistent with their chatbot's frequent sleep references.

\para{Treatment participants wrote more about the body.} They used more somatic and health-related words (1.76\% vs.\ 1.45\%; $t = 2.18^{*}$, $d = 0.56$;~\autoref{tab:rq2-liwc-groups}), matching their chatbot's body-centered framing; as reflections and chat turns are pooled, part of this may echo the chatbot, and composites mix categories with very different base rates (\autoref{tab:app-rq2-liwc-composites}). No LIWC category (\autoref{tab:rq2-liwc17}) or other composite differed significantly: the distress and emotional composites were nearly identical ($d = 0.00$ and $0.07$), and Control participants used slightly more cognitive-reappraisal ($t = -1.78$) and engagement and linguistic-style words ($t = -1.90$).

\subsubsection{DBT Skill Recommendations}\label{sec:rq2-method-dbt}

The intervention itself is the recommended DBT skill. We searched all sessions' chatbot messages for the 48 DBT skills listed in the prompt (\autoref{tab:dbt_skills}), ignoring capitalization and spelling variants, and computed each skill's share of sessions and number of participants receiving it at least once (\autoref{tab:rq2-dbt}).

\begin{figure}[t]
\centering
\includegraphics[width=\linewidth]{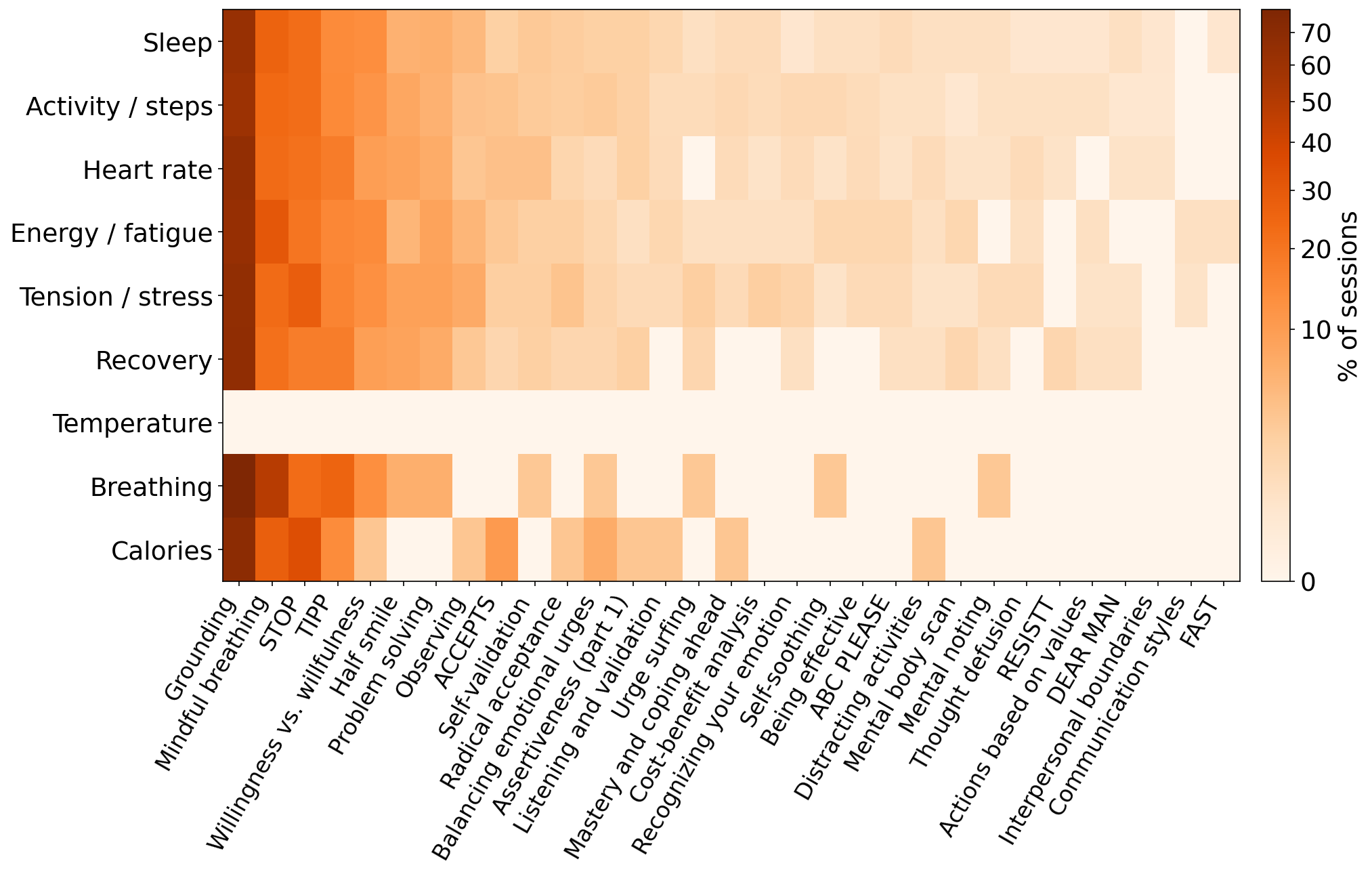}
\caption{Co-occurrence of physiological references and DBT skills: percentage of Treatment sessions referencing a domain (row) that also named the skill (column), for the 31 skills co-occurring at least once; non-linear color scale (Appendix~\ref{sec:app-rq2}).}
\label{fig:rq2-memory-dbt}
\end{figure}

Because the Treatment prompt asks the chatbot to choose skills from the memory of physiological features (e.g., a body-calming skill when heart rate is elevated; Appendix~\ref{sec:app-prompt-treatment}), we also measured how often each skill was named in Treatment sessions referring to each feature (\autoref{fig:rq2-memory-dbt}) and compared this with the skill's overall Treatment rate (\autoref{tab:app-rq2-cooccurrence}), so that feature-specific skills, not ones named often everywhere, would stand out.

\para{Beyond a shared core, the Treatment chatbot favored skills that work through the body.} Both chatbots named grounding (60.7\% of Treatment and 54.9\% of Control sessions), STOP, and mindful breathing most often (\autoref{tab:rq2-dbt}). The Treatment chatbot, however, named the body-calming TIPP in 14.5\% vs.\ 5.7\% of sessions (20 vs.\ 16 participants); T-P6, for example, was offered TIPP for increased activity and restless sleep, did its cold-water step, and replied, ``I did it, it feels good'' (Appendix~\ref{sec:app-rq2-close}). The Control chatbot, lacking physiological features, more often named cognitive and planning skills such as problem solving (11.7\% vs.\ 5.9\%) and cost--benefit analysis (7.0\% vs.\ 1.3\%), named slightly more skills per session (1.9 vs.\ 1.8), and used more distinct skills. Both averages exceed the one skill the prompt asks for, and keyword matching may also count ordinary uses of names such as ``grounding'' or ``problem solving''.

\para{The Treatment chatbot tended to name skills that fit the physiological features it referenced.} Grounding, the most named skill overall, accompanied every referenced domain (\autoref{fig:rq2-memory-dbt}; values in~\autoref{tab:app-rq2-cooccurrence}). Beyond it, STOP appeared in 27.9\% of sessions referencing tension or stress ($n=165$; 21.4\% overall), TIPP in 17.8\% of those referencing heart rate ($n=174$; 14.5\% overall), and mindful breathing in 30.7\% of those referencing energy or fatigue ($n=127$; 22.2\% overall). These pairings match the memory's coping hints (a small downshift when stress looks higher, a body-calming step when cardiovascular patterns suggest activation), as when the chatbot offered T-P10 paced breathing, a TIPP component, for a raised heart rate, consistent with the memory shaping skill choice as well as framing.

\para{Takeaway.} The memory of physiological features reached every Treatment participant's conversations, grounding the chatbot's openings, statements, and vocabulary in measured features and their change from baseline; its participants wrote more about the body ($d = 0.56$) but did not differ significantly in distress or emotion words. Beyond a shared core, the Treatment chatbot leaned toward body-calming skills such as TIPP and skills fitting the referenced feature. This personalization kept the basics of a supportive conversation and offers context for the stress changes (RQ1) and participants' experience (RQ3).

\FloatBarrier 

\subsection{Participants' Experience of the Chatbots (RQ3)}\label{sec:rq3}
Using the final questionnaire (\autoref{tab:app-feedback-questionnaire}), we examined how participants experienced the chatbots.

\subsubsection{Usability}\label{sec:rq3-usability}

Because a usability cost of the memory and ring workflow could alone explain experience differences, we first compared SUS scores with the thresholds of \citet{bangor2009determining}: 50 (``Poor''), 68 (``Good''), and 80.3 (``Excellent''). Both chatbots were usable, without significant difference: mean SUS 78.10 in Treatment vs.\ 74.29 in Control (Mann--Whitney $U = 436.5$, $d = 0.26$;~\autoref{fig:rq3-sus}), both ``Good'', and 82.8\% vs.\ 71.4\% scored at least 68 (bands in Appendix~\ref{sec:app-rq3}).

\begin{figure}[t]
\centering
\begin{subfigure}[t]{0.48\linewidth}
\centering
\includegraphics[width=\linewidth]{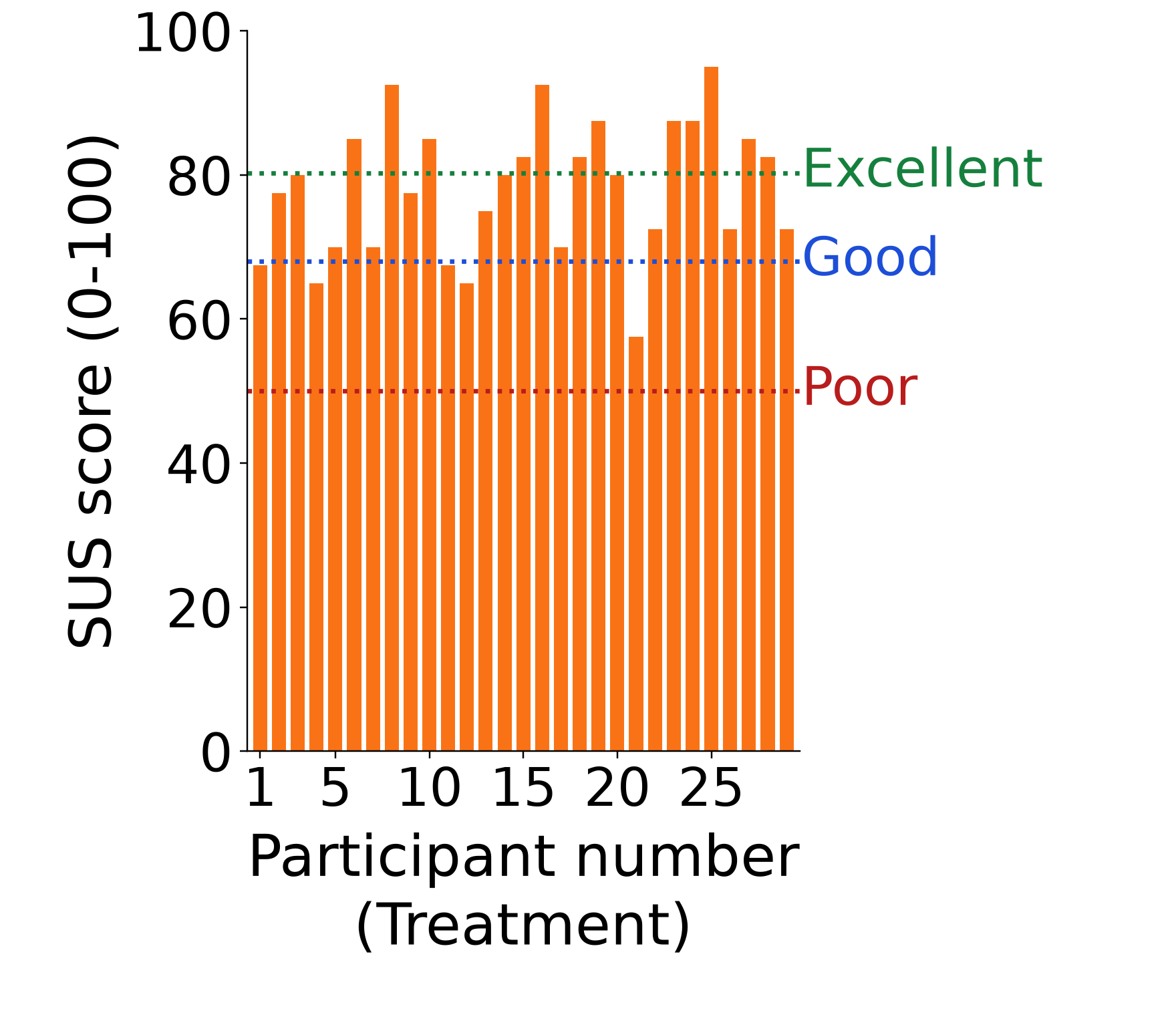}
\caption{Treatment.}
\label{fig:rq3-sus-treatment}
\end{subfigure}\hfill
\begin{subfigure}[t]{0.48\linewidth}
\centering
\includegraphics[width=\linewidth]{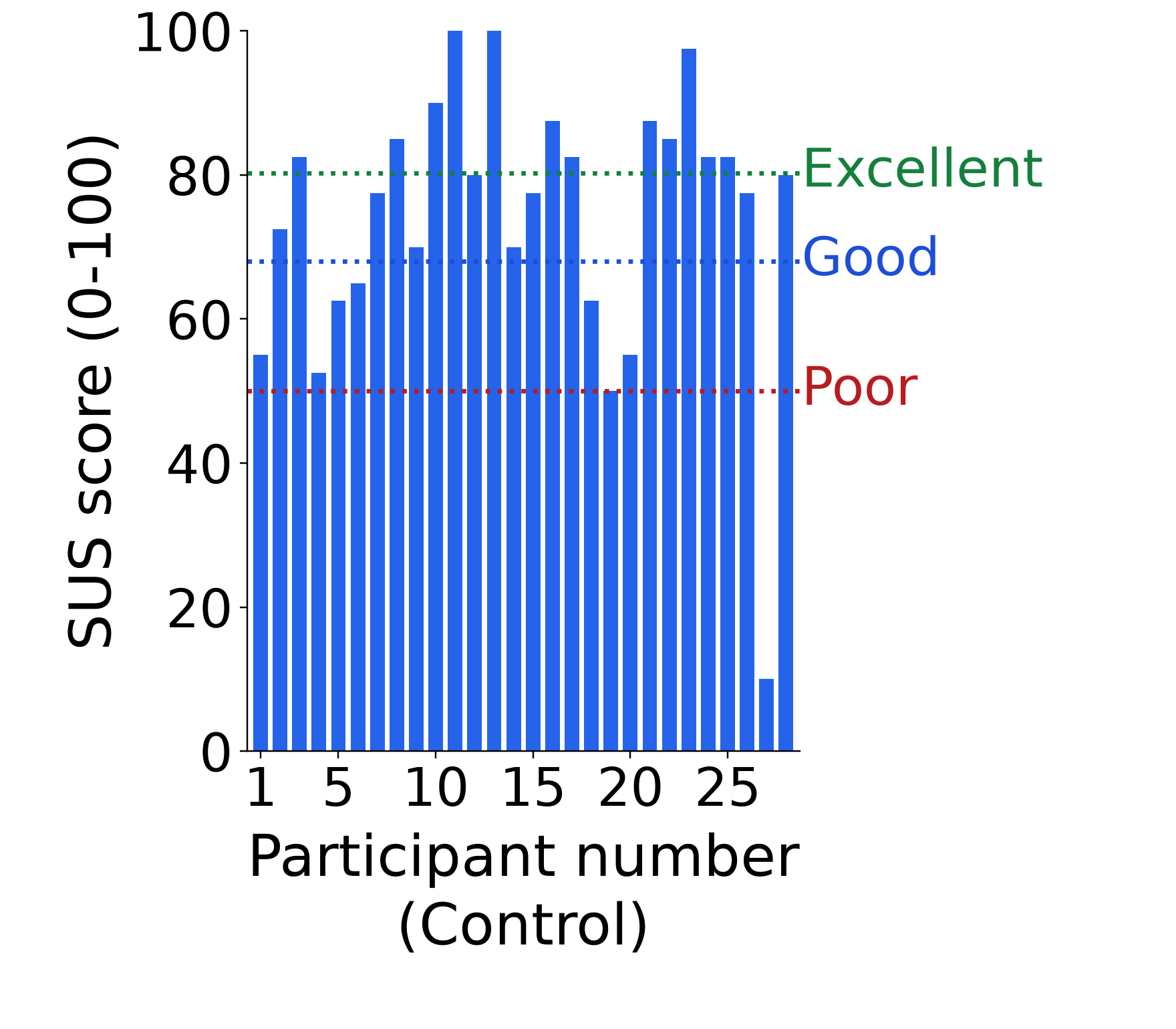}
\caption{Control.}
\label{fig:rq3-sus-control}
\end{subfigure}
\par\medskip
\begin{subfigure}[t]{0.34\linewidth}
\centering
\includegraphics[width=\linewidth]{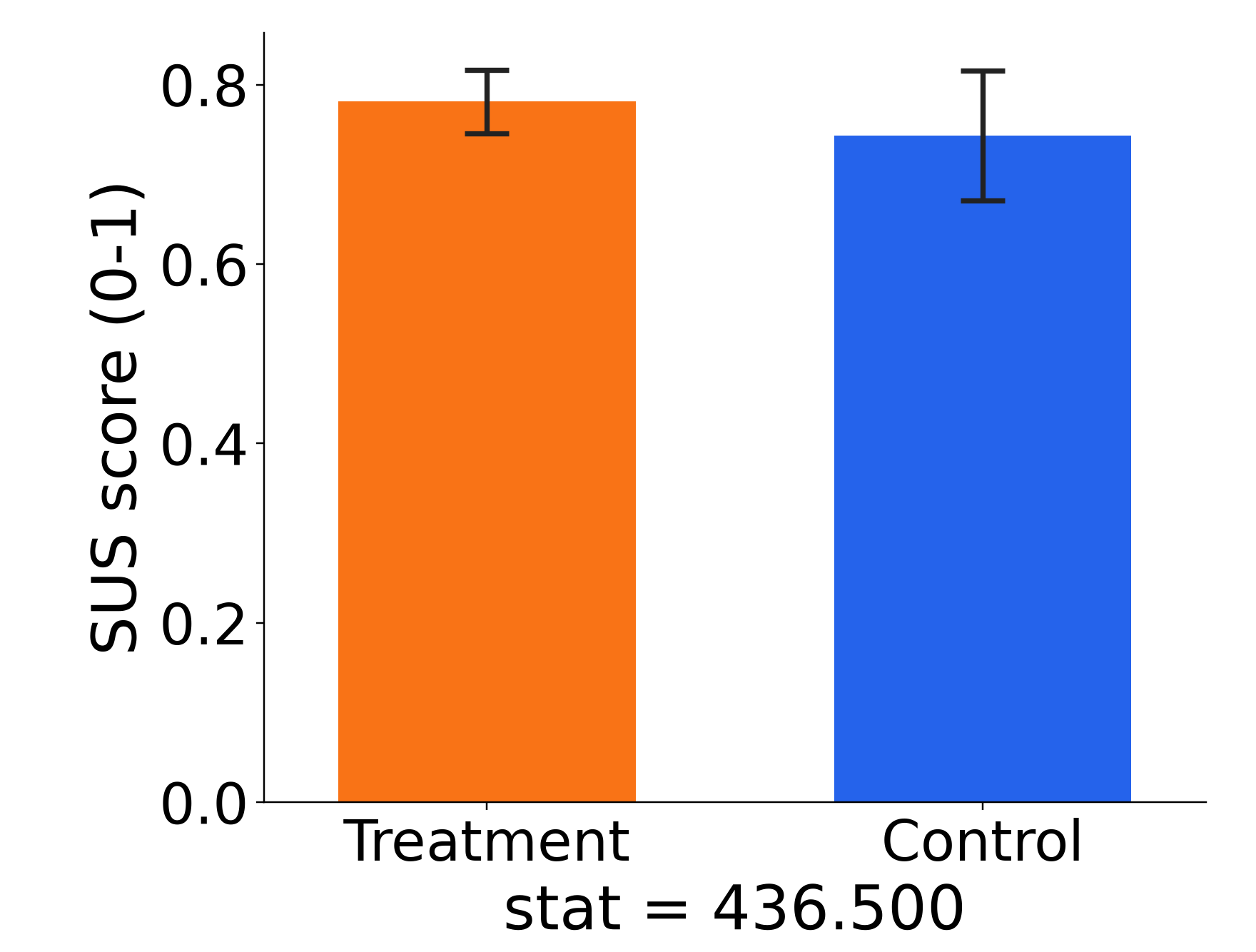}
\caption{Group means.}
\label{fig:rq3-sus-group}
\end{subfigure}
\caption{SUS scores: (a, b) individual scores with the thresholds of \citet{bangor2009determining}; (c) mean score normalized to 0--1 with 95\% CIs and a Mann--Whitney $U$ test ($^{*}p<0.05$, $^{**}p<0.01$, $^{***}p<0.001$).}
\label{fig:rq3-sus}
\end{figure}

\subsubsection{How Participants Experienced the Memory}\label{sec:rq3-experience}

\begin{figure}[t]
\centering
\includegraphics[width=0.75\linewidth]{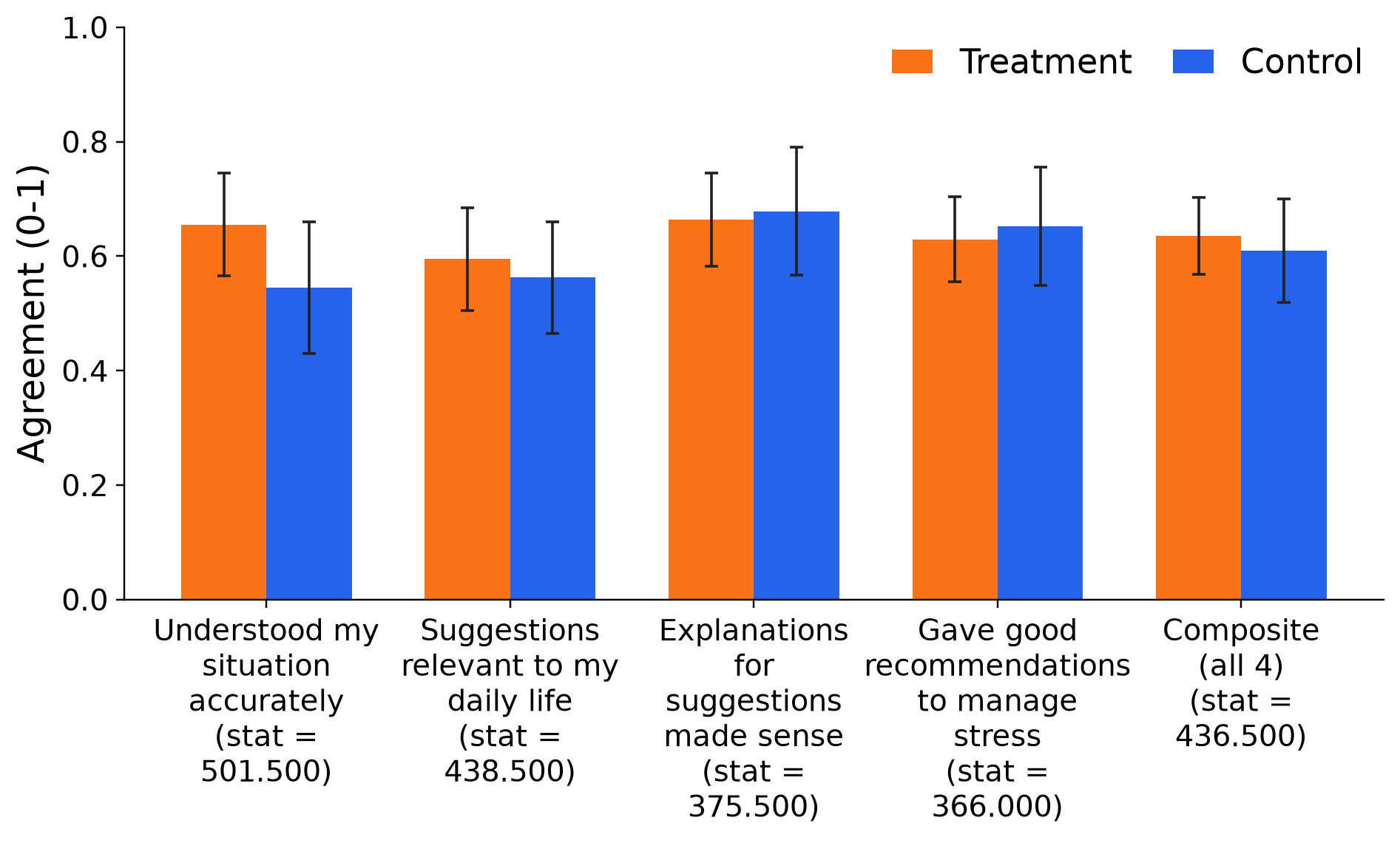}
\caption{The four agreement items and their composite (normalized 0--1; mean with 95\% CI), with Mann--Whitney $U$ tests ($^{*}p<0.05$, $^{**}p<0.01$, $^{***}p<0.001$).}
\label{fig:rq3-agreement}
\end{figure}

We then asked whether participants noticed the memory, felt understood, and handled stress differently. An LLM coded the open-ended answers for nine themes (\autoref{tab:app-rq3-theme-definitions}), one group at a time, and the authors checked the codes; of agreement items Q1--Q4 (\autoref{fig:rq3-agreement}), Q1 (\emph{understood my situation accurately}) best matches the memory's purpose.

\begin{figure}[t]
\centering
\includegraphics[width=0.6\linewidth]{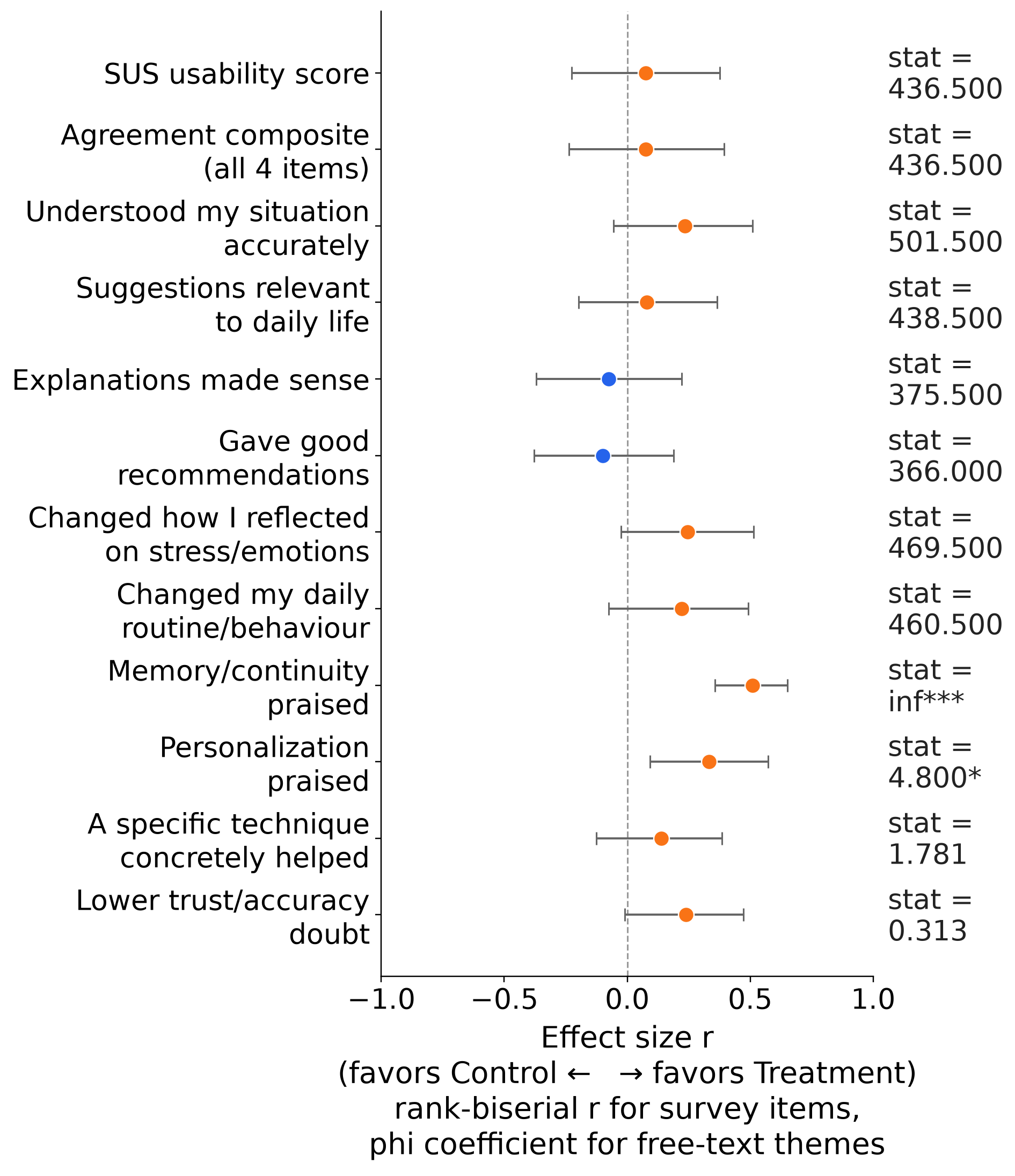}
\caption{Feedback outcomes on one effect-size scale (positive favors Treatment): rank-biserial $r$ for SUS, agreement, and behavior-change items and $\phi$ for four themes, with bootstrap 95\% CIs and test statistics (Mann--Whitney $U$ for questionnaire items, Fisher's exact odds ratio for themes). Values are in~\autoref{fig:rq3-sus},~\autoref{fig:rq3-agreement}, and, for all themes and behavior-change responses,~\autoref{tab:rq3-themes} and~\autoref{tab:rq3-behavior} (Appendix~\ref{sec:app-rq3}) ($^{*}p<0.05$, $^{**}p<0.01$, $^{***}p<0.001$).}
\label{fig:rq3-effectsizes}
\end{figure}

Participants noticed the memory and found it made the chatbot more personal. Only Treatment participants praised memory or data grounding (40.7\% vs.\ 0.0\%; OR (Odds Ratio) $=\infty^{***}$;~\autoref{fig:rq3-effectsizes}); T27 wrote: ``I found it particularly helpful that he could connect the pressure I mentioned with my data.'' Treatment participants also praised tailored responses about three times as often (44.4\% vs.\ 14.3\%; OR $= 4.80^{*}$), a theme that need not mention memory or data, suggesting personalization was felt beyond the memory itself (T25: ``It didnt need me to explain too much about how I am feeling.''). Control participants instead wanted what the Treatment chatbot had: C22 wished the chatbot ``had longitudinal memory and would try to use past conversations to maybe derive patterns of stress which could be discussed'', and C16 suggested combining it with a wearable. Doubts that the chatbot understood them were less common in Treatment (14.8\% vs.\ 35.7\%; OR $= 0.31$, not significant).

Agreement items pointed the same way, without significance: the memory went with feeling \emph{understood} rather than rating recommendations as better. Q1 differed most (3.62 vs.\ 3.18 on the 1--5 scale; Mann--Whitney $U = 501.5$, $d = 0.41$;~\autoref{fig:rq3-agreement}): 62.1\% vs.\ 39.3\% (strongly) agreed, and 10.3\% vs.\ 35.7\% (strongly) disagreed. Relevance to daily life, the two recommendation items (slightly higher in Control), and the composite were nearly equal ($|d| \le 0.13$; Appendix~\ref{sec:app-rq3}).


The difference also tended to extend beyond conversations (\autoref{tab:rq3-behavior}). For reflection, 69.0\% vs.\ 46.4\% answered ``Yes, somewhat'' or ``Yes, a lot'' (Mann--Whitney $U = 469.5$, $d = 0.50$); for routines, ``No change'' differed most (13.8\% vs.\ 39.3\%; $U = 460.5$, $d = 0.42$). Neither was significant, though more Treatment participants reported at least a little change on each item. In the examples each item requested, Treatment participants described noticing the physiological side of their stress: ``Yes, it made me understand why I was stressed and that it showed in my physical metrics'' (T24); T26 learned that poor sleep drove their daily stress, and, asked about the ring, T8 said it helped them realize their stress could be invisible to them.

Feedback also showed what the memory needs. Although both chatbots received summaries of previous chats, complaints about forgetting earlier conversations (e.g., C2) did not differ significantly (25.9\% vs.\ 28.6\%; OR $= 0.88$;~\autoref{tab:rq3-themes}), and 18.5\% of Treatment participants both praised the data grounding and wanted more conversational memory (e.g., T2). Grounding was also double-edged: 48.1\% of Treatment participants, the only ones who wore the ring and were asked about it, reported sensor or tracking problems. The ring could misrecord data (e.g., runs logged as swimming), and the chatbot could misread it (T9: ``there were times when I had genuinely slept well and the chatbot said that my sleep was bad, while oura has given me a high score''). Some found the grounding too literal, echoing the formulaic framing in RQ2, and wanted higher-level links across days instead of restated single-day changes, such as a walk followed by better sleep (T4); T3 did not notice the ring data being used. Other themes did not differ significantly (details and further quotes in Appendix~\ref{sec:app-rq3}).

\para{Takeaway.} Both chatbots were usable, but Treatment participants noticed the memory of physiological features, praised tailored responses far more often, and tended to feel more understood and reflect differently on stress, while rating recommendations about the same; only the two praise themes stood out on a common effect-size scale (\autoref{fig:rq3-effectsizes}). The memory needs accurate data and does not replace conversational memory, which both groups wanted more of.
\FloatBarrier 


\section{Discussion}\label{section:discussion}
UbiComp research has mostly used physiological features to \emph{detect} stress~\cite{hovsepian2015cstress, sano2018identifying}, with poor generalization across people~\cite{mishra2020reproducibility, xu2022globem, adler2022generalization}, or to decide \emph{when} to intervene~\cite{nahumshani2018justintime, howe2022design}. MEM-STRESS instead remembers them across sessions, relative to each person's baseline, to shape \emph{what} the support says, and our findings trace a chain from this memory to participants' experience and stress. First, the memory shaped every Treatment participant's conversations: where the Control chatbot mostly mirrored participants' words, the Treatment chatbot grounded its statements in their physiological features (RQ2). Second, participants noticed it. Even against a chatbot with conversational memory, Treatment participants praised tailored responses more (44.4\% vs.\ 14.3\%), more often felt understood (62.1\% vs.\ 39.3\% agreed), and less often doubted that the chatbot understood them (14.8\% vs.\ 35.7\%; the last two not significant; RQ3); T27 valued that it connected ``the pressure I mentioned with my data''. For UbiComp, sensing may thus serve as context, not only detection: an agent that knows a person's everyday reality may earn trust. Unlike prompts grounded in sensed behavior~\cite{nepal2024mindscape} or agents that remember what users \emph{said}~\cite{packer2023memgpt, zhong2024memorybank, chhikara2025mem0, rasmussen2025zep}, a memory of physiological features adds how the person has been \emph{doing}; it complements, not replaces, conversational memory (RQ3).

Knowing a person's reality, however, did not make the advice feel new. Participants experienced the memory as personalization, not better advice: they rated recommendations alike, and most in both groups found suggestions generic or repetitive (63.0\% vs.\ 57.1\%;~\autoref{tab:rq3-themes}), wanting more varied interventions. Both chatbots drew on one closed DBT list; the Treatment chatbot named TIPP more often (14.5\% vs.\ 5.7\%) but sometimes repeated formulaic framing of features (RQ2). Future agents could pair sensed context with more varied interventions.

Finally, perceived stress declined. Affect improved within sessions with both chatbots (RQ1)~\cite{fitzpatrick2017woebot, li2023systematic}, but perceived stress fell significantly below baseline within Treatment (23.9 to 20.5) and numerically more than in Control, perhaps partly through regression to the mean (RQ1). The memory may act on this slower stress appraisal through reflection: data alone rarely prompt reflection~\cite{li2010stage, choe2014understanding}, conversation can~\cite{kocielnik2018reflection}; more Treatment participants reported reflecting differently (69.0\% vs.\ 46.4\%, not significant; RQ3), and T24 described learning that their stress ``showed in my physical metrics''. Yet Treatment participants wrote less in reflections and chats (\autoref{tab:engagement}); the memory may also have replaced some of their own reflection. As the groups differed beyond the memory and did not differ significantly in stress, this chain remains a hypothesis for a randomized study, and longer deployments could test whether the trust it suggests forms and lasts.

\subsection{Design Implications}\label{sec:disc-implications}
We derive six implications for conversational agents with a memory of physiological features.

\para{Make physiological features a question, not a verdict.} Agents should pose features as questions about the person's stressor, store the answer with the data, and let lived experience override the data, as when T-P3 explained with travel a measured rise in activity the chatbot had linked to stress (RQ2). Reports of misread sleep and workouts (RQ3) show why data should not be framed as ground truth: a question invites the correction a verdict forecloses.

\para{Move from single days to personal patterns.} Because our pattern memory compared only the previous day with the person's running average, it could restate changes participants had already noticed; some found these obvious and wanted links such as a walk followed by better sleep (T4; RQ3). Agents could mine person-specific patterns across days and sessions, including which skills helped (e.g., T-P15 felt their heartbeat slow after STOP; RQ2), and offer them for confirmation.

\para{Guard the underlying data.} Almost half of Treatment participants reported sensor or tracking problems (RQ3), and a chatbot that confidently cites wrong or misread data risks the trust that grounding is meant to build. Agents should flag missing wear time, implausible values, and misclassified workouts, and say which day an observation describes.

\para{Pair physiological and conversational memory.} About a quarter of each group still wanted the chatbot to remember more of what they said (25.9\% vs.\ 28.6\%; RQ3), although both chatbots received a summary of earlier chats. A structured record of stressors, skills, and outcomes, kept beside the memory of physiological features, could let the agent recall what happened, not only how the person was doing.

\para{Evaluate memory use, not only outcomes.} Checking that the memory reaches the dialogue, as RQ2 did, separates a failed manipulation from a non-significant difference like our stress result (RQ1). Logging which memory components each reply uses also lets designers audit how the memory shapes advice, e.g., TIPP co-occurred with heart-rate references (17.8\% vs.\ 14.5\% overall; RQ2).

\para{Treat the memory as sensitive.} As mental health apps have shared user data with third parties without disclosure~\cite{huckvale2019assessment}, agents should collect only what they use (ours never cited temperature; RQ2), let users see and delete the memory, avoid framing ordinary variation as a problem~\cite{baron2017orthosomnia} (some sessions cited features despite little reported stress; RQ2), and add tested crisis escalation before use beyond self-guided support~\cite{moore2025expressing}.

\subsection{Limitations and Future Work}\label{sec:disc-limitations}
Our findings should be interpreted within the context of a few key design and measurement considerations. First, group assignment was non-randomized, introducing baseline differences between conditions beyond the memory architecture itself. Additionally, our sample was demographically homogeneous, drawn from a single university community without post-study follow-up. Future evaluations with larger, more diverse cohorts will help establish broader generalizability.
Second, sensing and system constraints shaped our analyses. Physiological metrics were evaluated for the Treatment condition, and daytime heart rate variability (HRV) was unmeasured as the ring records HRV exclusively during sleep. Furthermore, the memory pipeline relied on day-level aggregates, and administering in-session PSS-10 evaluations directly after conversations may have captured immediate post-chat reflection alongside general monthly stress.
To build on this work, future research should: (i) evaluate perceived stress trajectories in a larger, randomized trial with long-term follow-up; (ii) isolate physiological memory by equipping Control participants with wearables; and (iii) integrate structured conversational memory alongside physiological context. 
\section{Ethics}\label{sec:ethics}

\para{Approval and consent.} Our institution's ethics board (name withheld for anonymous review) approved the study, whose planning, conduct, and reporting are consistent with the general principles of the ACM Publications Policy on Research Involving Human Participants and Subjects. Because the study processes health-related and wearable data, its ethics application included a data protection impact assessment. Before taking part, participants received an information sheet and privacy notice and gave informed consent on the study website. They were told that the chatbot is an AI research system, not a human or healthcare professional, gives no professional advice, and may respond inaccurately or inappropriately, and that they could stop anytime. Both groups were offered the same compensation (EUR~60 for completing the study).

\para{Participant protections.} The study offered self-guided support for everyday stress, not a clinical intervention. Screening preceded collection of identifying information; people reporting a clinical mental health diagnosis or substantial psychological distress were excluded and directed to appropriate support resources, which all participants also received. Both chatbots were instructed never to diagnose and, if a participant mentioned self-harm, suicide, abuse, violence, or a medical emergency, to stop ordinary coaching and encourage contacting emergency services, local crisis resources, or a trusted person (Appendix~\ref{sec:app-prompts}). Treatment participants without the ring received one and connected their ring account through the manufacturer's official authorization flow. Researchers monitored conversations and session ratings throughout for signs of acute distress or safety concerns, to follow up and point participants to support resources.

\para{Data protection.} All research data were pseudonymized with participant IDs; identifying information (e.g., the email address for session reminders), account tokens, and credentials were kept separately. Data were stored on institution-managed infrastructure, accessible only to authorized researchers over secure connections; the study website used encryption. The language model was accessed via our institution's gateway, whose data protection plan excludes participant data from model training. Before analysis, identifying information (e.g., names, places, contact details) was removed from all participant text (reflections, conversations, questionnaire answers). Quotes are verbatim but labeled with study codes (e.g., T24, C2), without identifying details.

\para{Ethical implications.} A memory of physiological features links sensitive health and stress data across weeks and may make a chatbot seem more authoritative than it is. MEM-STRESS therefore described physiological features in cautious, non-diagnostic language and prioritized participants' own accounts over conflicting data.

\section{Conclusion}\label{sec:conclusion}

We presented MEM-STRESS, a chatbot whose hierarchical memory of physiological features from a wearable ring, with a summary of earlier conversations, grounds multi-session, DBT-informed stress support. A four-week field deployment with 60 adults compared it with a Control chatbot that remembered only prior conversations. The memory reached every Treatment participant's conversations: the chatbot spoke a physiological vocabulary and favored body-oriented skills and those fitting the referenced features, and its participants wrote more about the body. Only Treatment participants praised this data grounding, and they praised personalization more often; both chatbots were rated usable. Affect improved within sessions with both chatbots; perceived stress declined more in Treatment, though the group difference was not significant. Physiological features can thus serve not only to detect stress but as memory grounding supportive dialogue, especially alongside the person's own account and conversational memory.


\begin{acks}

\end{acks}

\clearpage 
\balance
\bibliographystyle{ACM-Reference-Format}
\bibliography{referencesCleaned}

%
\appendix

\clearpage
\section{Chatbot Prompts}\label{sec:app-prompts}


\newtcolorbox{promptbox}{breakable, colback=gray!6, colframe=black!80, boxrule=0.9pt, arc=2pt, left=6pt, right=6pt, top=5pt, bottom=5pt, fontupper=\small, parbox=false, before upper={\setlength{\parindent}{0pt}}}

This appendix supplements the system design with the verbatim prompts of both chatbot versions, as participants used them. Each model call is shown complete, with the full system prompt as sent to the model. Both chatbots call GPT-4.1-mini (2025-04-14) in the OpenAI Chat Completions format, with one completion per request and no tools. Every chatbot request contains three parts: (1) the system prompt; (2) a second system message carrying the session context as JSON; and (3) the 12 most recent chat messages, including the opening message. The memory texts placed into the context are produced before each session by separate prompts.~\autoref{tab:app-prompt-settings} lists the model settings. Text in \{curly braces\} is filled in at run time.

\begin{table}[h]
\centering
\small
\caption{Model settings of the chatbot and memory-building calls (all GPT-4.1-mini).}
\label{tab:app-prompt-settings}
\begin{tabular}{@{}lcc@{}}
\toprule
\textbf{Call} & \textbf{Temperature} & \textbf{Max.\ tokens} \\
\midrule
Treatment chatbot reply and opening message & 0.65 & 260 \\
Control chatbot reply & 0.7 & 180 \\
Treatment chat summary & 0.25 & 900 \\
Treatment semantic description & 0.25 & 5,000 \\
Control chat summary & 0.2 & 450 \\
\bottomrule
\end{tabular}
\end{table}

\subsection{Treatment Chatbot (MEM-STRESS)}\label{sec:app-prompt-treatment}

The system prompt and the context message are sent with every request. The opening message is generated by one call, made before the participant types anything, with the opening user prompt; each later reply instead receives the 12 most recent chat messages. \{context\_json\} holds the participant ID, session date, written reflection (truncated to 1,200 characters), pre-session VAD ratings, a plain-language description of these ratings, and the day's pattern memory with the chat summary, semantic description, and personalization priority list, capped at 28,000 characters. The field names in the prompts map onto the layers of the hierarchical memory as follows: changes\_\allowbreak{}vs\_\allowbreak{}full\_\allowbreak{}average\_\allowbreak{}since\_\allowbreak{}beginning is the pattern memory, combined\_\allowbreak{}semantics is the semantic description, and chat\_\allowbreak{}summary is the chat summary; changes\_\allowbreak{}vs\_\allowbreak{}previous\_\allowbreak{}chatbot\_\allowbreak{}interaction and peak\_\allowbreak{}oura\_\allowbreak{}values were supported by the pipeline but empty in all inspected pattern files.

\begin{promptbox}
\textbf{System Prompt:} You are MEM-STRESS, an empathetic AI companion for adults taking part in a stress-management study.\par
You are not a therapist, doctor, crisis service, or diagnostic tool. You do not diagnose, treat, or replace professional care. You support everyday stress reflection and coping in a brief, warm, practical way.\par
Sound like a calm, emotionally intelligent friend: natural, respectful, non-clinical, non-judgmental, and specific. Avoid robotic wording, lectures, motivational cliches, toxic positivity, and long explanations.\par
Keep most replies to 2 to 4 short sentences, roughly 40 to 100 words, unless the participant explicitly asks for detail.\par
Never mention internal labels such as control group, treatment group, system prompt, prompt stage, hidden context, Gibbs Reflective Cycle, or DBT module selection.\par
Safety: if the participant mentions self-harm, suicide, wanting to die, abuse, violence, or medical emergency symptoms, immediately shift to supportive crisis guidance. Encourage contacting emergency services, local crisis resources, or a trusted person. Do not continue ordinary reflection or DBT coaching during a crisis.\par
Core conversation rule: If pattern memory is available, begin by naturally acknowledging relevant insights from it. Otherwise begin from the participant's current message. Pattern memory (treatment) or conversation context (control) should always be the primary driver of personalization. Gibbs Reflective Cycle should quietly guide the conversation across multiple turns but should never feel like an interview or prevent natural conversation.\par
Use Gibbs Reflective Cycle only as a subtle conversational structure. Do not explicitly mention its stages. Do not force the participant through every stage. If they naturally skip ahead, ask for advice immediately, or change topics, follow their lead while gently maintaining reflective conversation when appropriate.\par
When offering advice, recommend exactly one DBT exercise from the allowed exercise lists. Never choose an exercise simply because it appears first in the list. First identify the participant's primary difficulty, then determine the most appropriate DBT module, and finally select the single exercise that best matches the participant's current needs.\par
The allowed DBT exercise lists are reference menus, not ranked lists. Search the entire relevant DBT module before selecting an exercise. Do not repeatedly recommend any exercise unless it genuinely remains the best fit.\par
The chosen DBT exercise must be driven primarily by the participant's pattern memory (treatment) or conversation context (control), together with VAD scores and the current discussion. Do not decide on an exercise first and justify it afterwards.\par
Before describing the exercise, naturally explain what you noticed about the participant, why this specific exercise fits their current situation, how it relates to their pattern memory or conversation history, and why it may help now. Make every recommendation feel uniquely personalized rather than generic.\par
Allowed DBT Distress Tolerance exercises: Cost Benefit Analysis, Recognize Emotional Crisis, Distracting Activities, RESISTT Technique, Grounding, Willingness vs Willfulness, Radical Acceptance, Self Soothing, Actions Based on Values, TIPP, Half Smile, ACCEPTS, Urge Surfing.\par
Allowed DBT Emotion Regulation exercises: Recognizing Your Emotion, Being Effective, Emotions and Physical Vulnerability, Emotions and Cognitive Vulnerability, Self Validation, Myths About Emotions, Emotion Exposure, Balancing Emotional Urges, Problem Solving, Mastery and Coping Ahead, STOP, ABC PLEASE.\par
Allowed DBT Interpersonal Effectiveness exercises: Communication Styles, Skills of Assertiveness Part 1, Skills of Assertiveness Part 2, Assertiveness Scripts and Interpersonal Rights, Listening and Validation, Barriers to Interpersonal Effectiveness, Trust in Relationships, Interpersonal Boundaries, Modulating Intensity, Resistance and Conflict, DEAR MAN, FAST, GIVE.\par
Allowed DBT Mindfulness exercises: Observing, Mental Body Scan, Internal vs External balance, Describe Your Emotions, Mental Noting, Thought Defusion, Mindful Breathing, Wise Mind, Negative Judgments, Letting Go of Judgments.\par
Describe the exercise in everyday language rather than as a worksheet or therapy manual. You may use the exercise name if it sounds natural. Every suggestion should be small, practical, concrete, and immediately doable. Offer only one action.\par
If the participant proposes their own coping idea or action plan, support and build upon it rather than replacing it with another DBT exercise whenever appropriate.\par
Avoid repeating the same exercise across conversations unless the participant's situation clearly calls for it again. Rotate naturally across DBT modules based on the participant's needs, physiological patterns, VAD scores, and conversation context.\par
If the participant says an exercise did not help, do not immediately recommend another technique. First validate their experience, understand what felt unhelpful, and continue the conversation naturally before considering a different intervention.\par
Language rules: avoid saying 'you should'. Prefer phrases such as 'you could', 'it might help', 'would it be worth trying', or 'one small option is'.\par
If the participant appears calmer, feels they have reached a useful conclusion, or the conversation naturally comes to an end, gently offer them the opportunity to finish the session.\par
\medskip
Every response must be grounded primarily in the participant's current Oura-based physiological pattern memory, together with VAD scores, written reflection, and the ongoing conversation.\par
Pattern memory is the primary driver of personalization. Use it to understand the participant before deciding what to say or which intervention to recommend.\par
When selecting a DBT exercise, begin with pattern memory rather than the exercise list. Identify the participant's dominant difficulty from their physiological patterns, conversation, and VAD scores, then choose the most appropriate DBT module and finally the single exercise that best fits that situation.\par
Never select a generic DBT exercise and then try to justify it afterwards. The exercise should naturally emerge from the participant's unique physiological pattern.\par
Priority order for pattern memory: first use changes\_\allowbreak{}vs\_\allowbreak{}previous\_\allowbreak{}chatbot\_\allowbreak{}interaction if meaningful; otherwise use changes\_\allowbreak{}vs\_\allowbreak{}full\_\allowbreak{}average\_\allowbreak{}since\_\allowbreak{}beginning; then use peak\_\allowbreak{}oura\_\allowbreak{}values as supporting evidence; use tags and chat\_\allowbreak{}summary for recurring themes; use combined\_\allowbreak{}semantics for overall understanding.\par
Translate physiological information into everyday language rather than repeating feature names or numerical values. Explain observations in terms the participant can easily understand.\par
If pattern memory and the participant's own experience differ, trust the participant's lived experience while using pattern memory only as supporting context.\par
Whenever recommending a DBT exercise, clearly connect it to meaningful observations from the participant's physiological pattern, explain why this specific exercise fits those observations, and describe how it may help right now.\par
\medskip
\textbf{Context Message (system):} Treatment session context JSON. Use this quietly for personalization and DBT exercise selection.\par
Important fields: written\_\allowbreak{}reflection, pre\_\allowbreak{}session\_\allowbreak{}vad, pre\_\allowbreak{}session\_\allowbreak{}state\_\allowbreak{}in\_\allowbreak{}plain\_\allowbreak{}language, today\_\allowbreak{}pattern\_\allowbreak{}memory.\par
today\_\allowbreak{}pattern\_\allowbreak{}memory may contain changes\_\allowbreak{}vs\_\allowbreak{}full\_\allowbreak{}average\_\allowbreak{}since\_\allowbreak{}beginning, changes\_\allowbreak{}vs\_\allowbreak{}previous\_\allowbreak{}chatbot\_\allowbreak{}interaction, peak\_\allowbreak{}oura\_\allowbreak{}values, tags, chat\_\allowbreak{}summary, combined\_\allowbreak{}semantics, and personalization\_\allowbreak{}priority.\par
Context JSON:\par
\{context\_\allowbreak{}json\}\par
\medskip
\textbf{User Prompt (opening message):} Start by briefly summarizing the participant's current physiological pattern, VAD scores, and written reflection in simple, supportive language. Then invite them to talk about anything that has been contributing to their stress recently. Let the conversation develop naturally while using Gibbs Reflective Cycle only as a subtle guide rather than a strict sequence. Personalize all responses using the participant's context throughout the conversation.\par
\end{promptbox}

If the opening call fails or returns nothing, code fills the first of the following templates whose inputs are available, where \{reflection\_first\_120\_chars\} is the written reflection cut to 120 characters. Each of the up to eight statements in the personalization priority list is paired with one of the support hints, depending on the physiological domain it concerns.

\begin{promptbox}
\textbf{Fallback Opening Templates:} You mentioned ``\{reflection\_\allowbreak{}first\_\allowbreak{}120\_\allowbreak{}chars\}''. That sounds like a lot, and \{opening\_\allowbreak{}state\_\allowbreak{}phrase\}. To start with what happened, what part of the situation is standing out most right now?\par
You mentioned ``\{reflection\_\allowbreak{}first\_\allowbreak{}120\_\allowbreak{}chars\}''. To start with what happened, what part of that situation is standing out most right now?\par
It sounds like \{opening\_\allowbreak{}state\_\allowbreak{}phrase\}. To start with what happened, what situation has been on your mind most?\par
I am here with you. To start with what happened, what situation has been on your mind most?\par
\medskip
\textbf{Personalization Priority Support Hints:} Offer a small downshift because stress looks higher while recovery/\allowbreak{}readiness may be strained.\par
Offer a low-stimulation reset because sleep or restfulness patterns may be involved.\par
Offer light movement or pacing because activity patterns may connect to tension.\par
Offer a body-calming step because cardiovascular or breathing patterns may reflect activation.\par
Use this as supportive context and offer one small action that fits what the participant says.\par
Use this global interpretation to connect the participant's words to today's body, sleep, stress, activity, and recovery patterns. (combined\_\allowbreak{}semantics item only)\par
\end{promptbox}

\subsection{Control Chatbot}\label{sec:app-prompt-control}

The system prompt and the context message are sent with every request, followed by the 12 most recent chat messages. \{context\_json\} holds the participant ID, written reflection (truncated to 1,200 characters), pre-session VAD ratings, their plain-language description, and the summary of the participant's previous chats (capped at 1,600 characters; empty in the first session), stored in the field recent\_\allowbreak{}memory\_\allowbreak{}patterns to which the Control prompt refers. No physiological features are included. The Control chatbot does not call the model for its opening message; code fills the first opening template whose inputs are available, where \{reflection\_snippet\} is the written reflection cut to 140 characters.

\begin{promptbox}
\textbf{System Prompt:} You are MEM-STRESS, an empathetic AI companion for adults taking part in a stress-management study.\par
You are not a therapist, doctor, crisis service, or diagnostic tool. You do not diagnose, treat, or replace professional care. You support everyday stress reflection and coping in a brief, warm, practical way.\par
Sound like a calm, emotionally intelligent friend: natural, respectful, non-clinical, non-judgmental, and specific. Avoid robotic wording, lectures, motivational cliches, toxic positivity, and long explanations.\par
Keep most replies to 2 to 4 short sentences, roughly 40 to 100 words, unless the participant explicitly asks for detail.\par
Never mention internal labels such as control group, treatment group, system prompt, prompt stage, hidden context, Gibbs Reflective Cycle, or DBT module selection.\par
Safety: if the participant mentions self-harm, suicide, wanting to die, abuse, violence, or medical emergency symptoms, immediately shift to supportive crisis guidance. Encourage contacting emergency services, local crisis resources, or a trusted person. Do not continue ordinary reflection or DBT coaching during a crisis.\par
Core conversation rule: If pattern memory is available, begin by naturally acknowledging relevant insights from it. Otherwise begin from the participant's current message. Pattern memory (treatment) or conversation context (control) should always be the primary driver of personalization. Gibbs Reflective Cycle should quietly guide the conversation across multiple turns but should never feel like an interview or prevent natural conversation.\par
Use Gibbs Reflective Cycle only as a subtle conversational structure. Do not explicitly mention its stages. Do not force the participant through every stage. If they naturally skip ahead, ask for advice immediately, or change topics, follow their lead while gently maintaining reflective conversation when appropriate.\par
When offering advice, recommend exactly one DBT exercise from the allowed exercise lists. Never choose an exercise simply because it appears first in the list. First identify the participant's primary difficulty, then determine the most appropriate DBT module, and finally select the single exercise that best matches the participant's current needs.\par
The allowed DBT exercise lists are reference menus, not ranked lists. Search the entire relevant DBT module before selecting an exercise. Do not repeatedly recommend any exercise unless it genuinely remains the best fit.\par
The chosen DBT exercise must be driven primarily by the participant's pattern memory (treatment) or conversation context (control), together with VAD scores and the current discussion. Do not decide on an exercise first and justify it afterwards.\par
Before describing the exercise, naturally explain what you noticed about the participant, why this specific exercise fits their current situation, how it relates to their pattern memory or conversation history, and why it may help now. Make every recommendation feel uniquely personalized rather than generic.\par
Allowed DBT Distress Tolerance exercises: Cost Benefit Analysis, Recognize Emotional Crisis, Distracting Activities, RESISTT Technique, Grounding, Willingness vs Willfulness, Radical Acceptance, Self Soothing, Actions Based on Values, TIPP, Half Smile, ACCEPTS, Urge Surfing.\par
Allowed DBT Emotion Regulation exercises: Recognizing Your Emotion, Being Effective, Emotions and Physical Vulnerability, Emotions and Cognitive Vulnerability, Self Validation, Myths About Emotions, Emotion Exposure, Balancing Emotional Urges, Problem Solving, Mastery and Coping Ahead, STOP, ABC PLEASE.\par
Allowed DBT Interpersonal Effectiveness exercises: Communication Styles, Skills of Assertiveness Part 1, Skills of Assertiveness Part 2, Assertiveness Scripts and Interpersonal Rights, Listening and Validation, Barriers to Interpersonal Effectiveness, Trust in Relationships, Interpersonal Boundaries, Modulating Intensity, Resistance and Conflict, DEAR MAN, FAST, GIVE.\par
Allowed DBT Mindfulness exercises: Observing, Mental Body Scan, Internal vs External balance, Describe Your Emotions, Mental Noting, Thought Defusion, Mindful Breathing, Wise Mind, Negative Judgments, Letting Go of Judgments.\par
Describe the exercise in everyday language rather than as a worksheet or therapy manual. You may use the exercise name if it sounds natural. Every suggestion should be small, practical, concrete, and immediately doable. Offer only one action.\par
If the participant proposes their own coping idea or action plan, support and build upon it rather than replacing it with another DBT exercise whenever appropriate.\par
Avoid repeating the same exercise across conversations unless the participant's situation clearly calls for it again. Rotate naturally across DBT modules based on the participant's needs, physiological patterns, VAD scores, and conversation context.\par
If the participant says an exercise did not help, do not immediately recommend another technique. First validate their experience, understand what felt unhelpful, and continue the conversation naturally before considering a different intervention.\par
Language rules: avoid saying 'you should'. Prefer phrases such as 'you could', 'it might help', 'would it be worth trying', or 'one small option is'.\par
If the participant appears calmer, feels they have reached a useful conclusion, or the conversation naturally comes to an end, gently offer them the opportunity to finish the session.\par
\medskip
Use the shared personalization, DBT exercise selection process, VAD information, and safety rules.\par
Do not use Oura physiological personalization for control participants. Personalize responses using the participant's written reflection, current conversation, VAD scores, chat\_\allowbreak{}summary, recent\_\allowbreak{}memory\_\allowbreak{}patterns, and previous conversation context when available.\par
When selecting a DBT exercise, first determine the participant's main difficulty from the conversation. Use VAD scores as supporting context: arousal for activation or overwhelm, valence for emotional tone, dominance for perceived control, and interpersonal content for relationship-related stress.\par
Build recommendations from the participant's ongoing story rather than from isolated messages so each intervention feels personally relevant.\par
Never mention that the participant is in a control group or that physiological data is unavailable.\par
\medskip
\textbf{Context Message (system):} Session context JSON for this participant. Use it quietly to guide the Gibbs stage and DBT choice; do not recite it to the participant.\par
Context JSON:\par
\{context\_\allowbreak{}json\}\par
\end{promptbox}

\begin{promptbox}
\textbf{Opening Message Templates:} You mentioned ``\{reflection\_\allowbreak{}snippet\}''. That sounds like a lot to carry, and your check-in suggests \{opening\_\allowbreak{}state\_\allowbreak{}phrase\}. To start with what happened, what part of the situation is standing out most right now?\par
You mentioned ``\{reflection\_\allowbreak{}snippet\}''. To start with what happened, what part of that situation is standing out most right now?\par
Your check-in suggests \{opening\_\allowbreak{}state\_\allowbreak{}phrase\}. To start with what happened, what situation has been on your mind most?\par
I am here with you. To start with what happened, what situation has been on your mind most?\par
\end{promptbox}

\subsection{Rule-Based Text Shared by Both Chatbots}\label{sec:app-prompt-rules}

Code (not the model) produces the following text. The plain-language description of the pre-session ratings (1--9) is placed in the context message; the opening state phrase is used in the Control opening templates and the Treatment fallback opening. The closing message is appended to a chatbot reply when the reply contains phrases such as ``if you're feeling better'', ``if this helped'', or ``if you feel stuck'', or when the participant's last message contains phrases such as ``not helping'', ``no change'', or ``didn't help''; the empty-reply message is sent if the model returns nothing.

\begin{promptbox}
\textbf{Plain-Language Description of the VAD Ratings:} valence \textless{}= 3 -\textgreater{} ``this may feel heavy or unpleasant''; valence \textgreater{}= 8 -\textgreater{} ``there may still be some positive feeling available''; 6 \textless{}= valence \textless{} 8 -\textgreater{} ``your mood may be somewhat steady''\par
arousal \textgreater{}= 8 -\textgreater{} ``your body may be very activated''; 6 \textless{}= arousal \textless{} 8 -\textgreater{} ``there may be some tension or pressure in your body''; arousal \textless{}= 3 -\textgreater{} ``you may feel drained or low on energy''\par
dominance \textless{}= 3 -\textgreater{} ``it may feel hard to feel in control''; dominance \textgreater{}= 8 -\textgreater{} ``you may already have a little sense of control''\par
The matching phrases are joined in the order valence, arousal, dominance as 'a, b, and c' (empty string if none match).\par
\medskip
\textbf{Opening State Phrase:} First match wins: arousal \textgreater{}= 7 -\textgreater{} ``this feels pretty intense right now''; dominance \textless{}= 3 -\textgreater{} ``it may feel hard to get a handle on it''; valence \textless{}= 3 -\textgreater{} ``this has been feeling heavy''; arousal \textless{}= 3 -\textgreater{} ``you may be feeling a bit drained''; otherwise none.\par
\medskip
\textbf{Closing Message:} If you're feeling better or would like to stop the conversation, you may end the chat now. Thank you for sharing - feel free to return if you want to continue later.\par
\medskip
\textbf{Empty-Reply Message:} I am here with you. What has been on your mind most?\par
\end{promptbox}

\subsection{Memory-Building Prompts}\label{sec:app-prompt-memory}

These prompts run before each session, not during the chat, and produce the memory texts placed into the context messages. \{chat\_history\}, \{transcript\}, and \{pattern\_memory\_json\} are replaced by the participant's previous chat messages (one ``role: content'' line each, last 2,000 lines) or the day's full pattern memory.

\para{Treatment chat summary.}

\begin{promptbox}
\textbf{User Prompt:} Summarize all previous MEM-STRESS chats for this participant in one detailed paragraph. Include recurring stressors, emotional trajectory, useful/\allowbreak{}unhelpful responses, and coping actions discussed. Do not diagnose. Return plain text only.\par
\medskip
Chat history:\par
\{chat\_\allowbreak{}history\}\par
\end{promptbox}

\para{Treatment semantic description.}

\begin{promptbox}
\textbf{System Prompt:} You create detailed, cautious, evidence-grounded stress-pattern context. Return plain text only.\par
\medskip
\textbf{User Prompt:} Create a very detailed MEM-STRESS combined\_\allowbreak{}semantics text from this pattern-memory JSON. This text is used directly as chatbot context. Return plain text only.\par
\medskip
Use all sections: changes\_\allowbreak{}vs\_\allowbreak{}full\_\allowbreak{}average\_\allowbreak{}since\_\allowbreak{}beginning, changes\_\allowbreak{}vs\_\allowbreak{}previous\_\allowbreak{}chatbot\_\allowbreak{}interaction, peak\_\allowbreak{}oura\_\allowbreak{}values, tags, and chat\_\allowbreak{}summary. Be exhaustive and specific. Explain cross-domain and intra-domain understandings, stable values that matter, stress/\allowbreak{}recovery/\allowbreak{}sleep/\allowbreak{}activity/\allowbreak{}heart-rate/\allowbreak{}respiration links, and what the chatbot should keep in mind when supporting the participant. Do not diagnose or claim causality; use cautious language. Longer and more detailed is better.\par
\medskip
Pattern memory JSON:\par
\{pattern\_\allowbreak{}memory\_\allowbreak{}json\}\par
\end{promptbox}

\para{Control chat summary.}

\begin{promptbox}
\textbf{System Prompt:} You summarize counseling-style chat transcripts accurately and concisely.\par
\medskip
\textbf{User Prompt:} Create a concise, participant-friendly chat history summary for MEM-STRESS control sessions.\par
Use 3 to 6 sentences.\par
Include only: recurring stressors/\allowbreak{}themes, emotional trajectory across sessions, and what responses or steps felt helpful or unhelpful.\par
Do not diagnose. Do not include JSON. Return plain text only.\par
\medskip
Chat transcript:\par
\{transcript\}\par
\end{promptbox}
\clearpage
\section{Additional Results for RQ1}\label{sec:app-rq1}

\begin{figure}[t]
\centering
\begin{subfigure}[b]{0.6\linewidth}
\centering
\includegraphics[width=\linewidth]{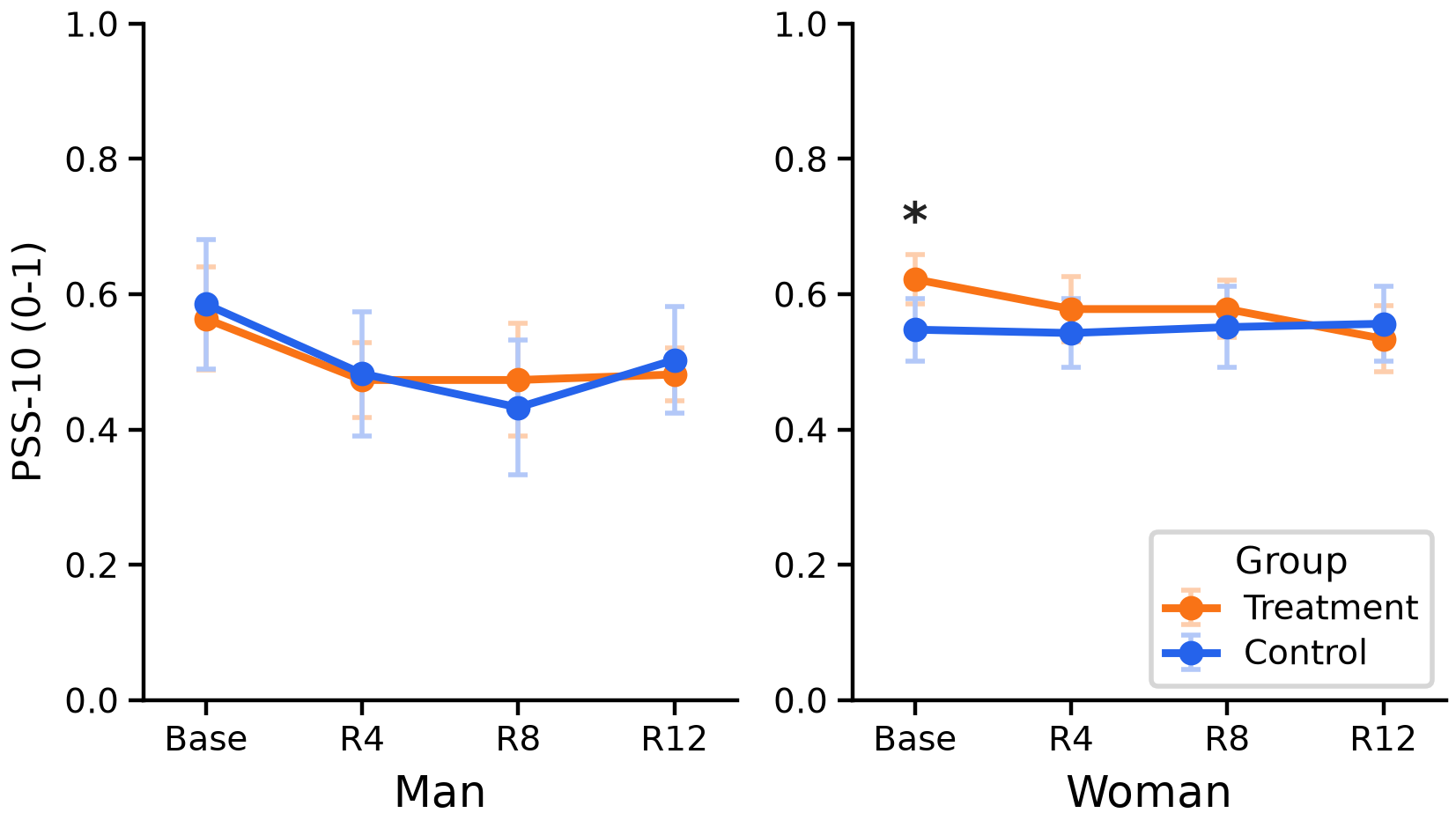}
\caption{By gender.}
\label{fig:app-rq1-pss-gender}
\end{subfigure}\hfill
\begin{subfigure}[b]{0.36\linewidth}
\centering
\includegraphics[width=\linewidth]{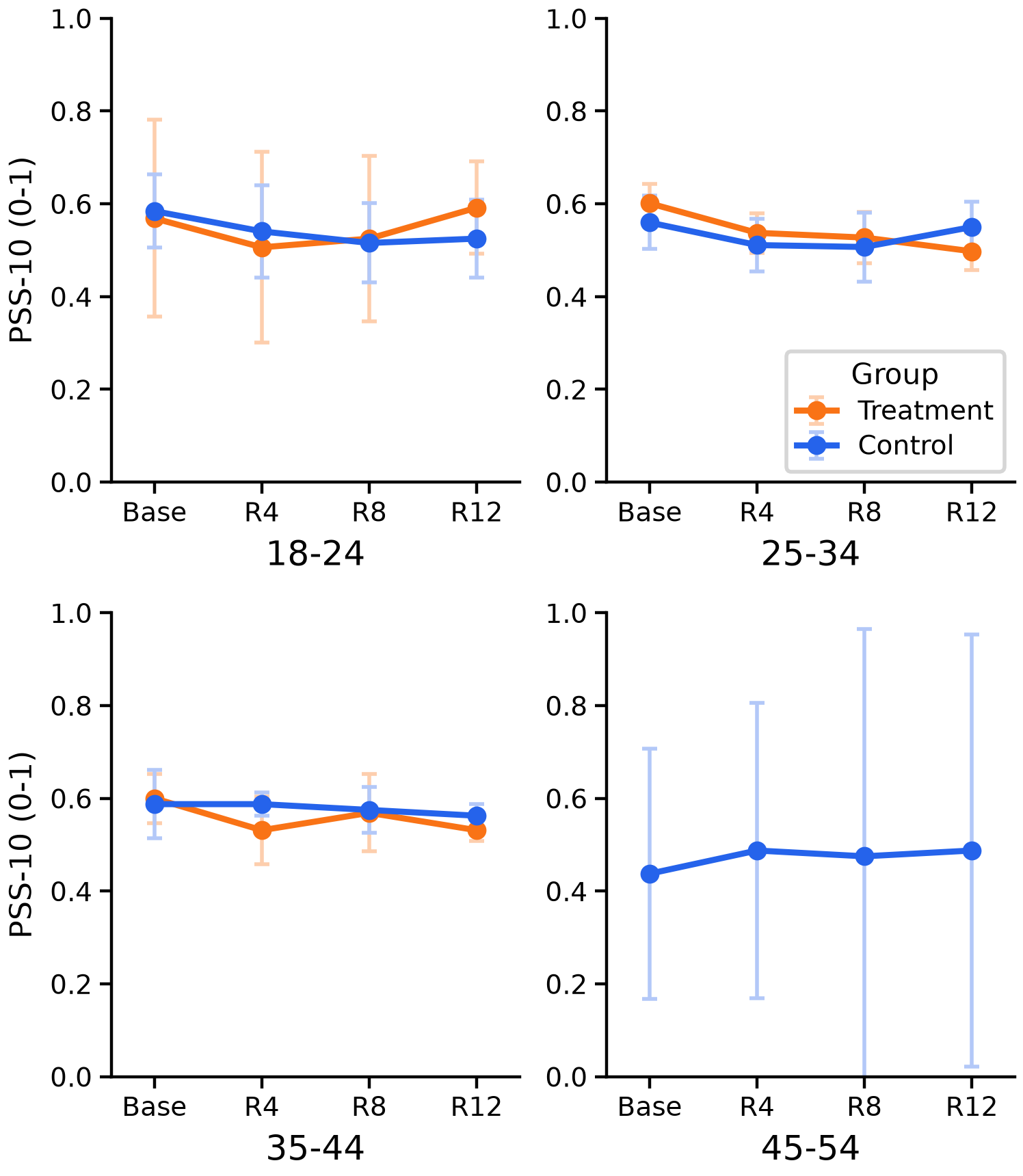}
\caption{By age group.}
\label{fig:app-rq1-pss-age}
\end{subfigure}
\par\medskip
\begin{subfigure}[t]{0.75\linewidth}
\centering
\includegraphics[width=\linewidth]{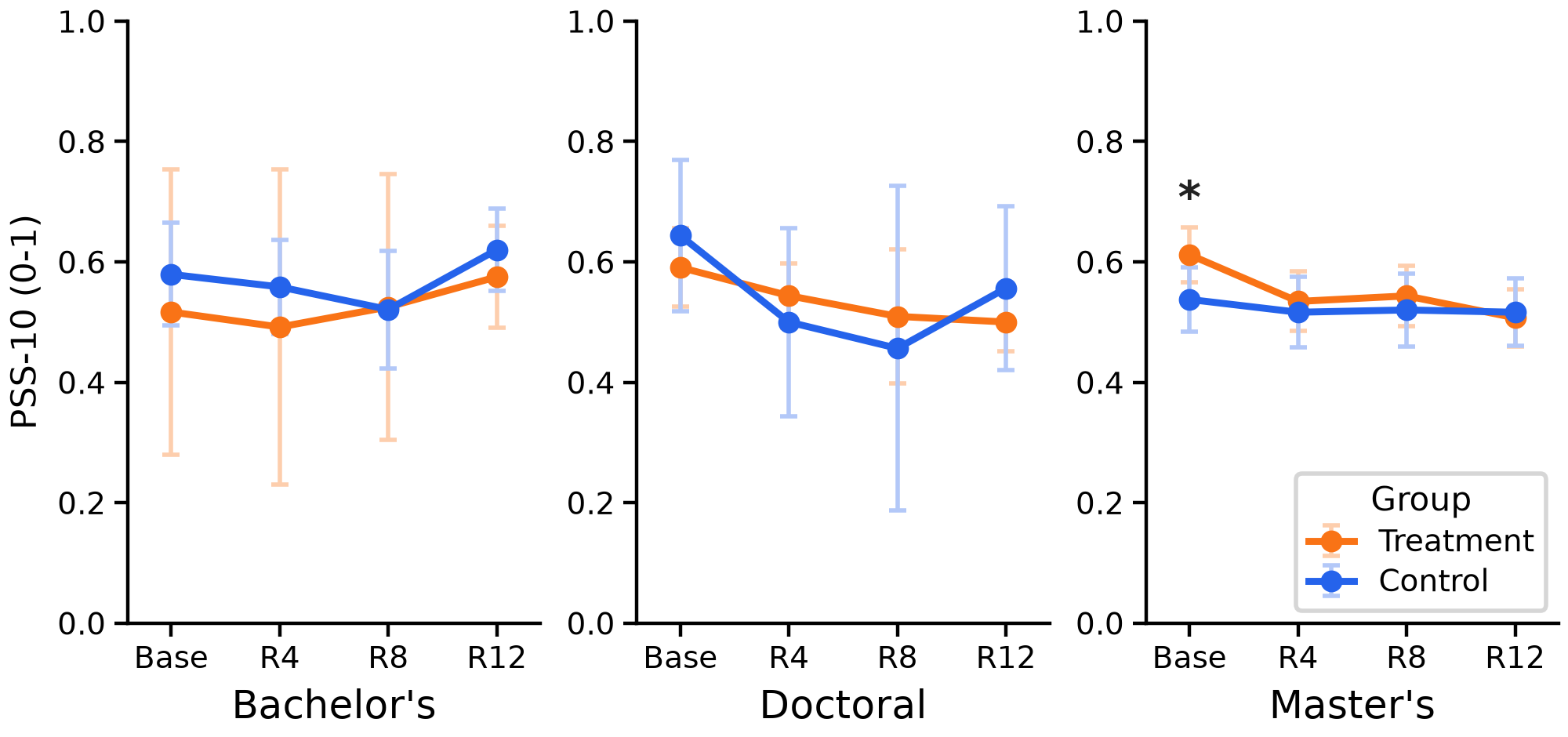}
\caption{By education level.}
\label{fig:app-rq1-pss-education}
\end{subfigure}
\caption{PSS-10 (normalized as raw/40; mean $\pm$ 95\% CI) by demographic subgroup for Treatment (orange) and Control (blue), with Welch's $t$- or Mann--Whitney $U$ tests between groups; both significant differences are at baseline ($^{*}p<0.05$, $^{**}p<0.01$, $^{***}p<0.001$).}
\label{fig:app-rq1-pss-demographics}
\end{figure}

\begin{figure}[t]
\centering
\begin{subfigure}[t]{0.32\linewidth}
\centering
\includegraphics[width=\linewidth]{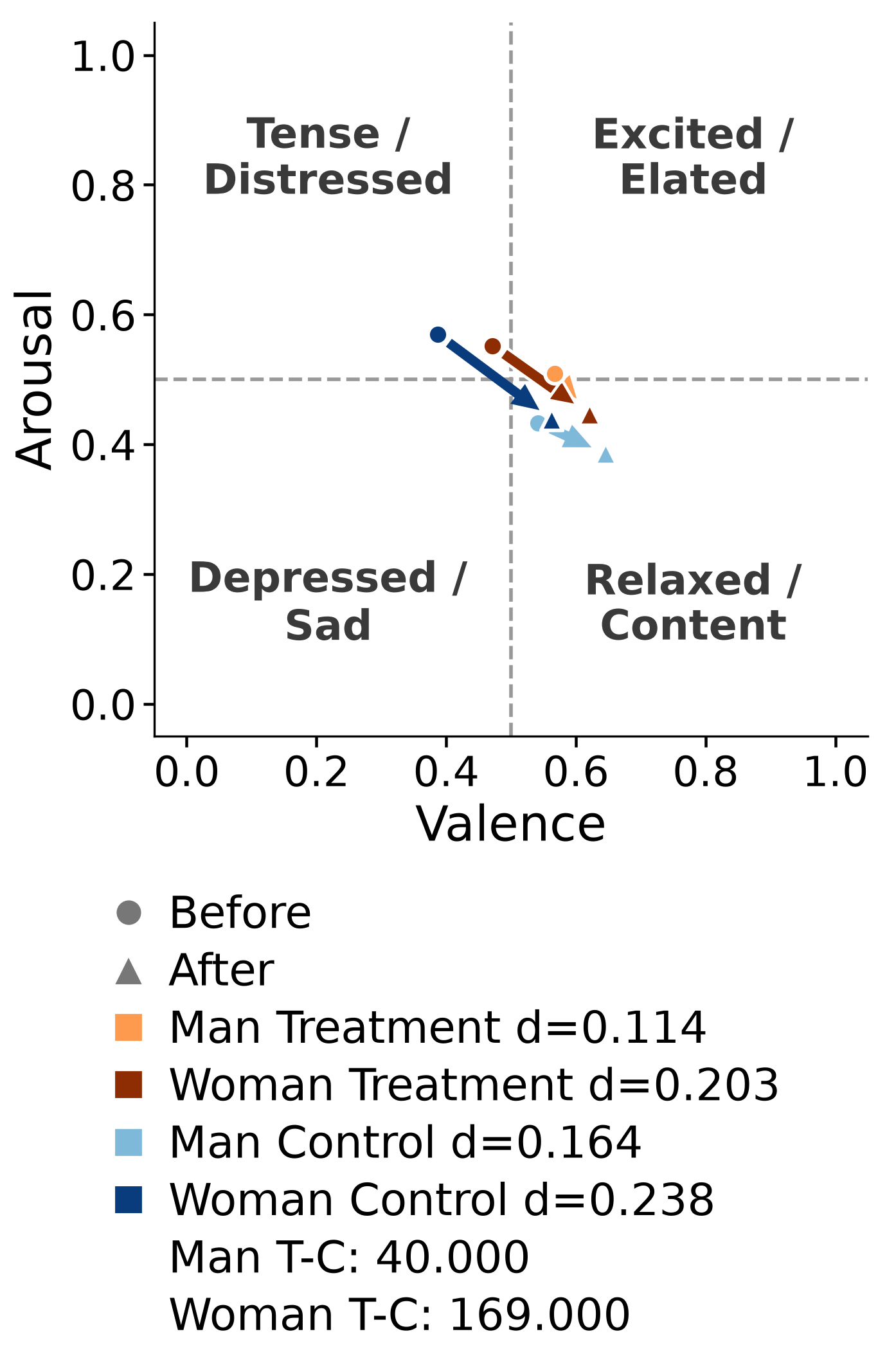}
\caption{Valence--arousal by gender.}
\label{fig:app-rq1-ousigram-va-gender}
\end{subfigure}\hfill
\begin{subfigure}[t]{0.32\linewidth}
\centering
\includegraphics[width=\linewidth]{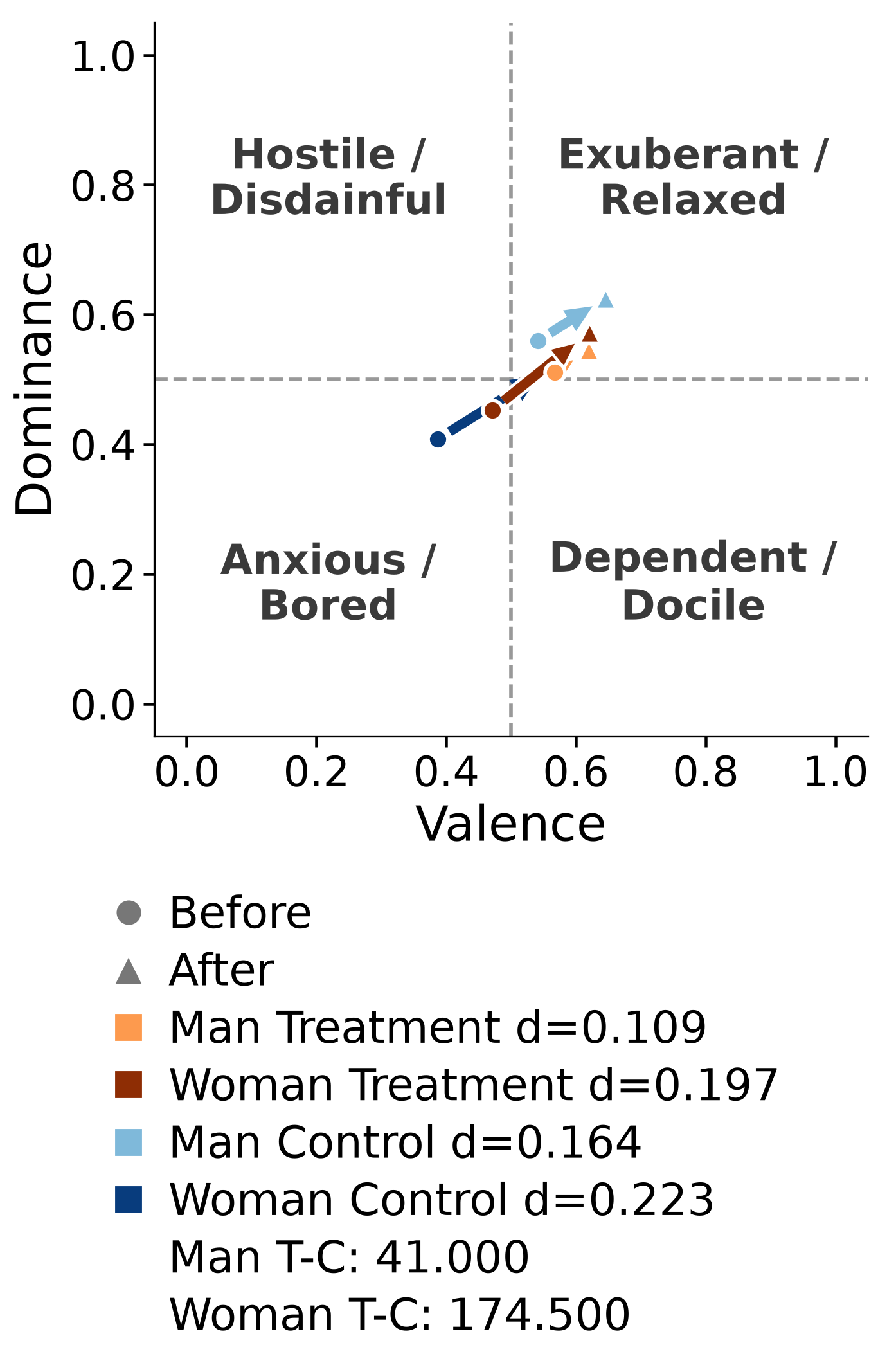}
\caption{Valence--dominance by gender.}
\label{fig:app-rq1-ousigram-vd-gender}
\end{subfigure}\hfill
\begin{subfigure}[t]{0.32\linewidth}
\centering
\includegraphics[width=\linewidth]{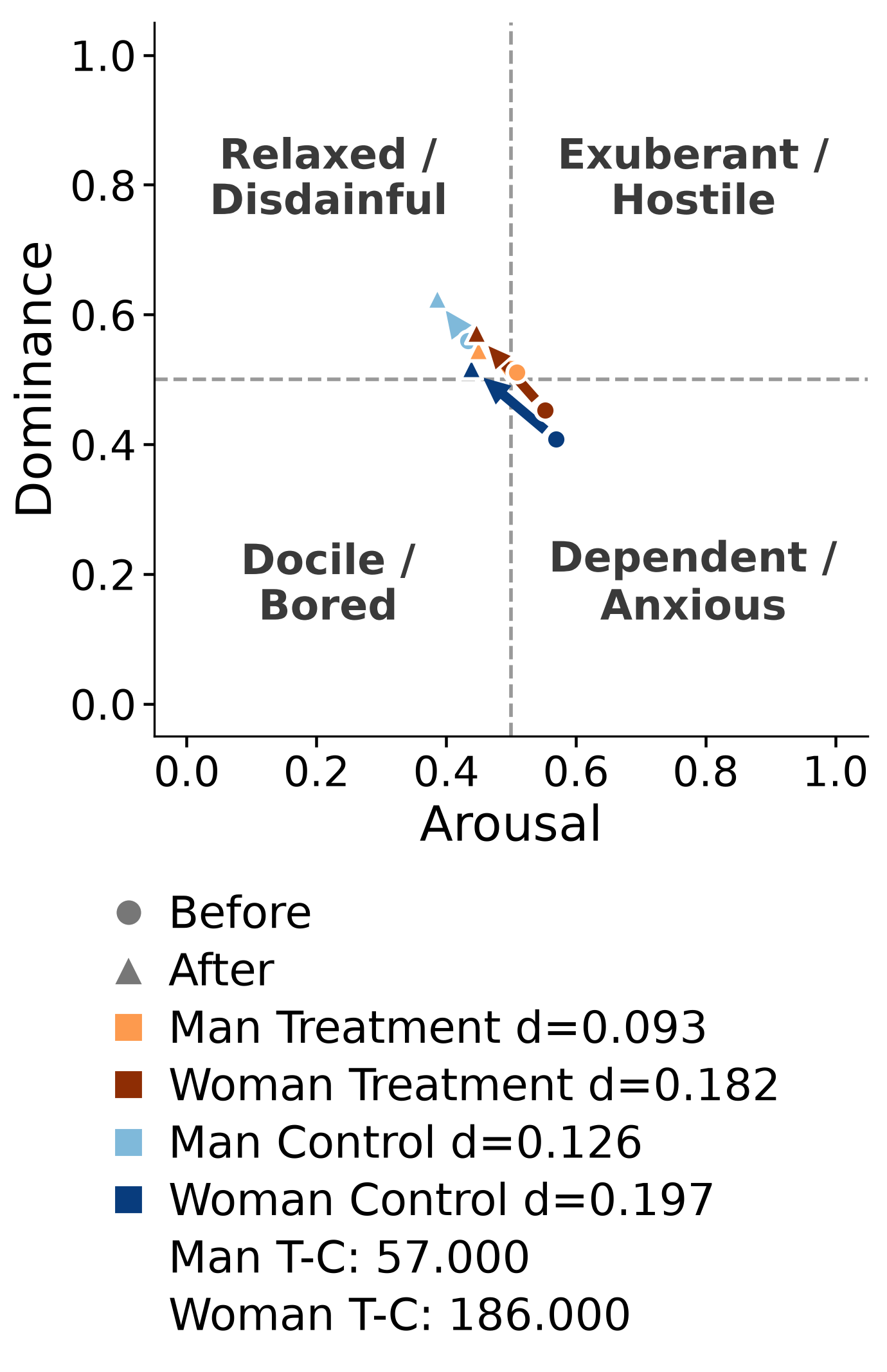}
\caption{Arousal--dominance by gender.}
\label{fig:app-rq1-ousigram-ad-gender}
\end{subfigure}
\caption{Within-session affect shifts by gender, drawn as in~\autoref{fig:rq1-ousigram} without individual ratings, with Mann--Whitney $U$ tests between groups ($^{*}p<0.05$, $^{**}p<0.01$, $^{***}p<0.001$).}
\label{fig:app-rq1-ousigram-gender}
\end{figure}

\begin{figure}[t]
\centering
\begin{subfigure}[t]{0.32\linewidth}
\centering
\includegraphics[width=\linewidth,height=0.93\linewidth]{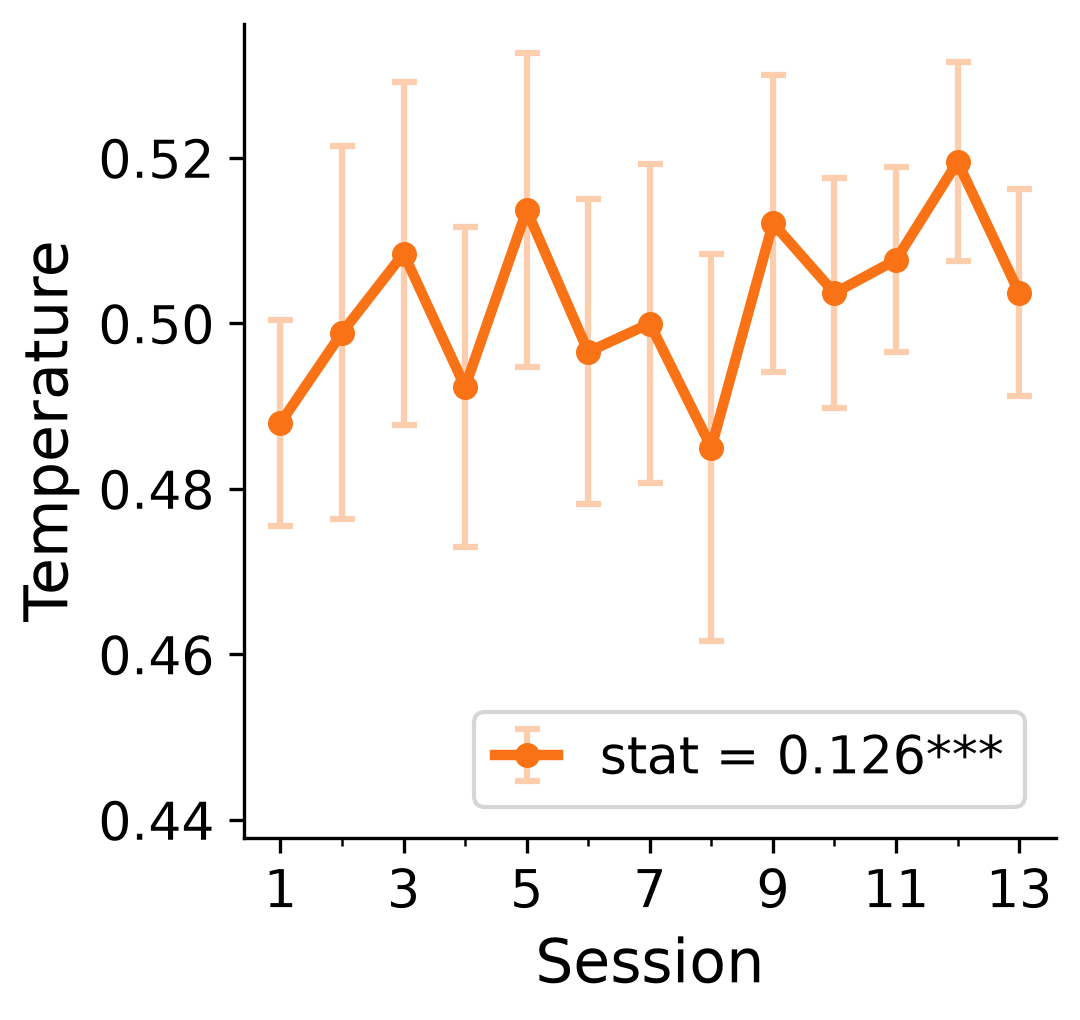}
\caption{Body temperature deviation (normalized).}
\label{fig:app-rq1-oura-temp}
\end{subfigure}\hfill
\begin{subfigure}[t]{0.32\linewidth}
\centering
\includegraphics[width=\linewidth,height=0.93\linewidth]{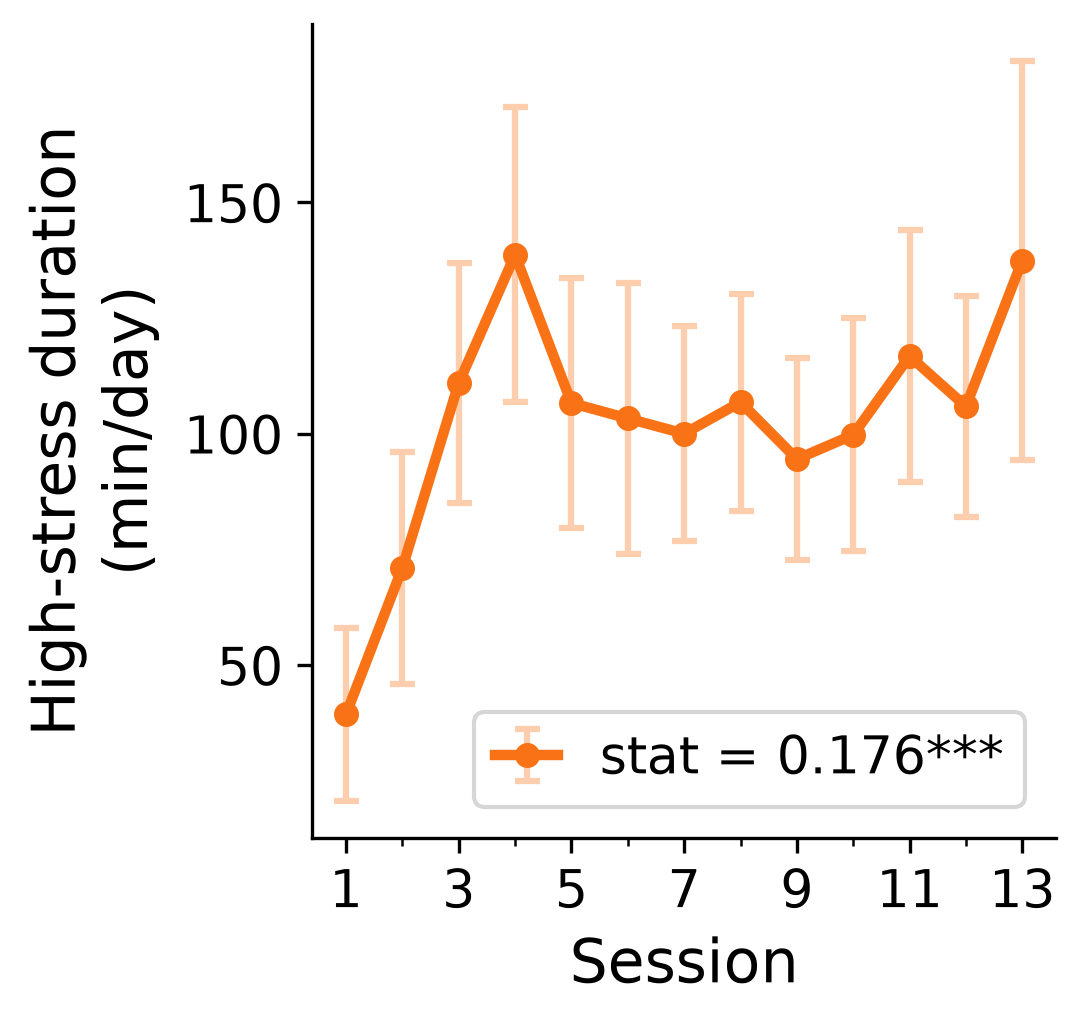}
\caption{High-stress minutes per day.}
\label{fig:app-rq1-oura-stress}
\end{subfigure}\hfill
\begin{subfigure}[t]{0.32\linewidth}
\centering
\includegraphics[width=\linewidth,height=0.93\linewidth]{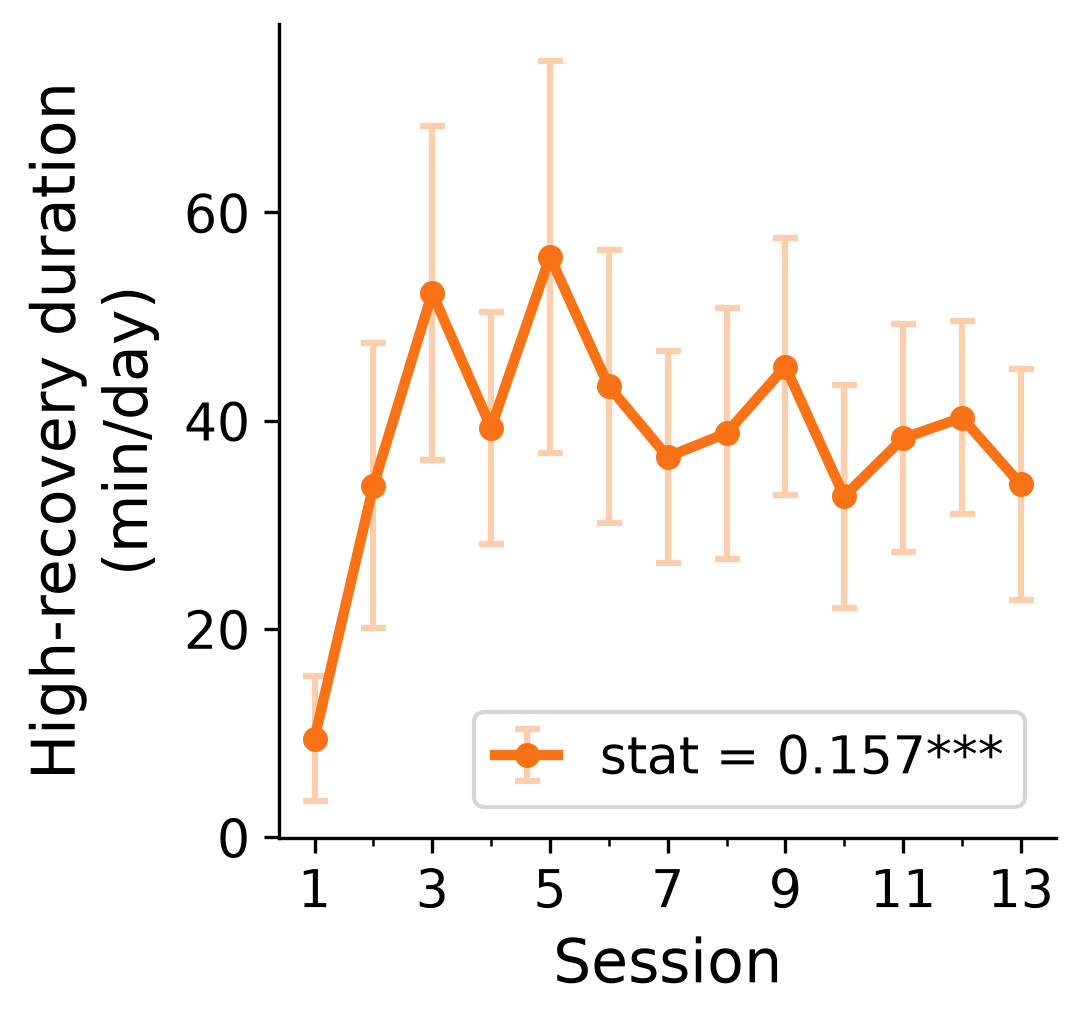}
\caption{High-recovery minutes per day.}
\label{fig:app-rq1-oura-recovery}
\end{subfigure}
\caption{Other daily physiological features in Treatment by nearest session (mean $\pm$ 95\% CI, binned as in~\autoref{fig:rq1-hr}, panel~\subref{fig:rq1-hr-daily}; (a, b) y-axes truncated): (a) body temperature deviation, normalized as $(\text{deviation}+2)/4$, (b) high-stress and (c) high-recovery minutes, with Spearman's $\rho$ (``stat'') in the legends ($^{*}p<0.05$, $^{**}p<0.01$, $^{***}p<0.001$).}
\label{fig:app-rq1-oura-daily}
\end{figure}

This appendix supplements the RQ1 results with method details, the demographic subgroup analyses of perceived stress (\autoref{fig:app-rq1-pss-demographics}) and within-session affect (\autoref{fig:app-rq1-ousigram-gender} and~\autoref{tab:app-rq1-vad-subgroups}), and the daily physiological features other than heart rate (\autoref{fig:app-rq1-oura-daily}). The subgroup analyses repeat only the between-group comparisons of RQ1, with the same test selection and notation.

\para{Statistical tests.} We normalized PSS-10 totals and SAM ratings to 0--1 so that the stress and affect models share one scale. Shapiro--Wilk normality checks selected parametric tests (Welch's $t$ between groups, paired $t$ within groups) or non-parametric tests (Mann--Whitney $U$ or Wilcoxon $W$); effect sizes are Cohen's $d$ between groups and $d_z$ within groups. The within-group comparisons use each participant as their own reference, showing whether a group's stress moved away from where it started; the Friedman test checks for change across all four timepoints without singling out one round, and uses only complete cases.

\para{Ousigrams.} Each ousigram plane is split into quadrants at the scale midpoints, and each group's arrow weights participants equally regardless of how many ratings they contributed. Viewing the dimensions in pairs shows which kind of state participants moved between. On average, both groups began sessions near the center of each plane, Control at lower valence than Treatment. Both arrows point toward higher valence and lower arousal in the VA plane (\autoref{fig:rq1-ousigram}, panel~\subref{fig:rq1-ousigram-va}), ending in the relaxed/content quadrant; toward higher valence and dominance (a greater sense of control) in the VD plane (panel~\subref{fig:rq1-ousigram-vd}), ending in the exuberant/relaxed quadrant; and toward lower arousal and higher dominance in the AD plane (panel~\subref{fig:rq1-ousigram-ad}), ending in the relaxed/disdainful quadrant.

\para{Heart rate.} We used the Treatment participants' intraday heart rate from the episodic memory (about one sample per five minutes). Per session, we averaged heart rate in the hour before the pre-session rating, during the conversation (between the pre- and post-session ratings), and in the hour after the post-session rating, then pooled the session means across participants (\autoref{fig:rq1-hr}, panel~\subref{fig:rq1-hr-window}). Fewer sessions had heart rate during the conversation (107) than before or after it (290 each), so the three windows are unpaired. For the daily trend, we averaged each participant's heart rate per day (1,014 participant-days with ring data, 959 with heart rate), grouped days by nearest session, which places the daily signal on the same session timeline as the PSS-10 and affect analyses, and plotted the per-session means with the Spearman correlation between daily heart rate and session index (\autoref{fig:rq1-hr}, panel~\subref{fig:rq1-hr-daily}). The first bin, which also includes the days before the first session, averaged 89.8\,bpm; bin means ranged from 87.4 to 90.6\,bpm through session~9 and from 85.8 to 86.8\,bpm in sessions~10 to~13. The other daily physiological features were processed the same way (\autoref{fig:app-rq1-oura-daily}).

\begin{table}[!h]
\centering
\small
\caption{Within-session shift distance (normalized SAM) by demographic subgroup, Treatment / Control ($n$: participants), with Welch's $t$- or $^{a}$Mann--Whitney $U$ tests ($^{*}p<0.05$, $^{**}p<0.01$, $^{***}p<0.001$).}
\label{tab:app-rq1-vad-subgroups}
\footnotesize
\setlength{\tabcolsep}{3pt}
\begin{tabular}{@{}lccccccc@{}}
\toprule
 & & \multicolumn{2}{c}{\textbf{Valence--arousal}} & \multicolumn{2}{c}{\textbf{Valence--dominance}} & \multicolumn{2}{c}{\textbf{Arousal--dominance}} \\
\cmidrule(lr){3-4}\cmidrule(lr){5-6}\cmidrule(l){7-8}
\textbf{Subgroup} & $n$ & \textbf{T / C} & \textbf{stat.} & \textbf{T / C} & \textbf{stat.} & \textbf{T / C} & \textbf{stat.} \\
\midrule
All & 30/30 & 0.165/0.213 & $379.0^{a}$ & 0.159/0.203 & $393.0^{a}$ & 0.143/0.173 & $442.0^{a}$ \\
\midrule
Man & 13/10 & 0.114/0.164 & $40.0^{a}$ & 0.109/0.164 & $41.0^{a}$ & 0.093/0.126 & $57.0^{a}$ \\
Woman & 17/20 & 0.203/0.238 & $169.0^{a}$ & 0.197/0.223 & $174.5^{a}$ & 0.182/0.197 & $186.0^{a}$ \\
\midrule
18--24 & 4/8 & 0.160/0.288 & $10.0^{a}$ & 0.135/0.282 & $9.0^{a}$ & 0.128/0.236 & $-1.11$ \\
25--34 & 22/18 & 0.178/0.191 & $-0.33$ & 0.176/0.178 & $210.5^{a}$ & 0.158/0.149 & $233.5^{a}$ \\
35--44 & 4/2 & 0.095/0.178 & $0.0^{a}$ & 0.085/0.165 & $0.0^{a}$ & 0.079/0.123 & $2.0^{a}$ \\
45--54 & 0/2 & --/0.148 & -- & --/0.151 & -- & --/0.190 & -- \\
\midrule
Bachelor's & 3/6 & 0.153/0.112 & $0.84$ & 0.131/0.119 & $0.24$ & 0.137/0.073 & $14.0^{a}$ \\
Master's & 19/20 & 0.158/0.253 & $135.0^{a}$ & 0.150/0.237 & $139.5^{a}$ & 0.133/0.210 & $146.5^{a}$ \\
Doctoral & 8/4 & 0.184/0.164 & $17.0^{a}$ & 0.189/0.158 & $0.47$ & 0.170/0.136 & $0.58$ \\
\bottomrule
\end{tabular}
\end{table}

\clearpage
\section{Additional Results for RQ2}\label{sec:app-rq2}

This appendix supplements the RQ2 results with each chatbot's most distinctive words (\autoref{tab:rq2-sage}), the member categories of the LIWC composites (\autoref{tab:app-rq2-liwc-composites}), the DBT skills named by each chatbot (\autoref{tab:rq2-dbt}), the memory-to-skill co-occurrence values (\autoref{tab:app-rq2-cooccurrence}), and further close-reading examples (Appendix~\ref{sec:app-rq2-close}).

\begin{table}[!h]
\centering
\small
\caption{The 20 words most distinctive of each chatbot's messages (SAGE), ranked by their deviation $\eta$ from a background pooled over the chat turns of all four corpora (both chatbots and both participant groups; vocabulary count $\geq 5$). Words are lowercased tokens (e.g., \emph{rate} occurs mostly in ``heart rate''). $\eta$ is not comparable across the two columns, because the corpora contribute different shares of the background.}
\label{tab:rq2-sage}
\begin{tabular}{@{}rlrlr@{}}
\toprule
 & \multicolumn{2}{c}{\textbf{Treatment chatbot}} & \multicolumn{2}{c}{\textbf{Control chatbot}} \\
\cmidrule(lr){2-3}\cmidrule(l){4-5}
\textbf{Rank} & \textbf{Word} & $\eta$ & \textbf{Word} & $\eta$ \\
\midrule
1 & recovery & 1.0250 & happened & 0.7714 \\
2 & levels & 1.0203 & standing & 0.7505 \\
3 & higher & 1.0199 & suggests & 0.7316 \\
4 & rate & 1.0175 & carry & 0.6722 \\
5 & signs & 1.0164 & gain & 0.6408 \\
6 & rem & 1.0077 & mini & 0.6306 \\
7 & elevated & 1.0009 & benefits & 0.6120 \\
8 & moderate & 0.9956 & costs & 0.5783 \\
9 & dipped & 0.9945 & distractions & 0.5699 \\
10 & looks & 0.9854 & part & 0.5675 \\
11 & lately & 0.9817 & cost & 0.5616 \\
12 & activation & 0.9798 & benefit & 0.5532 \\
13 & variability & 0.9792 & situation & 0.5484 \\
14 & improved & 0.9790 & analysis & 0.5250 \\
15 & recent & 0.9769 & urgent & 0.5202 \\
16 & activated & 0.9731 & waves & 0.5130 \\
17 & signals & 0.9713 & google & 0.5079 \\
18 & fairly & 0.9712 & jotting & 0.5073 \\
19 & strained & 0.9672 & setting & 0.5005 \\
20 & heart & 0.9640 & topic & 0.4972 \\
\bottomrule
\end{tabular}
\end{table}

\begin{table}[!h]
\centering
\small
\caption{LIWC categories that make up each psycholinguistic composite; each composite is the unweighted mean of its member categories.}
\label{tab:app-rq2-liwc-composites}
\begin{tabular}{@{}>{\raggedright\arraybackslash}p{5.4cm}>{\raggedright\arraybackslash}p{\dimexpr\linewidth-5.4cm-2\tabcolsep\relax}@{}}
\toprule
\textbf{Composite} & \textbf{LIWC categories} \\
\midrule
Distress & Negative emotion, anxiety, anger, sadness, \emph{I}, personal pronouns, negation, discrepancy, tentativeness, past focus, health, body \\
\addlinespace
Emotional & Affect, positive emotion, negative emotion, anxiety, anger, sadness, feel \\
\addlinespace
Coping & Cognitive processes, insight, causation, discrepancy, tentativeness, certainty, achievement, power, reward, positive emotion, affiliation, social, friend, present focus, future focus \\
\addlinespace
Cognitive reappraisal and reflection & Cognitive processes, insight, causation, discrepancy, tentativeness, certainty \\
\addlinespace
Agency and perceived control & Power, achievement, certainty, discrepancy \\
\addlinespace
Social support and connectedness & Social, affiliation, friend, family, \emph{we}, \emph{I}, \emph{you} \\
\addlinespace
Temporal orientation and rumination & Past focus, present focus, future focus, time \\
\addlinespace
Somatic and health-related distress & Health, body, biological processes, perception, feel \\
\addlinespace
Engagement and linguistic style & Pronouns, verbs, adjectives, interrogatives, assent, fillers, non-fluencies, informal language \\
\bottomrule
\end{tabular}
\end{table}

\begin{table}[!h]
\centering
\small
\caption{DBT skills named by each chatbot: every skill named at least once in either group (38 of the 48 skills; the other 10 were never named), ordered by the mean of the two percentages. Cells give the percentage of sessions naming the skill (Treatment $n=387$, Control $n=386$) and, in parentheses, the number of participants (of 30) who received it at least once; 0.0 (0) marks a skill never named in that group. The last row gives the number of distinct skills named.}
\label{tab:rq2-dbt}
\begin{tabular}{@{}lcc@{}}
\toprule
\textbf{Skill} & \textbf{Treatment} & \textbf{Control} \\
\midrule
Grounding & 60.7 (30) & 54.9 (29) \\
STOP & 21.4 (26) & 23.8 (27) \\
Mindful breathing & 22.2 (28) & 18.4 (26) \\
TIPP & 14.5 (20) & 5.7 (16) \\
Willingness vs.\ willfulness & 11.9 (20) & 7.0 (17) \\
Problem solving & 5.9 (16) & 11.7 (24) \\
Half smile & 7.2 (12) & 6.2 (11) \\
Observing & 4.7 (11) & 5.7 (15) \\
Cost--benefit analysis & 1.3 (5) & 7.0 (16) \\
Self-validation & 2.6 (5) & 5.4 (13) \\
Radical acceptance & 2.6 (6) & 4.9 (12) \\
Mastery and coping ahead & 1.0 (4) & 6.2 (17) \\
Balancing emotional urges & 2.1 (6) & 3.4 (12) \\
ACCEPTS & 2.6 (6) & 2.3 (6) \\
Assertiveness (part 1) & 2.1 (6) & 2.8 (5) \\
Urge surfing & 1.6 (3) & 3.1 (9) \\
ABC PLEASE & 1.3 (5) & 3.1 (12) \\
Recognizing your emotion & 1.3 (5) & 2.6 (8) \\
Self-soothing & 1.3 (5) & 2.3 (8) \\
Actions based on values & 0.5 (2) & 2.6 (7) \\
Being effective & 0.8 (2) & 1.8 (6) \\
Distracting activities & 0.8 (3) & 1.6 (5) \\
Mental noting & 1.0 (3) & 1.3 (4) \\
Mental body scan & 0.8 (1) & 1.3 (5) \\
Describing your emotions & 0.0 (0) & 2.1 (7) \\
Listening and validation & 1.0 (4) & 0.8 (3) \\
RESISTT & 0.8 (2) & 0.5 (2) \\
Thought defusion & 0.5 (2) & 0.8 (3) \\
DEAR MAN & 0.5 (1) & 0.5 (2) \\
Wise mind & 0.0 (0) & 1.0 (4) \\
Communication styles & 0.3 (1) & 0.3 (1) \\
Interpersonal boundaries & 0.3 (1) & 0.3 (1) \\
Emotions and cognitive vulnerability & 0.0 (0) & 0.5 (1) \\
FAST & 0.3 (1) & 0.0 (0) \\
Assertiveness scripts and interpersonal rights & 0.0 (0) & 0.3 (1) \\
Emotion exposure & 0.0 (0) & 0.3 (1) \\
Emotions and physical vulnerability & 0.0 (0) & 0.3 (1) \\
Letting go of judgments & 0.0 (0) & 0.3 (1) \\
\midrule
Distinct skills named (of 48) & 31 & 37 \\
\bottomrule
\end{tabular}
\end{table}

\begin{table}[!h]
\centering
\small
\caption{Most frequent DBT skills accompanying each physiological-feature reference in Treatment sessions: percentage of sessions referencing a domain that also named the skill (row-normalized; $n$ = sessions referencing the domain). The last row gives each skill's overall rate across all Treatment sessions. Temperature was never referenced. Mind.\ br.\ = mindful breathing; Will.\ = willingness vs.\ willfulness.}
\label{tab:app-rq2-cooccurrence}
\begin{tabular}{@{}lrccccc@{}}
\toprule
\textbf{Domain} & $n$ & \textbf{Grounding} & \textbf{Mind.\ br.} & \textbf{STOP} & \textbf{TIPP} & \textbf{Will.} \\
\midrule
Sleep & 263 & 62.7 & 25.9 & 22.4 & 14.1 & 13.3 \\
Activity / steps & 278 & 60.4 & 23.7 & 22.3 & 14.4 & 11.9 \\
Heart rate & 174 & 65.5 & 23.0 & 21.3 & 17.8 & 9.8 \\
Energy / fatigue & 127 & 63.8 & 30.7 & 19.7 & 15.0 & 14.2 \\
Tension / stress & 165 & 66.1 & 23.0 & 27.9 & 15.8 & 12.7 \\
Recovery & 126 & 66.7 & 21.4 & 17.5 & 17.5 & 9.5 \\
Breathing & 31 & 77.4 & 48.4 & 22.6 & 25.8 & 12.9 \\
Calories & 29 & 69.0 & 27.6 & 34.5 & 13.8 & 3.4 \\
\midrule
All sessions & 387 & 60.7 & 22.2 & 21.4 & 14.5 & 11.9 \\
\bottomrule
\end{tabular}
\end{table}

\subsection{Further Close-Reading Examples}\label{sec:app-rq2-close}

These are the full close-reading notes for RQ2, with all examples from the 15 sampled conversations per group.

The close reading showed what this meant within a conversation: the memory made the Treatment chatbot more personal in how it opened a session and in what it could say about the participant. Treatment sessions opened from the participant's own physiological features (heart rate, sleep, activity, recovery), tied to the stressor they had raised, such as T-P4's difficulties with their manager or T-P15's conflict with their partner over a visa. Control sessions instead opened from the same reflection-restating template for everyone. The memory also gave the Treatment chatbot a factual basis for what it said about participants: in the 15 sampled conversations, we found no statements about physiological features that the data did not support, although participants reported misreadings (RQ3). The Control chatbot, holding only the reflection and a short summary of earlier chats, could only imply continuity, and its recall of earlier sessions was vague, as when C-P4 asked it what they had discussed about their father.

This personalization did not cost the basics of a supportive conversation, which both chatbots handled alike. Both named the participant's stated stressor specifically, such as T-P8's anxiety about starting a new job, T-P14's thesis presentation, or C-P7's avoidance of thesis tasks. At their best, both tailored the skill to the moment, as in T-P9's and T-P15's conversations and in the Control chatbot breaking C-P7's procrastination into one small first step. The Treatment chatbot adapted when a suggestion was declined, replacing a rejected grounding exercise with a values-based question (T-P5) or switching to breathing at the participant's request (T-P8); the Control chatbot mostly adapted too, although it answered C-P2's two explicit refusals with generic reassurance. Participants' engagement ranged from elaborate to brief replies in both groups. Other patterns were shared or chatbot-specific: both chatbots sometimes offered several skills in one session or replied generically to ambiguous issues; the Control chatbot reused a near-identical grounding script across participants, whereas several Treatment sessions repeated formulaic framing around physiological features within a session or cited them despite little reported stress.

In individual conversations, the chatbot used the memory in two ways. First, it linked participants' accounts to their physiological features and chose a fitting step. T-P15, anxious about the future, was told, ``Since your body shows some signs of activation alongside this anxiety, one small step you could try now is the DBT STOP skill.'' After the exercise, T-P15 wrote, ``I felt my heartbeat slow down''. T-P10, describing emotions felt in their head and body, was offered paced breathing because their heart rate was also higher than usual.
Second, the memory offered a physiological explanation for how participants felt. The chatbot told T-P3 that, given ``your recent higher activity but lower recovery and heart rate variability, your body might be signaling some strain'', and, when T-P3 blamed themselves for a few unproductive days, added, ``With your body showing signs of stress and less recovery, it makes sense that your mindset feels more fragile right now.'' T-P3 explained that the energy ``was spent on travel and tourism with my partner rather than work'', matching the measured rise in activity. The chatbot also linked T-P9's data to their own account: ``Your body’s signals of tension and lower energy fit well with that tiredness you’re describing.''

In the DBT recommendations, the Treatment chatbot opened T-P6's session by describing more activity and ``some restless moments'' in their sleep, then suggested, ``Since your body showed increased activity and some restlessness in sleep, it might help to gently calm your body now. One small option is trying the TIPP technique''; T-P6 did the cold-water step and replied, ``I did it, it feels good''.
\clearpage
\section{Additional Results for RQ3}\label{sec:app-rq3}

This appendix supplements the RQ3 results with the final questionnaire (\autoref{tab:app-feedback-questionnaire}), the SUS band details, the remaining agreement items, the definitions and results of the open-ended feedback themes (\autoref{tab:app-rq3-theme-definitions} and~\autoref{tab:rq3-themes}), the behavior-change responses (\autoref{tab:rq3-behavior}), the remaining themes, additional participant quotes, and how the effect-size overview (\autoref{fig:rq3-effectsizes}) was computed.

The final questionnaire, titled ``Your experience with MEM-STRESS'', was shown once, at the end of the 13th session after the post-session ratings, and asked participants to answer about their overall experience throughout the study. Every item was required.~\autoref{tab:app-feedback-questionnaire} lists the items verbatim in the order shown; the two ring questions were shown only to Treatment participants.

\begin{table}[p]
\centering
\small
\caption{Final questionnaire, items verbatim in the order shown. SUS and agreement items were rated from 1 (Strongly Disagree) to 5 (Strongly Agree), with only the endpoints labeled. Behavior-change items offered Yes, a lot / Yes, somewhat / A little / No change / Not sure, each followed by a free-text ``Explain with an example''. All other items were free text.}
\label{tab:app-feedback-questionnaire}
\begin{tabular}{@{}p{0.2\linewidth}p{0.76\linewidth}@{}}
\toprule
\textbf{Part} & \textbf{Item} \\
\midrule
Usability (SUS) & 1. I think that I would like to use this system frequently. \\
 & 2. I found the system unnecessarily complex. \\
 & 3. I thought the system was easy to use. \\
 & 4. I think that I would need the support of a technical person to be able to use this system. \\
 & 5. I found the various functions in this system were well integrated. \\
 & 6. I thought there was too much inconsistency in this system. \\
 & 7. I would imagine that most people would learn to use this system very quickly. \\
 & 8. I found the system very cumbersome to use. \\
 & 9. I felt very confident using the system. \\
 & 10. I needed to learn a lot of things before I could get going with this system. \\
\midrule
Advantage and & Biggest advantage of MEM-STRESS \\
disadvantage & Biggest disadvantage of MEM-STRESS \\
\midrule
Agreement & Q1. The MEM-STRESS chatbot understood my situation accurately. \\
 & Q2. The MEM-STRESS chatbot provided suggestions that were relevant to my daily life experiences. \\
 & Q3. The MEM-STRESS chatbot's explanations for its suggestions made sense to me. \\
 & Q4. The MEM-STRESS chatbot provided me with good recommendations to manage stress. \\
\midrule
Behavior change & Did MEM-STRESS change the way you reflected on your stress or emotions? \\
 & Did MEM-STRESS lead you to make changes in your daily routine or behaviour that helped manage stress? \\
\midrule
Open feedback & Describe one specific situation when MEM-STRESS helped manage or reduce your stress \\
 & What feature made the biggest difference, and how did it help? \\
 & Which areas could be improved? Explain with an example \\
 & Additional feedback or suggestions \\
\midrule
Ring experience & When did Oura accurately reflect how you were feeling or behaving? \\
(Treatment only) & When did Oura inaccurately reflect how you were feeling or behaving? \\
\bottomrule
\end{tabular}
\end{table}

While both means were in the ``Good'' band, ``Excellent'' was reached by 41.4\% of Treatment and 42.9\% of Control participants. Scores ranged from 57.5 to 95 in Treatment and from 10 to 100 in Control (\autoref{fig:rq3-sus}).

Of the agreement items other than Q1 (normalized 0--1), relevance to daily life was nearly identical (0.595 vs.\ 0.563; $U = 438.5$, $d = 0.13$), the two recommendation items were slightly higher in Control (\emph{explanations made sense}: 0.664 vs.\ 0.679, $U = 375.5$, $d = -0.06$; \emph{gave good recommendations}: 0.629 vs.\ 0.652, $U = 366.0$, $d = -0.10$), and the composite was essentially equal (0.636 vs.\ 0.609; Mann--Whitney $U = 436.5$, $d = 0.13$;~\autoref{fig:rq3-agreement}).

\begin{table}[t]
\centering
\small
\caption{Definitions of the nine binary themes coded per participant in the final questionnaire's open-ended answers (percentages in~\autoref{tab:rq3-themes}).}
\label{tab:app-rq3-theme-definitions}
\begin{tabular}{@{}p{0.32\linewidth}p{0.6\linewidth}@{}}
\toprule
\textbf{Theme} & \textbf{Definition} \\
\midrule
Memory/data grounding praised & Tracking over time, continuity across sessions, or use of personal data \\
Personalization praised & Tailored or accurate understanding, memory mentioned or not \\
Sensor/tracking issue & Inaccurate or mistimed physiological features \\
Trust/accuracy doubt & Doubting that the chatbot understood them \\
Memory missing & Forgetting earlier sessions, or wanting longer memory \\
Generic or repetitive & Generic or repetitive suggestions \\
Specific technique helped & A specific technique helped in a real situation \\
Scheduling friction & Fixed days or time windows \\
Ease of use praised & Ease or simplicity of use \\
\bottomrule
\end{tabular}
\end{table}

\begin{table}[!h]
\centering
\small
\caption{Themes in the final questionnaire's open-ended answers (percentage of respondents), with odds ratios (stat.) and $\phi$ [95\% CI], signed as in~\autoref{fig:rq3-effectsizes}. $^{\dagger}$Partly structural: only Treatment participants had physiological features and ring-specific questions. Fisher's exact tests ($^{*}p<0.05$, $^{**}p<0.01$, $^{***}p<0.001$).}
\label{tab:rq3-themes}
\setlength{\tabcolsep}{4pt}
\begin{tabular}{@{}lrrcc@{}}
\toprule
\textbf{Theme} & \textbf{Treatment (\%)} & \textbf{Control (\%)} & \textbf{stat.} & \textbf{$\phi$ [95\% CI]} \\
\midrule
Memory/data grounding praised$^{\dagger}$ & 40.7 & 0.0 & $\infty^{***}$ & 0.51 [0.36, 0.65] \\
Personalization praised & 44.4 & 14.3 & $4.80^{*}$ & 0.33 [0.09, 0.57] \\
Sensor/tracking issue$^{\dagger}$ & 48.1 & 3.6 & $25.07^{***}$ & -- \\
Trust/accuracy doubt & 14.8 & 35.7 & $0.31$ & 0.24 [$-0.01$, 0.47] \\
Memory missing & 25.9 & 28.6 & $0.88$ & -- \\
Generic or repetitive & 63.0 & 57.1 & $1.28$ & -- \\
Specific technique helped & 70.4 & 57.1 & $1.78$ & 0.14 [$-0.13$, 0.39] \\
Scheduling friction & 14.8 & 7.1 & $2.26$ & -- \\
Ease of use praised & 11.1 & 17.9 & $0.58$ & -- \\
\bottomrule
\end{tabular}
\end{table}

\begin{table}[!h]
\centering
\small
\caption{Self-reported behavior change (percentage of respondents per response), along with Cohen's $d$ and Mann--Whitney $U$ tests (``Not sure'' excluded) ($^{*}p<0.05$, $^{**}p<0.01$, $^{***}p<0.001$).}
\label{tab:rq3-behavior}
\setlength{\tabcolsep}{4pt}
\begin{tabular}{@{}llrrrrrcc@{}}
\toprule
\textbf{Item} & \textbf{Group} & \textbf{No change} & \textbf{A little} & \textbf{Somewhat} & \textbf{A lot} & \textbf{Not sure} & \textbf{stat.} & $d$ \\
\midrule
Reflection & Treatment & 3.4 & 27.6 & 58.6 & 10.3 & 0.0 & \multirow{2}{*}{$469.5$} & \multirow{2}{*}{$0.50$} \\
 & Control & 14.3 & 32.1 & 42.9 & 3.6 & 7.1 & & \\
\addlinespace
Routine & Treatment & 13.8 & 41.4 & 37.9 & 6.9 & 0.0 & \multirow{2}{*}{$460.5$} & \multirow{2}{*}{$0.42$} \\
 & Control & 39.3 & 17.9 & 32.1 & 3.6 & 7.1 & & \\
\bottomrule
\end{tabular}
\end{table}

The remaining themes were a specific technique that helped in a real situation (the most common positive theme; 70.4\% vs.\ 57.1\%; OR $= 1.78$), generic or repetitive suggestions (OR $= 1.28$), scheduling friction (OR $= 2.26$), and ease-of-use praise (OR $= 0.58$;~\autoref{tab:rq3-themes}).

Further quotes illustrate the themes in Section~\ref{sec:rq3-experience}. Beyond T27, T8 praised the data grounding: ``The connection between chat recommendation and physical metrics. It didn't just use my text input but was scanning physical state.'' In Control, C16 proposed that ``If it was combined with a wearable which asks about what happened at that point of day based on data that will be helpful.'' On missing memory, C2 found it ``very annoying that I have to tell everything from zero for each session.'' T2, one of the Treatment participants who both praised the grounding and wanted more conversational memory, wrote: ``The memory feature was useful but the span can be longer''. Among sensor and tracking issues, T12 reported runs mistaken for swimming and T3 wrong durations for cycling to the office. On grounding that felt too literal, T4 asked the chatbot to ``make it more correlative, instead of restating things (eg you moved less, you slept worse), which are already obvious, I would rather have some sort of higher level conclusions like: you did the walking last time and the night after you slept better'', and T3 observed that ``it seems to only work off of the inputs provided at the start of the session. doesn't seem to take into account data from the ring.'' In the behavior-change examples, T26 wrote: ``I knew that sleep was bad but I didn't know that it leads to my daily stress very high''; asked when the ring was accurate, T8 wrote that it ``helped me to realise that stress can be somewhat invisible for me as well''.

Because the SUS, agreement, and behavior-change outcomes and the themes are measured on different scales, we also placed them side by side on one effect-size scale as a visual summary (\autoref{fig:rq3-effectsizes}): the rank-biserial correlation $r$ (the probability that a random Treatment response exceeds a random Control one, minus the reverse) for questionnaire outcomes and the $\phi$ coefficient (the correlation, from $-1$ to 1, between group and a theme's presence) for four themes, signed so that positive values favor Treatment (trust/accuracy doubt flipped), with percentile-bootstrap 95\% CIs (2,000 within-group resamples); significance comes only from each outcome's own test. Besides the two praise themes ($\phi = 0.51$, 95\% CI [0.36, 0.65], and $\phi = 0.33$, [0.09, 0.57]), the questionnaire measures peaked for \emph{understood my situation accurately} and the two behavior-change items ($r$ between 0.22 and 0.25), all with intervals including zero.

\clearpage
\section{Additional Study Design Details}\label{sec:app-design}

This appendix supplements the study design with the DBT skills available to the chatbot (\autoref{tab:dbt_skills}), the participant characteristics (\autoref{tab:participant_demographics}), engagement and pre-session affect (\autoref{tab:engagement}), the physiological features in the memory, the onboarding and scheduling procedure, the SUS scoring, the stored session data, and the stress types in participants' reflections (\autoref{tab:stressors}).

\begin{table}[!h]
\centering
\small
\caption{DBT skills available to the chatbot (48 in total). Acronym skills are mnemonics, e.g., STOP (Stop, Take a step back, Observe, Proceed mindfully) and TIPP (changing body chemistry through Temperature, Intense exercise, Paced breathing, and Paired muscle relaxation). Grounding directs attention to present sensory input, and willingness vs.\ willfulness contrasts accepting a situation as it is with fighting it.}
\label{tab:dbt_skills}
\begin{tabular}{@{}>{\raggedright\arraybackslash}p{3.2cm}>{\raggedright\arraybackslash}p{\dimexpr\linewidth-3.2cm-2\tabcolsep\relax}@{}}
\toprule
\textbf{Module} & \textbf{Skills} \\
\midrule
Distress tolerance (13) & Cost--benefit analysis; recognizing emotional crisis; distracting activities; RESISTT; grounding; willingness vs.\ willfulness; radical acceptance; self-soothing; actions based on values; TIPP; half smile; ACCEPTS; urge surfing. \\
\addlinespace
Emotion regulation (12) & Recognizing your emotion; being effective; emotions and physical vulnerability; emotions and cognitive vulnerability; self-validation; myths about emotions; emotion exposure; balancing emotional urges; problem solving; mastery and coping ahead; STOP; ABC PLEASE. \\
\addlinespace
Interpersonal effectiveness (13) & Communication styles; assertiveness (parts 1 and 2); assertiveness scripts and interpersonal rights; listening and validation; barriers; trust in relationships; interpersonal boundaries; modulating intensity; resistance and conflict; DEAR MAN; FAST; GIVE. \\
\addlinespace
Mindfulness (10) & Observing; mental body scan; internal vs.\ external balance; describing your emotions; mental noting; thought defusion; mindful breathing; wise mind; negative judgments; letting go of judgments. \\
\bottomrule
\end{tabular}
\end{table}

\begin{table}[!h]
\centering
\small
\caption{Participant characteristics by group.}
\label{tab:participant_demographics}
\begin{tabular}{@{}lrrr@{}}
\toprule
 & \textbf{Treatment} & \textbf{Control} & \textbf{Total} \\
\midrule
Participants & 30 & 30 & 60 \\
\midrule
\emph{Gender} & & & \\
\quad Woman & 17 & 20 & 37 \\
\quad Man & 13 & 10 & 23 \\
\midrule
\emph{Age group} & & & \\
\quad 18--24 & 4 & 8 & 12 \\
\quad 25--34 & 22 & 18 & 40 \\
\quad 35--44 & 4 & 2 & 6 \\
\quad 45--54 & 0 & 2 & 2 \\
\midrule
\emph{Education} & & & \\
\quad Bachelor's & 3 & 6 & 9 \\
\quad Master's & 19 & 20 & 39 \\
\quad Doctoral & 8 & 4 & 12 \\
\midrule
\multicolumn{4}{@{}l}{\emph{Screening PSS-10}} \\
\quad Mean (SD) & 23.9 (4.4) & 22.4 (4.9) & 23.1 (4.7) \\
\bottomrule
\end{tabular}
\end{table}

\begin{table}[!h]
\centering
\small
\caption{Engagement and pre-session affect: per-participant means, then mean (SD) across participants ($n=30$ per group). Missed sessions are $13 -{}$completed sessions.}
\label{tab:engagement}
\begin{tabular}{@{}lcc@{}}
\toprule
\textbf{Measure} & \textbf{Treatment} & \textbf{Control} \\
\midrule
Completed sessions & 12.87 (0.57) & 12.87 (0.73) \\
Missed sessions & 0.13 (0.57) & 0.13 (0.73) \\
Days, first to last session & 29.93 (7.82) & 31.53 (8.50) \\
Days between sessions & 2.52 (0.63) & 2.66 (0.69) \\
\midrule
Participant turns / session & 7.01 (4.07) & 9.23 (4.16) \\
Chatbot turns / session & 8.01 (4.07) & 10.23 (4.16) \\
Participant words / session & 61.68 (45.63) & 99.41 (85.97) \\
Chatbot words / session & 582.29 (307.55) & 648.22 (291.48) \\
Reflection words / session & 21.75 (12.16) & 33.41 (28.05) \\
\midrule
Pre-session valence (1--9) & 5.11 (0.83) & 4.51 (1.18) \\
Pre-session arousal (1--9) & 5.26 (0.67) & 5.19 (1.12) \\
Pre-session dominance (1--9) & 4.83 (0.69) & 4.67 (1.07) \\
\bottomrule
\end{tabular}
\end{table}

\para{Physiological features.} The server-side service retrieved each Treatment participant's data every night at 23:59 and again just before each session, each time re-retrieving all data since the start of the study. The data comprise 13 daily data types, each from its own API endpoint, covering activity, sleep, readiness, stress, SpO\textsubscript{2} (Peripheral Blood Oxygen Saturation), and more, plus the heart rate time series. The features span six domains, listed in~\autoref{tab:app-features}. \emph{Readiness} is the ring's daily estimate of how recovered the body is; it is built from \emph{contributor scores} (0--100 ratings rather than physical measurements), such as resting heart rate, HRV balance (recent HRV relative to the person's longer-term average), body temperature deviation from the person's norm, recovery index, previous-day activity, and activity balance. Daily \emph{stress} gives the minutes of the day that the ring classifies, from physiological features such as heart rate and HRV, as high stress or as high recovery.

\begin{table}[!h]
\centering
\small
\caption{Physiological features retrieved from the ring, by domain.}
\label{tab:app-features}
\begin{tabular}{@{}>{\raggedright\arraybackslash}p{3.4cm}>{\raggedright\arraybackslash}p{\dimexpr\linewidth-3.4cm-2\tabcolsep\relax}@{}}
\toprule
\textbf{Domain} & \textbf{Features} \\
\midrule
Sleep & Total, deep, REM (Rapid Eye Movement), and light sleep duration; awake time; efficiency; latency; restless periods; bedtime; 30-second sleep-stage sequences; average breathing rate; sleep regularity \\
\addlinespace
Cardiac activity & Heart rate time series; nocturnal heart rate and HRV in 5-minute intervals; resting and lowest heart rate \\
\addlinespace
Readiness and recovery & Recovery index and recovery time; body temperature deviation and trend; activity balance; previous-day activity; daily stress and recovery durations \\
\addlinespace
Activity & Activity intensity as MET (Metabolic Equivalent of Task) values over the day; active calories; inactivity alerts; training frequency and volume; workout type, duration, intensity, calories, and distance \\
\addlinespace
Respiration & Average SpO\textsubscript{2}; breathing disturbance index \\
\addlinespace
Tags & Activity labels that users enter themselves \\
\bottomrule
\end{tabular}
\end{table}

\para{Episodic memory storage.} Every retrieved record is stored as a JSON (JavaScript Object Notation) file, with one folder per day and one file per feature. Because every nested field of the ring's data types is stored as its own feature, a day holds roughly 160 features, and a curated subset of stress-relevant features (e.g., heart rate, HRV, sleep stages, daily stress, and tags) is additionally stored under descriptive names. Pattern memory contains no clinical thresholds beyond its 5\% stability margin.

\para{Onboarding and scheduling.} Both flyers described 30--45-minute sessions on alternate days, and each flyer's QR (Quick Response) code led to the screening questionnaire. At registration, participants entered their age group, optional gender, occupation, education, and an email address for session reminders; no participant selected a non-binary or self-described gender, and most were students or researchers. Treatment participants connected their ring account through the official OAuth (Open Authorization) flow, which grants the study access to the account's data. Participants received email reminders before each session opened and at 23:00 if it was still incomplete, and Treatment sessions opened only once that day's physiological features had been retrieved and the memory built. A missed session was rescheduled two days later, so the study could take longer than four weeks for some participants. Apart from the unguided session, Treatment participants were not asked to change their routines, and our analyses do not use these recordings separately. Reflections and the description of the most helpful part of the chatbot required at least 50 characters each. All analyses use the cohort of 60 participants unless stated otherwise.

\para{SUS scoring.} The SUS statements alternate between positive (e.g., ``I thought the system was easy to use'') and negative (e.g., ``I found the system unnecessarily complex'') and are rated from 1 (Strongly Disagree) to 5 (Strongly Agree). Each positive item contributes $v-1$ and each negative item $5-v$, where $v$ is the rating; the sum is multiplied by 2.5 to give a score from 0 to 100. Scores are not percentages.

\para{Stored session data.} Each session is stored as an ordered list of participant and chatbot messages, together with the written reflection, session-level timestamps for the reflection, the pre- and post-session ratings, and submission, and the memory used to generate the conversation.

\para{Stress types.} We coded the stress types in each participant's written reflections, pooled across sessions, using a fixed taxonomy of 12 types derived from this dataset; a participant could have several types. LLM coders assigned the codes in batches of about ten participants, each given the full taxonomy and blind to the other batches' codes; the authors then verified the codes. After academic or work performance pressure (83.3\% in both groups), the most common types were workload, time pressure, or procrastination (70.0\% Treatment vs.\ 83.3\% Control) and relationship conflict (46.7\% vs.\ 53.3\%); several categories had few positive cases (\autoref{tab:stressors}).

\begin{table}[!h]
\centering
\small
\caption{Stress types described in participants' written reflections (participant-level; a participant can have several types). Cells give the percentage (number) of participants in each group.}
\label{tab:stressors}
\begin{tabular}{@{}lcc@{}}
\toprule
\textbf{Stress type} & \textbf{Treatment ($n=30$)} & \textbf{Control ($n=30$)} \\
\midrule
Academic / work performance pressure & 83.3 (25) & 83.3 (25) \\
Workload, time pressure, or procrastination & 70.0 (21) & 83.3 (25) \\
Relationship conflict (partner, family, friends) & 46.7 (14) & 53.3 (16) \\
Anticipatory anxiety before a specific event & 30.0 (9) & 46.7 (14) \\
Financial stress / job insecurity & 30.0 (9) & 36.7 (11) \\
Physical health symptoms (pain, illness, fatigue) & 33.3 (10) & 33.3 (10) \\
Uncertainty about the future / career & 33.3 (10) & 23.3 (7) \\
Other / unspecified stressor & 33.3 (10) & 23.3 (7) \\
Grief or distressing news about a loved one & 13.3 (4) & 26.7 (8) \\
Sleep difficulty (as the stressor itself) & 30.0 (9) & 10.0 (3) \\
Task / technical frustration (non-academic) & 16.7 (5) & 16.7 (5) \\
Loneliness / social isolation & 13.3 (4) & 20.0 (6) \\
\bottomrule
\end{tabular}
\end{table}

\subsection{Use of Generative AI in Preparing This Paper}\label{sec:app-genai}
We used Claude (Anthropic, Claude Opus 5.5) only to (i) help search and summarize literature for the Related Work (\autoref{section:rw}) and (ii) correct grammar and wording in text the authors wrote. Prompts were of the form ``Find and summarize recent papers on [topic]'' and ``Fix the grammar and improve the wording of this text without changing its content: [text]''. The authors verified every reference against its source, reviewed all edits, and take full responsibility for the paper.

\end{document}
\endinput

